\RequirePackage{fix-cm}
\documentclass[natbib,smallextended]  {svjour3}       
\journalname{\ssr}
\usepackage{txfonts}
\usepackage{graphicx}           
\usepackage{amssymb}            
\usepackage{mathrsfs}           
\usepackage[usenames]{color}    
\usepackage{array}              
\usepackage{longtable}          
\usepackage{natbib}             
\bibpunct{(}{)}{;}{a}{}{,}      %

\newcommand{\referee}[1]{#1}

\newcommand{\tFV}{finite volume}             \newcommand{\TFV}{Finite volume}                   \newcommand{\sFV}{FV}
\newcommand{\tFI}{helicity-flux integration} \newcommand{\TFI}{Helicity-flux integration}       \newcommand{\sFI}{FI}
\newcommand{\tTM}{discrete flux-tubes}       \newcommand{\TTM}{Discrete flux-tubes}             \newcommand{\sTM}{DT}

\newcommand{\tGV}{DeVore-Valori}             \newcommand{\TGV}{DeVore\_GV}                      \newcommand{\sGV}{DeVore\_GV}
\newcommand{\tSA}{DeVore-Anfinogentov}       \newcommand{\TSA}{DeVore\_SA}                      \newcommand{\sSA}{DeVore\_SA}
\newcommand{\tKM}{DeVore-Moraitis}           \newcommand{\TKM}{DeVore\_KM}                      \newcommand{\sKM}{DeVore\_KM}
\newcommand{\tJT}{Coulomb-Thalmann}          \newcommand{\TJT}{Coulomb\_JT}                     \newcommand{\sJT}{Coulomb\_JT}
\newcommand{\tSY}{Coulomb-Yang}              \newcommand{\TSY}{Coulomb\_SY}                     \newcommand{\sSY}{Coulomb\_SY}
\newcommand{\tGR}{Coulomb-Rudenko}           \newcommand{\TGR}{Coulomb\_GR}                     \newcommand{\sGR}{Coulomb\_GR}
\newcommand{\tCB}{connectivity-based}        \newcommand{\TCB}{Connectivity-based}              \newcommand{\sCB}{CB}
\newcommand{\tTE}{twist-number}              \newcommand{\TTE}{Twist-number}                    \newcommand{\sTE}{TN}
\newcommand{\tFT}{field-line helicity}                    \newcommand{\sFT}{FL}

\newcommand{\TLL}  {Low and Lou}                   \newcommand{\sLL}  {LL}
\newcommand{\TTD}  {Titov and D\'emoulin}          \newcommand{\sTD}  {TD}
\newcommand{\TJst} {MHD-emergence stable}          \newcommand{\sJst} {MHD-st}
\newcommand{\TJun} {MHD-emergence unstable}        \newcommand{\sJun} {MHD-un}
\newcommand{\TJdiv}{MHD-emergence stable div($\vB$)}\newcommand{\sJdiv}{MHD-st-div($\vB$)}

\newcommand{\eq}[1]{Eq.~(\ref{eq:#1})}
\newcommand{\Eq}[1]{Equation~(\ref{eq:#1})}
\newcommand{\eqs}[2]{Eqs.~(\ref{eq:#1},\ \ref{eq:#2})}
\newcommand{\eqss}[2]{Eqs.~(\ref{eq:#1} - \ref{eq:#2})}
\newcommand{\appx}[1]{Appendix~\ref{s:#1}}
\newcommand{\sect}[1]{Sect.~\ref{s:#1}}
\newcommand{\Sect}[1]{Section~\ref{s:#1}}
\newcommand{\tab}[1]{Table~\ref{t:#1}}
\newcommand{\app}[1]{Appendix~\ref{s:#1}}
\newcommand{\fig}[1]{Fig.~\ref{f:#1}}
\newcommand{\Fig}[1]{Figure~\ref{f:#1}}

\newcommand{\BE}{\begin{equation}}
\newcommand{\EE}{\end{equation}}
\newcommand{\BA}{\begin{eqnarray}}
\newcommand{\EA}{\end{eqnarray}}

\newcommand{\Nabla}{\vec{\nabla}}

\newcommand{\rmd}{{\rm d}}
\newcommand{\dz}{\, \rmd z}
\newcommand{\dxdy}{\, \rmd x \rmd y}
\newcommand{\dS}{\rmd \vec{S}}
\newcommand{\dV}{\, \rmd \mathcal{V}}
\newcommand{\surf}{{\partial \mathcal{V}}}
\newcommand{\vol}{\mathcal{V}}
\newcommand{\intz}{\int_{z}^{z_2}}
\newcommand{\ints}{\int_{\surf}}
\newcommand{\intv}{\int_{\vol}}
\newcommand{\curlA}{\Nabla \times \vA}
\newcommand{\curlAp}{\Nabla \times \vAp}
\newcommand{\divA}{\Nabla \cdot \vA}
\newcommand{\divAp}{\Nabla \cdot \vAp}

\newcommand{\curlB}{\Nabla \times \vB}
\newcommand{\divB}{\Nabla \cdot \vB}
\newcommand{\divBp}{\Nabla \cdot \vBp}

\newcommand{\vA}{\vec{A}}

\newcommand{\Ay}{A_{\rm y}}
\newcommand{\Az}{A_{\rm z}}
\newcommand{\Apz}{A_{\rm p,z}}
\newcommand{\vAp}{\vA_{\rm p}}

\newcommand{\vb}{\vec{b}}
\newcommand{\bx}{b_{\rm x}}
\newcommand{\by}{b_{\rm y}}
\newcommand{\vB}{\vec{B}}
\newcommand{\Bx}{B_{\rm x}}
\newcommand{\By}{B_{\rm y}}
\newcommand{\Bz}{B_{\rm z}}
\newcommand{\vBp}{\vB_{\rm p}}
\newcommand{\vBns}{\vB_{\rm ns}}
\newcommand{\vBs}{\vB_{\rm s}}
\newcommand{\vBd}{\vB_{\rm \delta}}

\newcommand{\vr}{\vec{r}}
\newcommand{\vJ}{\vec{J}}
\newcommand{\vn}{\vec{n}}
\newcommand{\vv}{\vec{v}}
\newcommand{\vz}{\vec{z}}
\newcommand{\vx}{\vec{x}}
\newcommand{\vX}{\vec{X}}
\newcommand{\vY}{\vec{Y}}
\newcommand{\hatn}{\hat{\vn}}
\newcommand{\hatz}{\hat{\vz}}

\newcommand{\E}{E}                    
\newcommand{\Ep}{E_{\rm p}}           
\newcommand{\Eps}{E_{\rm p,s}}        
\newcommand{\EJs}{E_{\rm J,s}}        
\newcommand{\Emix}{E_{\rm mix}}       
\newcommand{\EdivBJ}{E_{\rm J,ns}}    
\newcommand{\EdivBp}{E_{\rm p,ns}}    
\newcommand{\Efree}{E_{\rm free}}     
\newcommand{\Ediv}{E_{\rm div}}       
\newcommand{\Ens}{E_{\rm ns}}         

\newcommand{\HH}{\mathscr{H}}
\newcommand{\Hdef}{\mathscr{H}_\vol}
\newcommand{\HdefJ}{\mathscr{H}_{\vol,J}}
\newcommand{\HdefJP}{\mathscr{H}_{\vol,JP}}

\newcommand{\Hv}{H_\vol}

\newcommand{\Hj}{H_{\rm J}}
\newcommand{\Hself}{\mathscr{H}_{\rm twist}}

\newcommand{\avfi}[1]{\langle\,|f_i#1|\,\rangle}
\newcommand{\sj}{\sigma_J}
\newcommand{\epsN}{\epsilon_{\rm N}}
\newcommand{\epsM}{\epsilon_{\rm M}}
\newcommand{\epsE}{\epsilon_{\rm E}}

\newcommand{\eg}{\textit{e.g.}, }
\newcommand{\ie}{\textit{i.e.}, }

\newcommand{\articlepath}{./}
\newcommand{\strtable}{\renewcommand{\arraystretch}{1.2}} 
\newlength{\imsize}

\newcommand{\aap}{{\it Astron. Astrophys.}}

\newcommand{\apj}{{\it Astrophys. J.}}
\newcommand{\apjl}{{\it Astrophys. J. Lett.}}

\newcommand{\grl}{{\it Geophys. Res. Lett.}}

\newcommand{\physrep}{{\it Physics Reports}}

\newcommand{\solphys}{{\it Solar Phys.}}

\newcommand{\ssr}{{\it Space Sci. Rev.}}

\begin{document}
\title{Magnetic helicity estimations in models and observations of the solar magnetic field. Part~I: Finite volume methods}
\author{
   Gherardo~Valori     
   \and
   Etienne~Pariat     
   \and
   Sergey~Anfinogentov
   \and
   Feng~Chen       
   \and
   Manolis~K.~Georgoulis 
   \and
   Yang~Guo        
   \and
   Yang~Liu        
   \and
   Kostas~Moraitis   
   \and
   Julia~K.~Thalmann 
   \and
   Shangbin~Yang       
}
\institute{G.~Valori \at
   University College London, Mullard Space Science Laboratory, Holmbury St. Mary, Dorking, Surrey, RH5 6NT, U.K.   \email{g.valori@ucl.ac.uk} 
 \and E.~Pariat \at
   LESIA, Observatoire de Paris, PSL Research University, CNRS, Sorbonne Universit\'e, UPMC Univ. Paris 06, Univ. Paris Diderot, Sorbonne Paris Cit\'e, 92190 Meudon, France
 \and S.~Anfinogentov \at
   Institute of Solar-Terrestrial Physics SB RAS 664033, Irkutsk, P/O box 291; Lermontov st, 126a, Russia
 \and F.~Chen \at
   Max-Plank-Institut f\"ur Sonnensystemforschung, 37077 G\"ottingen, Germany 
 \and  M.K.~Georgoulis \at
   Research Center for Astronomy and Applied Mathematics of the Academy of Athens, 4 Soranou Efesiou Street, 11527 Athens, Greece
 \and Y.~Guo \at
   School of Astronomy and Space Science, Nanjing University, Nanjing 210023, China
 \and Y.~Liu \at 
   Stanford University HEPL Annex, B210, Stanford, CA 94305-4085 USA
 \and K.~Moraitis \at
   Research Center for Astronomy and Applied Mathematics of the Academy of Athens, 4 Soranou Efesiou Street, 11527 Athens, Greece
 \and J.K.~Thalmann \at
   Institute of Physics/IGAM, Univeristy of Graz, Universit\"atsplatz 5/II, 8010 Graz, Austria
 \and S.~Yang  \at
   Key Laboratory of Solar Activity, National Astronomical Observatories, Chinese Academy of Sciences, Beijing 100012, China
}
\titlerunning{Magnetic helicity: Finite volume methods}
\authorrunning{G.~Valori et al.}
\date{Received ***; accepted ***}

\begin{abstract}
{
Magnetic helicity is a conserved quantity of ideal magneto-hydrodynamics characterized by an inverse turbulent cascade.
Accordingly, it is often invoked as one of the basic physical quantities driving the generation and structuring of magnetic fields in a variety of 
\referee{astrophysical and laboratory plasmas.}
We provide here the first systematic comparison of \referee{six}    existing methods for the estimation of 
\referee{the helicity of magnetic fields known in a finite volume.}
All 
\referee{such}
methods are reviewed, benchmarked, and compared with each other, and specifically tested for accuracy and sensitivity to errors.
To that purpose, we 
\referee{consider} four groups of numerical tests, ranging from solutions of the three-dimensional, force-free \referee{equilibrium}, to magneto-hydrodynamical numerical simulations.
Almost all methods are found to produce the same value of magnetic helicity within few percent in all tests.
In the more solar-relevant and realistic of the tests employed here, the simulation of an eruptive flux rope, the spread in the computed values obtained by all but one method is 
\referee{only} 3\%, 
\referee{indicating}
the reliability and mutual consistency of such methods 
\referee{in appropriate parameter ranges}.
However, methods show differences in the sensitivity to numerical resolution and to errors in the solenoidal property of the input fields.
In addition to finite volume methods, we also 
\referee{briefly discuss} a method that estimates helicity from the field lines' twist, and one that exploits the field's value at one boundary and a coronal minimal connectivity instead of a pre-defined three-dimensional magnetic-field solution.
}
\end{abstract}
 \maketitle

\section{Introduction}\label{s:intro}
The volume integral
\BE
 \HH(t) \equiv \intv \left (\vA\cdot\vB \right )\dV \, 
\label{eq:Hgen}
\EE
is the helicity of the vector field $\vB=\curlA$ in a given volume $\vol$, with $\vA(\vx,t)$ representing the corresponding space- and time-dependent vector potential.
If the field consists of a discrete collection of flux tubes, $\HH$ is the winding number \citep{Moffatt1969} expressing their degree of mutual linkage.
By extension, \eq{Hgen} is a measure of the entangling (``knottedness'') of the field's streamlines.

For a given vector potential $\vA$, the addition of the gradient of a (sufficiently regular but otherwise arbitrary) scalar function, \ie the transformation  $\vA \longrightarrow \vA +\Nabla \psi$, does  not change the resulting $\vB$.
This property of the definition of $\vB$ is called gauge-invariance.
Due to this freedom in the gauge, $\HH$ is not uniquely defined, since
\BE
 \HH(\vA +\Nabla \psi)= \HH(\vA) + \ints \left( \psi \vB \right ) \cdot \dS - \intv \psi \left(\divB\right) \dV \, ,
 \label{eq:variance}
\EE
where $\dS=\rmd S\hatn$, with $\rmd S$ being the infinitesimal element of the bounding surface $\surf$ of the volume $\vol$, and $\hatn$ the outward-oriented normal to $\surf$.
Hence, $\HH$ is not gauge-invariant unless two conditions are met: 
First, the vector field $\vB$ must be solenoidal, as implied by its definition as curl of $\vA$, and, second, the volume's bounding surface $\surf$ must be a flux surface of $\vB$, \ie $(\vB\cdot\hatn)|_\surf=0$. 
When applied to a magnetic field $\vB$, the solenoidal requirement is satisfied by virtue of Maxwell equations, although possibly only to a finite extent in numerical experiments, and $\surf$ is a flux surface if no magnetic field line is threading  the boundary.
This latter requirement is rarely 
\referee{satisfied} in natural systems, and makes \eq{Hgen} of little interest for practical use.

On the other hand, helicity has the fundamental property of being  strictly conserved in ideal magneto-hydrodynamics \citep[MHD, ][]{Woltjer1958}.
Since MHD evolution in the absence of dissipation preserves the topology of the magnetic field, the field lines' winding numbers, or, more generally, magnetic helicity, cannot be changed during the evolution. 
Even more appealing is the fact that magnetic helicity, contrary to magnetic energy, is very well conserved in non-ideal dynamics as well \citep{1984GApFD..30...79B}, as expected theoretically because it cascades to large scales rather than to the small, dissipative scales \citep[see, \eg][]{Frisch1975}.
Thanks to these properties, helicity has the possibility of being used as a constraint for the magnetic field evolution.
\referee{For isolated systems, conservation of helicity effectively restricts the allowed time-evolution to helicity-conserving paths in phase-space, which, for instance, yielded the so-called Taylor hypothesis on magnetic field relaxation \citep[see, \eg][]{1986RvMP...58..741T}.
In the solar context, helicity conservation is involved in magnetic field dynamos \citep[see, \eg][]{Brandenburg2005}, as well as a potential trigger of CMEs \citep[see, \eg][]{Rust1994}}
\subsection{Relative magnetic helicity}\label{s:intro_Hdef}
In order to overcome the limitations of \eq{Hgen}, \cite{1984JFM...147..133B} and \cite{Finn85} introduced the relative magnetic helicity
\BE
\Hdef\equiv\intv \left (\vA+\vAp\right )\cdot \left (\vB-\vBp \right ) \dV \, ,
\label{eq:Hdef}
\EE
of a magnetic field $\vB$ with respect to a reference magnetic field $\vBp=\curlAp$.
Even though \eq{Hdef} allows for 
\referee{an arbitrary} reference field, here we adopt the usual choice of the electric current-free (potential) field for $\vBp$.
This choice has the following motivation: in order for $\Hdef$ in \eq{Hdef} to be gauge invariant, the input and potential fields, $\vB$ and $\vBp$, respectively, must be solenoidal,  and such that 
\BE
 \left . \large(\hatn \cdot \vB \large )\right|_\surf=
 \left . \large(\hatn \cdot \vBp\large )\right|_\surf \, . \label{eq:Bp_bc}
\EE
With such a prescription, the potential field that is chosen as a reference is uniquely defined and represents the minimal energy state for a given distribution of the normal component of the field on the boundaries \cite[see, \eg][]{2013A&A...553A..38V}.
Moreover, $\surf$ needs not to be a flux surface for $\Hdef$ to be gauge-invariant, and the definition \eq{Hdef} can be applied to arbitrary field distributions.

\cite{2003eclm.book..345B} derived a  useful decomposition of \eq{Hdef} as
\BE
\Hdef=\HdefJ+\HdefJP
 \label{eq:Hdec}
\EE
where
\BA
\HdefJ  &\equiv& \phantom{2}  \intv  \left (\vA-\vAp\right )\cdot \left (\vB-\vBp \right ) \dV \, , \label{eq:HdefJ} \\
\HdefJP &\equiv&          2   \intv             \vAp        \cdot \left (\vB-\vBp \right ) \dV \, . \label{eq:HdefJP}
\EA
The two definitions in \eqs{HdefJ}{HdefJP} have the property of being \textit{separately} gauge-invariant under the same assumptions guaranteeing  the gauge-invariance of $\Hdef$.
The first term $\HdefJ$ corresponds to the general definition of helicity \eq{Hgen}, but this time of the current-carrying part of the field only, $(\vB-\vBp)=\Nabla \times (\vA-\vAp)$. 
By construction, such a field has no normal component on the boundary, \ie has $\surf$ as  a flux surface.
The second term $\HdefJP$ has no intuitive interpretation, but is a sort of mutual helicity that  basically takes care of the flux threading $\surf$ (via the transverse component of $\vAp$), and is gauge-invariant because only the current-carrying part of $\vB$ appears.

The conservation properties of the relative magnetic helicity were numerically tested by \cite{Pariat2015}, confirming that helicity is a very well conserved quantity even in presence of very strong dissipation. 
\referee{The equation regulating the relative magnetic helicity variation rate due to dissipation and flux through the boundaries is also derived by \cite{Pariat2015}.}
In the particular case of the simulation of a jet eruption examined in that article, relative helicity is conserved more than one order of magnitude better than free energy. 

In the particular case of applications to the solar atmosphere, one additional complication is that the magnetic field cannot be measured in the solar corona with the resolution 
necessary for the computation of magnetic helicity.
The magnetic field is instead inferred by inverting spectropolarimetric measurements of emission from lower, higher-density layers of the atmosphere, yielding basically two-dimensional maps of the field vector mostly at photospheric heights, see \eg \cite{2000RvGeo..38....1L}.
Therefore, in order to compute the helicity, it is first necessary to introduce a model of the solar corona based on the observed photospheric field values. 
In this work we exclude addressing this problem directly by using numerical models of magnetic fields, in this way testing the different helicity methods in a strictly controlled environment.

In summary, magnetic helicity is a fundamental quantity of plasma physics that is almost exactly  conserved in most conditions.
This can be relevant in fusion plasmas, as well as in astrophysical ones.
In this article we focus on solar applications, but the conclusions derived are general enough to be extended  to other fields. 

In the following  we refer to the relative magnetic helicity \eq{Hdef} computed with respect to the potential field defined by the boundary condition \eq{Bp_bc} simply as helicity, and we consider $\vol$ to be a single, rectangular, 3D, finite volume.
More general formulations are possible, \eg \cite{Longcope2008} introduces a procedure for defining reference fields on multiple sub-volumes, possibly covering the entire open space. 
We also refer to the discussion and references in  \cite{Prior2014} for a physical interpretation of $\HH$ in \eq{Hgen} under several different gauges in the presence of open field lines threading opposite faces of $\vol$.

\subsection{Overview of the methods for the estimation of helicity}\label{s:intro_methods}
\begin{table*}
\referee{
\caption{Synoptic view of helicity computation methods, their properties and formulation, as described in \sect{intro_methods}. The subset of methods actually tested in this paper is listed in \tab{methods_here}.
\label{t:methods}
}
}
 \centering
 \framebox[0.49\textwidth]{
  \parbox{0.48\textwidth}{\centering \textcolor{blue}{\textbf{\TFV{} (\sFV{})}} \vspace{4pt}\\
    $
    \Hdef=\intv (\vA+\vAp ) \cdot ( \vB-\vBp  ) \dV
    $
    \vspace{3pt}
    \\ see \eq{Hdef}
    \begin{itemize}\small
         \item Requires $\vB$ in $\vol$  \eg from MHD simulations or NLFFF
         \item Compute $\Hdef$  at one time
         \item May employ different gauges (see \tab{methods_here})
    \end{itemize}\normalsize
     
  }
 }
 \framebox[0.49\textwidth]{
  \parbox{0.48\textwidth}{\centering \textcolor{blue}{\textbf{\TFI{} (\sFI{})}}  \vspace{4pt}\\
    $
        \frac{\rmd  \Hdef}{\rmd t}=2 \ints \left[ (\vAp\cdot\vB ) v_n -
                                                        \left (\vAp\cdot\vv_{\rm t}  \right ) B_n
                                                 \right] \rmd S
    $
    \vspace{8pt}
    \\ 
    \begin{itemize}\small
         \item Requires time evolution of vector field on $\surf$
         \item Requires knowledge or model of flows on $\surf$
         \item Valid for a specific set of gauge and assumptions, see \cite{Pariat2016tmp}
         \vspace{0pt}
    \end{itemize}\normalsize
  }
 }
 \\
 \framebox[0.985\textwidth]{
  \parbox{\textwidth}{ \centering \textcolor{blue}{\textbf{\TTM{} (\sTM{})}} \vspace{4pt}\\
     $
        \HH\simeq \sum_{i=1}^{M}                          \mathcal{T}_i     \Phi_i^2 +
                  \sum_{i=1}^{M} \sum_{j=1, j\neq i}^{M}  \mathcal{L}_{i,j} \Phi_i \Phi_j   \, ,
     $
     \vspace{3pt}
     \\ see \eq{H_tubes} \vspace{3pt}\\
   \parbox{0.48\textwidth}{\centering \textcolor{blue}{\textbf{\TTE{} (\sTE)}}  \vspace{4pt}\\
     $
       \HH \simeq \mathcal{T} \Phi^2
     $
     \vspace{3pt}
     \\ see \eq{H_TE}
     \begin{itemize}\small
           \item Estimation of the twist contribution to $\HH$
           \item Requires $\vB$ in $\vol$ 
           \item Requires a flux-rope-like structure for computing the twist $ \mathcal{T}$
     \vspace{12pt}
     \end{itemize}\normalsize
   }
   \parbox{0.48\textwidth}{\centering \textcolor{blue}{\textbf{\TCB{} (\sCB{})}}  \vspace{4pt}\\
     $
       \HH =
            A \sum_{i=1}^M \alpha _i \Phi _i^{2 \delta} +
            \sum _{l,m=1}^M \mathcal{L}_{lm} \Phi _l \Phi _m
     $
     \vspace{3pt}
     \\ see \eq{Hm}
     \begin{itemize}\small
           \item Requires the vector field on photosphere at one time
           \item Models the corona connectivity as a collection of $M$ force-free flux tubes 
           \item Minimal connection length principle

     \end{itemize}\normalsize
   }
  }
 }
\end{table*}
Several methods of helicity estimation are currently available.
A practical categorization, according to decreasing levels of required input information, results into
\begin{itemize}
 \item \tFV{} (\sFV{})  
 \item \tTE{} (\sTE{})
 \item \tFI{} (\sFI{})
 \item \tCB{} (\sCB{})
\end{itemize}
methods. 
In practical applications, some assumption about the unknown coronal magnetic field need to be made, and the above groups of methods essentially differ in the nature of this assumption  
\referee{and in the correspondent helicity definition.}
\referee{A synoptic view of the available methods for the estimation of magnetic helicity in finite volumes is presented in \tab{methods}.}

\textbf{\TFV{}  (\sFV{})} methods entirely rely on external techniques, such as nonlinear force-free field extrapolations or MHD simulations, to produce numerical models of the coronal magnetic field, \cite[see, \eg][]{Wiegelmann2015}. 
\referee{The ``finite volume'' characterization indicates that the methods are designed to provide the helicity value in a bounded volume, typically one employed in a 3D numerical simulation, as opposed to methods that estimate the helicity in a semi-infinite domain.}
The helicity in a given volume at a given time can be  directly computed if the  magnetic field is known at each location in $\vol$ at that time.
Therefore, \sFV{} methods are direct implementation of \eq{Hdef} which requires only the computation of the vector potentials for a given discretized magnetic field $\vB$ in $\vol$,  see \eg  \cite{2011SoPh..272..243T, 2012SoPh..278..347V, Yang2013,2013A&A...553A..43A,Rudenko2014, Moraitis2014}.
Despite the apparent straightforward task that such methods have, differences in the gauge, in the implementations, and in the sensitivity to the input discretized magnetic field may impact on the accuracy of the helicity estimation.
To test the accuracy of \tFV{} methods is the main focus of this article, and such methods are extensively described in \sect{methods}.

The \tFT{} method by \cite{Russell2015} is also using the full 3D vector magnetic field in a volume as an input to the method.
Rather than producing a single number for the value of \eq{Hdef} in $\vol$, the \tFT{} method provides the value of helicity associated to a single flux tube, and follows its evolution in time. 
In this sense, the \sFT{} method is a powerful investigating tool for studying the distribution of helicity in numerical simulations, especially those involving reconnection processes. 
Given its peculiar and focussed applications, we do not discuss this method further, and 
\referee{refer} the reader to the above-mentioned article. 

Similarly to \sFV{} methods, the \textbf{\tTE{} (\sTE{})} method \citep[see, \eg][]{Guo2010} requires in input the 3D discretized magnetic field vector. 
The method also assumes the presence of a flux rope in the coronal volume, and proceeds by relating the twist of that structure with helicity.
The level of approximation of the true helicity value that is implied by such a technique is assessed here for the first time.
Application of this method to observations can be found in \cite{2013ApJ...779..157G}.

\textbf{\TFI{} (\sFI{})} methods, do not make any assumption about the coronal field, but rather assume that the helicity accumulated in a given volume is the result of the helicity flux through the volume boundaries, from a given point in time onward, see, \eg  Eq.~(5) in \cite{1984JFM...147..133B} for negligible dissipation.
Such an estimation requires the knowledge of the time evolution of magnetic and velocity fields on the bounding surface of the considered volume (see \eg Eq.~(16) in \cite{1999PPCF...41B.167B}. The vector potential of the potential field appearing there can be derived from the field distribution on the boundaries).
Under these assumptions, in the case of negligible dissipation, no information on the  magnetic field inside the volume is necessary.

In practical applications, such methods follow the time evolution of the photospheric field and assume that the flux of helicity through  that boundary accumulates in the coronal field, see \eg \cite{2001ApJ...560..476C}.
Since only the flux is computed, \sFI{}  methods can only estimate the variation of accumulated helicity with respect to an unknown initial state.
Methods that exploit this approach appear in \cite{Nindos2003,2005A&A...439.1191P,LaBonte2007,Liu2012}, among others, but direct comparisons between them do not yet exist.

A method of computing the helicity that also uses only the distribution of magnetic field on the bottom boundary is the \textbf{\tCB{} (\sCB{})} method \citep{Georgoulis_2012}.
The method is based on modeling the unknown connectivity of the coronal field with a collection of 
slender force-free flux tubes, each with different constant force-free parameter $\alpha\equiv(\curlB)/\vB$.
More specifically, the \sCB{}  method takes in input the photospheric observations and models the coronal field as a single (linear) or a collection of (nonlinear) flux tube(s) as an integral step of the helicity computation itself.
The set of flux tubes is obtained assuming that the line connectivity is, globally, the shortest possible, mimicking the compact character of the more flare-productive active regions.
In this way, the \tCB{}  method requires as input only the knowledge of the magnetic field distribution at the photosphere, at each time.
Different flux tubes have a different value of the force-free parameter, hence the characterization of the method as ``nonlinear'', despite the simplification of neglecting the braiding  between different flux tubes that is used in summing up the helicity and energy contributions of individual flux tubes.
Therefore, the \sCB{} method is an approximate, nonlinear method that is meant  to produce a lower-limit estimation of the true helicity associated to a flux-balanced coronal field in a very fast way.
In this sense, the \sCB{} method  does not share the same purpose of \tFV{}  and \tFI{}  methods, which, in the ideal situation, are in principle capable of obtaining the true value of helicity in a volume, at the price of requiring more information in input.

\referee{Both the \sTE{} and the \sCB{} methods are based on the representation of the magnetic field as a collection of a discrete number of finite-sized flux tubes, as opposite to a continuous three-dimensional field. We therefore categorized both methods as \textbf{\tTM{} (\sTM{})} methods, see \tab{methods}.}

\subsection{Systematic comparison of methods}\label{s:intro_comparison}
\Sect{intro_methods} briefly introduced the methods currently available for the estimation of $\Hdef$.
Many of those have been independently tested and already applied to observations. 
However, the accuracy, mutual consistency, and sensitivity of these methods are not sufficiently tested, while a systematic comparison of  methods in both their theoretical as well as practical aspects is necessary.
This work is the first one of a series of papers undertaking such a task.
Beside benchmarking, the ultimate goal of this series is no less than to assess if and how can helicity be meaningfully used as a tracer of the evolution of the magnetic field in magnetized plasmas.
To that purpose, we designed a collection of progressively more complex and realistic discretized test fields, starting with 3D solutions of the force-free equations, proceeding then to time-dependent MHD simulations of flux emergence, and finally to applications to real solar observations. 

The present article is focussed on \sFV{} methods, whereas in \cite{Pariat2016tmp} we mostly focus on \sFI{} methods and the comparison with the \sCB{} method.
Subsequent  papers are dedicated to the \sTE{} method \referee{\citep{Guo2016tmp}}, and to applications to observed solar active regions.
The results of this and of subsequent articles presented in this section are the direct outcome of the International Team on magnetic helicity hosted  at ISSI-Bern in the 2014-16 period\footnote{See the web page 
\texttt{http://www.issibern.ch/teams/magnetichelicity/} for more information.}.
As a reference for future testing, we make available the data cube of each employed test field and the complete results of the analysis in tabular form, of which the plots presented here are a subset.
That material can be found in  the section Publications of the ISSI website given in the footnote.

\subsection{This article}\label{s:intro_content}
\begin{table*}
\referee{
\caption{Summary of the methods employed in this article, their short label, their categorization according to \tab{methods}, the section of this article where they are described, and their main bibliographic reference.
\label{t:methods_here}
} 
}
 \begin{tabular}{ l@{\quad} l@{\quad}  l@{\quad}  l@{\quad} l@{\quad} }
  Method &Label & Category &  Section & Reference \\
 \hline
 \tJT{}  & \sJT{}  &  \TFV{} &  \sect{JT}        & \cite{2011SoPh..272..243T} \\
 \tSY{}  & \sSY{}  &  \TFV{} &  \sect{SY}        & \cite{Yang2013}            \\
 \tGR{}  & \sGR{}  &  \TFV{} &  \sect{GR}        & \cite{Rudenko2014}         \\
 \tGV{}  & \sGV{}  &  \TFV{} &  \sect{GV}        & \cite{2012SoPh..278..347V} \\
 \tKM{}  & \sKM{}  &  \TFV{} &  \sect{KM}        & \cite{Moraitis2014}        \\
 \tSA{}  & \sSA{}  &  \TFV{} &  \sect{SA}        & Not available              \\
 \TTE{}  & \sTE{}  &  \TTM{} &  \sect{TE_method} & \cite{Guo2010}             \\
 \TCB{}  & \sCB{}  &  \TTM{} &  \sect{CB_method} & \cite{Georgoulis_2012}    
\end{tabular}
\end{table*}
In view of the large scope of the project outlined in \sect{intro_comparison}, it is necessary in the first place to determine the respective limits, field of applications, and  precision of each of the existing \sFV{} methods, and check if \sFV{} methods can be reliably used to benchmark other, more approximate  methods.

More in detail, we wish to properly quantify the reliability of $\Hdef$ estimations when the field is known in the volume $\vol$.
Such a reliability can be tested by quantifying the sensitivity and robustness of the estimations with respect to resolution, energy and helicity properties of the input field, and sensitivity to violation of the solenoidal constraint by the discretized field.

In order to compare and benchmark existing methods against the above properties in a representative variety of relevant setting, we consider test cases that confront the methods with increasing complexity, uncertainties, and realism.
We consider strongly-controlled-environment, equilibrium test-cases such as the \cite{1990ApJ...352..343L} and \cite{1999A&A...351..707T} solutions of the force-free equations.
Such tests provide basic benchmarking as they differ for helicity content, resolution, and accuracy of the solenoidal property.
Then, two series of snapshots from  MHD numerical simulations of flux emergence resulting in stable \citep{Leake2013d} and unstable \citep{Leake2014} configurations are considered. 
The flux emergence test cases are also meant to build a bridge toward \sFI{} methods studied in \cite{Pariat2016tmp}, which is more focussed in understanding how the helicity information is modified when the knowledge of the magnetic field is limited (typically to the photosphere only).

In addition to \sFV{} methods, in this article we also consider 
\referee{ two \sTM{} mehtods, namely}
the \tTE{} (\sTE{}) and the \tCB{} (\sCB{}) methods.
\referee{These methods were already used to obtain estimates of the magnetic helicity in several observational studies \citep{2013ApJ...779..157G,2007ApJ...671.1034G,Georgoulis_2012,2012ApJ...759L...4T,Tziotziou2013,Tziotziou2014,Moraitis2014}. Benchmarking \sTM{} methods together with the \sFV{} ones enables the reader to better interpret results of past and future studies applying such methods.
From a different point of view, the }
\sTE{} method is included because it requires the same information as the other \tFV{} methods, \ie the full knowledge of the magnetic field in the entire considered volume.
The \sCB{} method is included 
because, despite requiring only the photospheric vector magnetogram, it can use any available information of the coronal connectivity. 
Also, similarly to \sFV{} methods, the \tCB{} method can be applied to a single time snapshot, rather  than requiring a series of temporal snapshots, which is the case for the \sFI{} methods.
\referee{A list of the methods tested in this article is given in \tab{methods_here}.}

The methods applied in this article are described in \sect{methods}, whereas the numerical magnetic fields used as tests are discussed in \sect{tests}. 
\Sect{metrics} summarizes the diagnostic tools used for the comparing the results.
The main results of the comparison are then discussed in \sect{results}, with specific discussion of the dependence on resolution presented in \sect{resolution} and of the dependence on the solenoidal property given in \sect{divergence}. 
An overview of the \sFV{} methods results is given in \sect{discussion}, whereas the 
\referee{the comparative tests with \sTM{} methods are} presented in \sect{resultsothers}. 
Finally our conclusions are summarized in \sect{conclusions}.

\section{Helicity computation methods}\label{s:methods}
\TFV{} methods require the knowledge of the magnetic field $\vB$ in the entire volume $\vol$, and differ essentially in the way in which the vector potentials are computed.
The methods presented here compute vector potentials employing either the Coulomb gauge ($\divA=0$) or the DeVore gauge \citep[$\Az=0$,][]{2000ApJ...539..944D}. 
Due to the gauge-invariant property of \eq{Hdef}, the employed gauge should be irrelevant for the helicity value. 
It may have, however, consequences on the number and type of equations to be solved for that purpose. 
Methods using the Coulomb gauge differ in the way in which the magnetic fields and the corresponding vector potentials are split into potential and current-carrying parts. 
Hence, they differ to some extent in the equations that they solve.
Methods applying the DeVore gauge are applications of the method in \cite{2012SoPh..278..347V} that differ only in the details of the numerical implementation.
\referee{In the following, different \sFV{} methods are identified by the gauge they employ (DeVore or Coulomb), followed by the initial of the author of the reference article describing its implementation (\eg \sGR{} labels the Coulomb method described in \cite{2011SoPh..270..165R}), see \tab{methods_here}.}

All the \sFV{} methods considered in this article, except for 
\referee{the \sGR{} method}, define the reference potential field as $\vBp=\Nabla \phi$, with $\phi$ being the scalar potential, solution of
\BA
  \nabla^2 \phi =0 \, ,                \label{eq:dV_phi}     \\
  \left . \left (\hatn \cdot \Nabla \phi \right ) \right |_\surf = 
  \left . \left (\hatn \cdot \vB\right)\right |_\surf \, , \label{eq:dV_bcphi}
\EA
such that the constraint \eq{Bp_bc} is satisfied.
Errors in solving \eqs{dV_phi}{dV_bcphi} are a first source of inaccuracy for the methods.

\subsection{Methods employing the Coulomb gauge}\label{s:coulomb}
Vector potentials in the Coulomb gauge satisfy
\BA
   \nabla^2\vAp =0 \, ,    \label{eq:C_Ap}                  \\
   \divAp=0 \, , \label{eq:C_divAp}               \\
   \left . \hatn\cdot\left(\nabla \times \vAp\right ) \right |_\surf=
   \left ( \hatn \cdot \vB \right ) |_\surf                            \, ,   \label{eq:C_bcApn}
\EA
for the vector potential of the potential field, and 
\BA
   \nabla^2 \vA =-\vJ \, ,    \label{eq:C_A}                  \\
   \divA=0           \, , \label{eq:C_divA}               \\
   \left . \hatn\cdot\left(\nabla \times \vA\right ) \right |_\surf=
   \left ( \hatn \cdot \vB \right ) |_\surf                            \, ,    \label{eq:C_bcAn}
\EA
for the vector potential of the input field, where $\vJ=\Nabla\times\vB$.
The conditions \eq{C_bcApn} and \eq{C_bcAn} are the translation into vector potential representation of \eq{Bp_bc}.
The accuracy of Coulomb methods depend on the accuracy in solving numerically the above Laplace and Poisson problems.
This includes the accuracy in fulfilling numerically the gauge condition, \ie the solenoidal property of the vector potentials $\vAp$ and $\vA$.

From the computational point of view, the numerical effort required to solve for the vector potentials consists, in general, in the solutions  of \eqss{C_Ap}{C_bcApn} and \eqss{C_A}{C_bcAn}, \ie of six 3D Poisson/Laplace problems, one for each Cartesian component of the vector potentials $\vAp$ and $\vA$.

\subsubsection{\textbf{\TJT}}\label{s:JT}
In order to solve  \eqss{C_Ap}{C_bcApn} and \eqss{C_A}{C_bcAn}, appropriate additional boundary conditions for $\vA$ and $\vAp$ on the boundaries of the 3D-rectangular computational domain need to be specified. 
For this purpose, the method of  \cite{2011SoPh..272..243T}, decomposes $\vA$ into a current-carrying and a potential (current-free) part, in the form $\vA=\vA_{\rm c}+\vAp$. 
The reproduction of the input fields' flux at the volume's boundaries, $\surf$, is entirely dedicated to $\vAp$  (obeying \eqss{C_Ap}{C_bcApn}). 
The electric current distribution in $\vol$ and on $\surf$, on the other hand, are delivered by $\vA_{\rm c}$ (\eqs{C_A}{C_divA} with $\vA_{\rm c}$ instead of $\vA$ and \eq{C_bcAn} replaced by $ \left . \hatn\cdot\left(\nabla \times \vA_{\rm c}\right ) \right |_\surf=0$).

In particular, the tangential components of $\vAp$ on a particular face, $f$, of the 3D computational domain are specified  to be the 2D stream function of a corresponding Laplacian field, $\phi_f$, in the form $\vA_{{\rm p}, \bf t}=-\hatn \times\Nabla_t\phi_f$, where $\Nabla_t$ is the 2D-gradient tangential operator on the face $f$. 
The Laplacian field itself is gained by substituting in \eq{C_bcApn} and seeking the solution of the derived 2D Laplace problem  $\nabla^2_t\phi_f=-\hatn\cdot\vB$, for which boundary conditions on the four edges of each face $f$ need to be specified.
This approach of defining $\vAp$ on $\surf$ is in principle used by all Coulomb methods considered in the present study,  but the specific way in which the 2D Laplace problems are formulated is different. 

\cite{2011SoPh..272..243T} use Neumann conditions in the form $\partial_n\phi_f=c_f$, where $c_f$ is a constant along a particular face and $\partial_n$ is the derivative in the direction normal to the edge of the face $f$ (see their Sect.\ 2.1 for details).
The different $c_f$ are constructed in such a way that the total outflow through the volume's bounding surface $\surf$ is minimized. 
In this way, a vanishing tangential divergence ($\nabla_t\cdot\vec{A}_{{\rm p},\bf t}=0$) is enforced on $\surf$, and following Gauss' theorem, \eqss{C_Ap}{C_bcApn} are approximately fulfilled. 

%
Another difference of the applied Coulomb methods is how the current-carrying vector potential $\vA_{\rm c}$ is calculated and its solenoidality enforced. 
\cite{2011SoPh..272..243T}  solves \eq{C_A} for $\vec{A}_{\rm c}$ numerically, similar to \cite{2013SoPh..283..369Y}, just with differing boundary conditions. 
In the \sJT{} case,  $\nabla_t\cdot\vec{A}_{\rm c, \bf t}=0$ on $\surf$ is explicitly enforced in order to fulfill  the Coulomb gauge for $\vA_{\rm c}$.


The method discussed in \cite{2011SoPh..272..243T} is implemented in C. 
The Poisson and Laplace problems are solved numerically using the Helmholtz solver in Cartesian coordinates of the Intel\textsuperscript{\textregistered} Mathematical Kernel Library. 

\subsubsection{\textbf{\TSY}}\label{s:SY}
The \sSY{} method is described in \cite{2013SoPh..283..369Y, Yang2013}.
In the original formulation of  \cite{2013SoPh..283..369Y}, the method requires a balanced magnetic flux through each of the side boundaries of the volume.
This restriction has been further removed in \cite{Yang2013}.
In order to solve \eqs{C_Ap}{C_bcApn}  and \eqs{C_A}{C_bcAn}, the \sSY{} method additionally enforces the boundary condition $(\hatn \cdot \vAp)|_\surf=(\hatn \cdot \vA)|_\surf=0$ at all boundaries. 
Then, the transverse vector potential at the boundaries and the vector potential at the edges is obtained by using Gauss’ theorem. 
After obtaining the boundary values, \cite{2013SoPh..283..369Y,Yang2013} firstly resolve the Laplace \eq{C_Ap} and the Poisson \eq{C_A} to obtain  an initial guess of the solution, $\vAp’$ and $\vA’$. 
These preliminary solutions satisfies \eq{C_bcApn} and \eq{C_bcAn}, but not the Coulomb gauge condition.
The \sSY{} method then uses a divergence-cleaning technique based on the Helmholtz vector decomposition to iteratively impose the Coulomb constraint to the vector potentials, without modifying their values at the boundaries.
Comparing with the \sJT{} method, in \sSY{} are the vector potentials that are decomposed, rather than the boundary contributions. 
The method is implemented in Fortran; Poisson and Laplace problems are solved numerically using the Helmholtz solver in Cartesian coordinates of the IMSL\textsuperscript{\textregistered} (International Mathematics and Statistics Library).

\subsubsection{\textbf{\TGR}}\label{s:GR}
The \sGR{} method is described in \cite{2011SoPh..270..165R}.
A distinctive feature of the algorithm is that the \sGR{} method defines the reference potential field in terms of vector potential $\vBp=\curlAp$, rather than using \eqs{dV_phi}{dV_bcphi}.
The corresponding boundary value problem, \eq{C_Ap},  is solved with the constraint \eq{C_divAp} and the boundary condition $(\vAp \cdot \hatn ) |_\surf=0$.
The Laplace problem is divided into six sub-problems, one for each side of $\vol$. 
\referee{
 Such  a splitting  of the Laplace problem is correct only if the total magnetic flux is zero (balanced) on       each side of the box independently. To satisfy this requirement, a compensation field
 $\vBp^\mathrm{m} = {\Nabla}\times \vAp^\mathrm{m}$
 is introduced. It is built as a field of 5 magnetic monopoles located outside of the box. Positions and charge of the monopoles are selected such as to compensate unbalanced flux on each side of the volume independently.  The modified magnetic field $\vB'= \vB
  - \vBp^\mathrm{m} $  has zero total flux on each face independently  and can be correctly used as a boundary condition for the sub-problems
 \begin{equation}\label{eq_sa:sub-problem_bc}
 \hatn\cdot \left(\Nabla \times \vAp^{f_i}\right)\Big\vert_{f_j} = \delta_{ij}(\hatn\cdot \vB')\mid_{f_j} \, ,
 \end{equation}
}
where $\vAp^{f_i}$ is the  vector potential of sub-problem solution corresponding to the side $f_i$.

\referee{
After solving all  sub-problems, the} full solution is then obtained by summation of the solutions of the six  sub-problems $\vec{A}^{f_i}$ and the vector potential of a compensation field, $\vec{A}^\mathrm{m}$, as
\BE
\vAp = \vAp^\mathrm{m}  + \sum_{i =1}^{6}\vAp^{f_i} \, .
\label{eq:GR_subproblems}
\EE
The field described by the first term of \eq{GR_subproblems}, instead, is flux balanced on each side of the box independently.

Instead of solving numerically the Poisson problem \eqss{C_A}{C_bcAn}, the \sGR{} methods adopts a decomposition similar to the one in \sJT{} method, \ie $\vA=\vA_{\rm c}+\vAp$, but the vector potential of the current-carrying part of the field is computed as
\BE
 \vA_{\rm c}(\vr)=-\frac{1}{4\pi}\int_\vol\frac{\vr -\vr'}{|\vr-\vr'|^3} \times \left ( \vB -\vBp \right ) \dV \, .
 \label{eq:GR_Ac}
\EE
In contrast to other methods, the solutions of the Laplace and Poison problems for the $\vec{A}^{f_i}$ components  are derived analytically as decompositions into a set of orthonormal basis functions. 
The detailed description of the strategy for solving these equations can be found in the original paper by \cite{2011SoPh..270..165R}.

In the current implementation the method is relatively demanding in terms of running time. 
Therefore, it is applied here only to a subset of test cases.

\subsection{Methods employing the DeVore gauge}\label{s:devore}
Using DeVore gauge $\Az=0$ \citep{2000ApJ...539..944D}, \cite{2012SoPh..278..347V} derived the expression for the vector potential of the magnetic field $\vB$ in the finite volume $\vol=[x_1,x_2]\times[y_1,y_2]\times[z_1,z_2]$ as
\BE
  \vA=\vb +\hatz \times \intz \vB \dz' \, ,   \label{eq:dV_A}
\EE
where the integration function  $\vb(x,y)=\vA(z=z_2)$ obeys to
\BE
  \partial_x \by -\partial_y \bx=\Bz(z=z_2) \, ,     \label{eq:dV_a}
\EE
and $b_{\rm z}=0$.
The particular solution of \eq{dV_a} employed here is 
\BA
  \bx=-\frac{1}{2} \int_{y_1}^{y} \Bz(x,y',z=z_2) \, \rmd y' \,            ,  \label{eq:dV_bx} \\
  \by=\phantom{-} \frac{1}{2} \int_{x_1}^{x} \Bz(x',y,z=z_2) \, \rmd x' \, ,  \label{eq:dV_by}
\EA
but see \cite{2012SoPh..278..347V} for alternative options.
The above equations are applied in the computation of the vector potential of the potential field too  by substituting $\vB$ with $\vBp$ everywhere in \eqss{dV_A}{dV_by}.
In particular, using \eqs{dV_bx}{dV_by} for both $\vAp$ and $\vA$ implies $\vAp=\vA$ at $z=z_2$, although this is not necessarily required by the method.

The DeVore gauge can be exactly imposed also in numerical applications, which is generally not the case for the Coulomb gauge.
On the other hand, since $\Az=0$, then
\BE
 \Bx=-\partial_z \Ay = \partial_z \intz \Bx \dz' \, ,
\EE
where \eq{dV_a} and (\ref{eq:dV_bx}, \ref{eq:dV_by}) where used; a similar expression holds for  $\By$.
Hence, the accuracy of DeVore method in reproducing  $\Bx$ and $\By$ from $\vA$ of \eq{dV_A} depends only on how accurately the relation
\BE
   \partial_z \intz  = 
\mbox{ identity} \,
   \label{eq:dV_integral}
\EE
is verified numerically.
On the other hand, even when \eq{dV_integral} is 
\referee{obeyed to acceptable accuracy}, one can easily show that, for a non-perfectly solenoidal field $\vBns$, it is 
\BE
\vBns -\Nabla \times \vA=\hatz\intz \left( \Nabla \cdot \vBns \right ) \dz' \, ,
\EE
as derived in Eq.~(B.4) of  \cite{2012SoPh..278..347V}.
Hence, the accuracy in the reproduction of the $z$-component of the field  depends on the solenoidal level of the input field (and on how accurately  \eq{dV_a} is solved).  

All DeVore-gauge methods discussed in this study employs \eqs{dV_phi}{dV_bcphi} and \eqss{dV_A}{dV_by}, but they differ in the way integrals are defined, and in the way the solution to \eq{dV_phi} is implemented. 
The computationally most demanding part of the method is the solution of the 3D scalar Laplace equation for the computation of the potential field, \eq{dV_phi}.
This makes DeVore methods computationally appealing since they require very little computation time.

\subsubsection{\textbf{\TGV{}}} \label{s:GV}
\TGV{} is the original implementation described in \cite{2012SoPh..278..347V}, where the requirement \eq{dV_integral} is enforced by defining the $z$-integral operator as the numerical inverse operation of the second order central differences operator, see Section~4.2 in \cite{2012SoPh..278..347V}.
The Poisson problem for the determination of the scalar potential $\phi$ in \eqs{dV_phi}{dV_bcphi} is solved numerically using the Helmholtz solver in the proprietary Intel\textsuperscript{\textregistered} Mathematical Kernel Library (MKL). 

Following Eq.~(39) in \cite{2012SoPh..278..347V}, the DeVore gauge for the potential field can be reduced to the Coulomb gauge. 
We checked the effect of this gauge choice in the tests below, and found no significant difference with the standard DeVore gauge.
The DeVore-Coulomb gauge for the potential field is thus no further discussed here. 

\subsubsection{\textbf{\TKM{}}} \label{s:KM}
\TKM{} is described in \cite{Moraitis2014}. \
This implementation has two main differences with the one of \sGV{}. 
The first is in the solver of Laplace's equation.
\sKM{} uses the routine HW3CRT that is included in the freely available FISHPACK library \citep{Swartztrauber79}. 
A test, however, with the corresponding Intel MKL solver revealed minor differences in the solutions obtained with the two routines, and a factor of $\leq 2$ more computational time required by the FISHPACK solver.
The second and most important difference with the \sGV{} method is in the numerical calculation of integrals and derivatives in \eqss{dV_A}{dV_by}.
In \sKM{} integrations are made with the modified Simpson's rule of error estimate $1/N^4$ \citep{Press92}, with $N$ being the number of integration points, and, in the special case $N=2$, with the trapezoidal rule instead. 
In addition, differentiations are made using the appropriate (centered, forward or backward) second-order numerical derivative, without trying to numerically realize \eq{dV_integral}.
Finally, \eqss{dV_A}{dV_by} in \sKM{} are used in the same way for both the potential and the reference fields.

\subsubsection{\textbf{\TSA}}\label{s:SA}
\sSA{} follows the general scheme of the \sGV{} method  with two differences. 
The first one is that the \eqs{dV_phi}{dV_bcphi} for the potential $\phi$ are solved in Fourier space separately for all faces of the box.
In particular, the problem is divided into six sub-problems using
\BE
\phi = \phi^c + \sum_{i=1}^{6}\phi_i \, ,
\label{eq:SA_subproblems}
\EE
where $\phi^c$ is the 3D scalar potential of the compensation field $\vB^\mathrm{c}=\Nabla\phi_c$ and $\phi_i$ are 3D solutions for the potential field with the normal component given on $i^{st}$ side of $\vol$ and vanishing boundary conditions on the other sides of $\vol$.
The individual Laplace problems for each $\phi_i$ are then solved in Fourier space following the general scheme of the potential and linear force-free field extrapolation employing the fast Fourier transform by \cite{1981A&A...100..197A}.
For the application here, the original extrapolation algorithm is modified to take into account the imposed boundary conditions.This method of solving  \eqs{dV_phi}{dV_bcphi} will be described in a dedicated forthcoming paper.

The second difference with the \sGV{} method is that  \eq{dV_A} is modified by introducing a new integration function  $\vec{c}$ that is computed using $B_z$  from any level $z_r$ inside the data cube. 
In particular, by addition and subtraction to \eq{dV_A}, one has
\BE
\vec{A} =\vec{b} +\vec{\hat{z}} \times \left ( \int_z^{z_2} \vec{B} dz^\prime + \int_{z_r}^{z_2} \vec{B} dz^\prime  - \int_{z_r}^{z_2} \vec{B} dz^\prime \right )  \, ,
\label{eq:SA_A}
\EE
which can be re-casted as
\BE
\vec{A} = \vec{c} +\vec{\hat{z}} \times \left( \int_z^{z_2} \vec{B} dz^\prime   - \int_{z_r}^{z_2} \vec{B} dz^\prime \right)  \, ,
\label{eq:vpot_from_zr} \\
\EE
where we have defined 
\BE
 \vec{c}=\vb + \vec{\hat{z}} \times \int_{z_r}^{z_2} \vec{B} dz^\prime \, .
 \label{eq:c_coeff}
\EE
Taking the $x$- and $y$-derivatives of \eq{c_coeff}, and using \eq{dV_a} and $\divB=0$, one derives 
\BE
\partial_x c_y - \partial_y c_x = B_z(z = z_r)  \, .
\label{eq:equation_for_c}
\EE
	
The solution of \eq{equation_for_c} is then analogous to \eqs{dV_bx}{dV_by}, where $B_z(z~=~z_2)$ is replaced by $B_z(z~=~z_r)$. 
Tests using the \sLL{} case of \sect{LL} shown that the minimal error in $\vAp$ and $\vA$ is obtained for $z_r = (z_2-z_1)/2$.
The vector potential is finally computed following \eq{vpot_from_zr}.

This scheme coincides with the original one of \sGV{} if $z_r$ is taken at the  top boundary of the box, \ie for $\vec{c}(z_r = z2) = \vb$.

\subsection{\TTM{} methods}\label{s:othermethods}
\cite{1984JFM...147..133B} and \cite{Demoulin2006a} have shown that the relative magnetic helicity can be approximated as the summation of the helicity of $M$ flux tubes:
\BE
\HH\simeq \sum_{i=1}^{M}                          \mathcal{T}_i     \Phi_i^2 + 
          \sum_{i=1}^{M} \sum_{j=1, j\neq i}^{M}  \mathcal{L}_{i,j} \Phi_i \Phi_j   \, ,
\label{eq:H_tubes}
\EE
where $\mathcal{T}_i$ denotes the twist and writhe of magnetic flux tube $i$ with flux $\Phi_i$, and $\mathcal{L}_{i,j}$ is the linking
number between two magnetic flux tubes $i$ and $j$ with fluxes $\Phi_i$ and $\Phi_j$, respectively. 
The first and second term on the right hand side of \eq{H_tubes} represents the self and mutual helicity, respectively.
With the approximation of discrete magnetic flux tubes, the physical quantity of the magnetic helicity is related to the topological concept of the writhe, twist and linking number of curves, and the magnetic flux associated with those curves. 
The formulae of computing these topological quantities for both close and open curves have been derived in \cite{Berger2006} and \cite{Demoulin2006a}.
\referee{For the purpose of our comparison it must be noticed that \tTM{} methods do not provide the vector potentials and potential fields in the considered volumes like \sFV{} methods. Therefore, the comparison between \sTM{} methods and \sFV{} methods is necessarily restricted to the helicity values only.}
The \tTE{} method and \tCB{} method presented in this section adopt different assumptions in the helicity formulae and the magnetic field models to compute the magnetic helicity.

\subsubsection{\textbf{\TTE}} \label{s:TE_method}
The \sTE{} method is described in \cite{Guo2010} and \cite{,2013ApJ...779..157G}.
This method is aimed at computing the helicity of a highly twisted magnetic structure, such as a magnetic flux rope.
A magnetic flux rope is considered as \referee{an isolated,} single flux tube such that only the self magnetic helicity is computed. 
The helicity contributed by the writhe is also omitted \referee{assuming that} the flux rope is 
not highly kinked. 
With these two assumptions, the magnetic helicity of a single highly twisted structure is simplified as
\BE
\Hself \simeq \mathcal{T} \Phi^2 \, ,
\label{eq:H_TE}
\EE
where $\mathcal{T}$ is the twist number of the considered magnetic flux rope with flux $\Phi$. 
In order to estimate $\mathcal{T}$, the formula derived in \cite{Berger2006} to compute the twist number of a sample curve referred to an axis is employed. 
Practically, the axis can be determined by the symmetry of a magnetic configuration or by other assumptions, such as requiring it to be horizontal and to follow the polarity inversion line \citep{Guo2010}. 
The boundary 
\referee{of the flux rope} is determined by the quasi-separatrix layer (QSL)
that is found to wrap the flux rope \citep{2013ApJ...779..157G}.
Then the twist density, $\rmd \mathcal{T}/\rmd s$, at an arc length $s$ is:
\BE
\frac{\rmd \mathcal{T}}{\rmd s} = \frac{1}{2 \pi} \vec{T} \cdot \vec{V} \times \frac{\rmd \vec{V}}{\rmd s} \, .
\label{eq:twist_TE}
\EE
Two unit vectors are used in \eq{twist_TE}:  $\vec{T}(s)$, that is tangent to the axis curve, and $\vec{V}(s)$, that  is normal to $\vec{T}$ and pointing from the axis curve to the sample curve. 
By integrating the equation along the axis curve the total twist number is derived. 
\Eq{twist_TE} is suitable for smooth curves in arbitrary geometries without self intersection. 
Since it makes no assumption about the magnetic field, it can be applied to both force-free and non-force-free magnetic field models.

\subsubsection{\textbf{\TCB}}\label{s:CB_method}
The \sCB{}  method was introduced by \cite{2012ApJ...759....1G} and was used by a number of studies thereafter. 
In principle, the method requires only the full (vector magnetic field) photospheric boundary condition to self-consistently estimate a lower limit of the free energy and the corresponding relative helicity. 

A key element of the method is the discretization of a given, continuous photospheric flux distribution into a set of partitions with known spatial extent and flux content. 
Each partition is then treated as the collective footprint of one or more flux tubes. 
To map the relative locations of these footprints, one either infers or calculates the coronal magnetic connectivity that distributes the partitioned magnetic flux into opposite-polarity connections, treated thereafter as discrete magnetic flux tubes. 
The flux content of these connections, with both ends within the photospheric field of view (FOV), constitutes the magnetic connectivity matrix corresponding to the given photospheric boundary condition. 

The unknown coronal connectivity is either inferred by any explicit solution of the volume magnetic field or calculated with respect to the existing photospheric boundary condition. 
In the first case, individual field-line tracing associates connected flux with photospheric partitions, providing the magnetic connectivity matrix upon summation of individual field-line contributions. 
Obviously, only magnetic field lines entirely embedded in the finite volume are taken into account. 
In the second case, a simulated-annealing method is used to absolutely and simultaneously minimize the flux imbalance (hence achieving connections between opposite-polarity partitions) and the (photospheric) connection length.  
This criterion is designed to emphasize photospheric magnetic polarity inversion lines by assigning higher priority to connections alongside them. 
The converged simulated-annealing solution, that provides the connectivity matrix, is unique for a given photospheric partition map. 
More information and examples are provided in  \cite{2012ApJ...759....1G}  and  \cite{2012ApJ...759L...4T,Tziotziou2013}.

The connectivity matrix in a collection of partitions of both polarities will reveal a number of $M$  discrete, assumed slender, flux tubes with flux contents $\Phi_i$; $i \equiv \{ 1, ..., M \}$. 
The respective force-free parameters $\alpha _i$ are assumed constant for a given flux tube but vary between different tubes, thus implementing the nonlinear force-free (NLFF) field approximation. 
Force-free parameters for each flux tube are the mean values of the force-free parameters of the tubes' respective footprints, each calculated by the relation $\alpha _i={{4 \pi} \over {c}} {{I_i} \over {\mathcal{F}_i}}$; $i \equiv \{ 1, ..., M' \}$ for $M'$ magnetic partitions, where $I_i$ is the total electric current of the $i$-partition and $\mathcal{F}_i$ its flux content. 
The total current is calculated by applying the integral form of Amp\'{e}re's law along the outlining contour of the partition. 

Knowing $\mathcal{F}_i$, $\alpha _i$, and the relative positions of each flux tube's footpoints, \cite{2012ApJ...759....1G} showed that a lower limit of the free magnetic energy for a collection of $M$ flux tubes is 

\BE
E_{c_{CB}} \equiv E_{c_{(CB; self)}} + E_{c_{(CB; mutual)}} = 
A \lambda ^2 \sum_{i=1}^M \alpha _i^2 \Phi _i^{2 \delta} + {{1} \over {8 \pi}} 
\sum _{l=1}^M \sum _{m=1; l \ne m}^M \alpha _l \mathcal{L}_{lm} \Phi _l \Phi _m\;\;,
\label{eq:Ec}
\EE

where $A$, $\delta$ are known fitting constants, $\lambda$ is the length element (the pixel size in observed photospheric magnetograms), and $\mathcal{L}_{lm}$ is the mutual-helicity parameter for a pair $(l, m)$ of flux tubes. 
This parameter is inferred geometrically, by means of trigonometric interior angles for the relative positions of the two pairs of flux-tube footpoints. 
The locations of point-like footpoints of the slender flux tubes coincide with the flux-weighted centroids of the respective flux partitions. 
As included in \eq{Ec}, the parameter  $\mathcal{L}_{lm}$ does not include braiding between the two flux tubes, that can be found only by the explicit knowledge of the coronal connectivity. 
Additional complexity via braiding will only add to the free energy $E_{c_{CB}}$ in \eq{Ec}. 
Therefore, the above $E_{c_{CB}}$ is already a lower limit of the actual $E_c$, assuming only "arch-like" (i.e., one above or below the other) flux tubes that do not intertwine around each other. 
In addition, \eq{Ec} does not include an unknown free-energy term that is due to the generation, caused by induction, of potential flux tubes around the collection of non-potential ones \citep{Demoulin2006a}.
Such a term would again contribute to the mutual term of the free energy. 

The corresponding self-consistent relative helicity is, then, 

\BE
H_{m_{CB}} \equiv H_{m_{(CB; self)}} + H_{m_{(CB; mutual)}} = 
8 \pi A \lambda ^2 \sum_{i=1}^M \alpha _i \Phi _i^{2 \delta} + 
\sum _{l=1}^M \sum _{m=1; l \ne m}^M \mathcal{L}_{lm} \Phi _l \Phi _m\;\;.
\label{eq:Hm}
\EE

From \eqs{Ec}{Hm} we identically have $E_{c_{CB}} \equiv 0$ for potential flux tubes ($\alpha _i =0$; $i \equiv \{ 1,...,M \}$). For $H_{m_{CB}}=0$ in this case, we further require 
$\sum _{l=1}^{M_p} \sum _{m=1; l \ne m}^{M_p} \mathcal{L}_{lm_P} \Phi _l \Phi _m =0$, where $\mathcal{L}_{lm_P} \ne \mathcal{L}_{lm}$ is the mutual-helicity factor for a collection of $M_p \ne M$ collection of potential flux tubes. 
As \cite{Demoulin2006a} discuss, this can be the case for a flux-balanced potential-field boundary condition. 
In practical situations of not-precisely flux-balanced magnetic configurations, however, one may approximate $H_{m_{CB; mutual}} =0$, in case {\it all} $\alpha _i$; $i \equiv \{ 1,...,M \}$ are zero within uncertainties $\delta \alpha _i$, which are fully defined in this analysis. 
More generally, one may use the ``energy-helicity diagram'' correlation of \cite{2012ApJ...759L...4T,Tziotziou2014} to infer $| H_{m_{CB}}| \propto E_{c_{CB}}^{0.84 \pm 0.05}$ for $E_{c_{CB}} \longrightarrow 0$. 

\section{Analysis metrics}\label{s:metrics}
Apart from extremely simplistic magnetic fields, the analytical computation of the relative magnetic helicity in a non-magnetically bounded system is highly non-trivial. 
Even with simple natural-world-relevant models relative magnetic helicity cannot be analytically estimated.
Similarly, the exact value of $\Hv$ in the finite volume of the 3D discretized magnetic fields used here as tests is, in general, not known. 
Hence, we need to provide indirect accuracy metrics to judge the examined methods. 

The main goal of the analysis presented below is to compare the helicity values that are obtained employing the potential field and vector potentials computed with the methods described in \sect{coulomb} and \sect{devore}.
Since the helicity of $\vB$ defined by \eq{Hdef} involves the corresponding potential field $\vBp$, as well as the vector potentials for $\vBp$ and $\vB$, this basically implies providing a quantitative estimation of the accuracy of such fields.
To that purpose, we introduce normalized quantities and metrics as follows:

For each discretized magnetic field, we define $\Hv$ as the helicity defined in \eq{Hdef} normalized to $\Phi^2$, 
\BE
  \Hv \equiv \Hdef / \Phi^2 \, ,
\label{eq:H}
\EE
where
\BE
  \Phi(\vB)=\frac{1}{2}\int_{\rm z=0} \large | \Bz(x',y',z=0) \large |\dxdy 
\label{eq:flux}
\EE
is half of the unsigned flux through the bottom boundary, corresponding to the injected flux for an exactly flux-balanced configuration.
\referee{In that normalization, a uniformly twisted flux rope with field lines having $N$ turns has an helicity equal to $N$ \citep[see \eg][]{2009AdSpR..43.1013D}.}
In computing the helicity values with different methods we refrain from using simplifications of \eq{Hdef} coming from the specific gauge in use,  in this way keeping the comparison as general as possible.
Hence, for each \sFV{} method and for each test case, the value of $\Hv$ as defined by \eq{H} is obtained by computing separately the four volume contributions of \eq{Hdef} and normalizing them to $\Phi^2$.
We reserve the calligraphic symbol $\HH$ for  non-normalized helicities.

The numerically obtained $\Hv$ values depend in principle on many factors. 
In the first place, different methods may have a different level of accuracy in computing the vector potentials of the test and potential fields, depending on the strategy applied to solve the relevant equations, see \sect{methods}.
Second, the reference potential field is uniquely defined by the requirement that $\Hv$ is gauge invariant, yielding to \eq{Bp_bc}. 
However, without violating that requirement, the potential field can equivalently be computed both as the curl of the vector potential, as in some methods of \sect{coulomb}, or as the gradient of the scalar potential.
Numerically, the derived potential fields may not be identical for different methods.
Finally, since helicity estimation methods are  developed for application to research-relevant dataset, all employed tests are defined on discretised grids of moderate- to high-resolution, and therefore violate the solenoidal property to some extent \cite[cf.][]{2013A&A...553A..38V}.
Different methods might be affected differently by small violations of the solenoidal property of the test field.
Substantial violations of the solenoidal property are not considered here since the very definition of $\Hv$ is devoid of meaning in that case.  

We report in Tables~\ref{t:lltd}-\ref{t:stun_div} the complete listing of all employed metrics for all \referee{\sFV{}} methods and \referee{test} cases considered in this study. 
On the other hand, in the next sections we provide concise  summaries of the tables' values for subsets of test cases and/or  methods. 
To this purpose, we compute the mean of the relevant $\Hv$ values, and a relative spread around it, defined as the standard deviation of the $\Hv$ values distribution over the mean.
In addition, in order to discern among different factors influencing $\Hv$ values, several diagnostic metrics are here introduced.

\subsection{Accuracy of vector potentials}\label{s:fom}
The vector potentials required in the helicity computation of \eq{H} must reproduce the correspondent magnetic fields as accurately as possible.
In order to compare two vector fields $\vX$ and $\vY$ in $\vol$ we employ the metrics 
\BA
\epsN =1-\frac{\sum_i|\vX_i-\vY_i|}{\sum_i|\vX_i|} \, , \label{eq:En}\\
\epsE =\frac{\sum_i|\vY_i|^2}{\sum_i|\vX_i|^2}   \, , \label{eq:eps}
\EA
which are, respectively, the complement of the normalized vector error and the energy ratio, introduced by \cite{2006SoPh..235..161S}\footnote{We use the notation  $\epsN$ rather than $1-E_{\rm N}$ to avoid confusion with energy symbols}.
Both are unity if $\vX_i=\vY_i$ in all grid points $i$ in $\vol$.
The metrics are applied to the pair $(\vX=\vBp,\vY=\curlAp)$ or  $(\vX=\vB,\vY=\curlA)$, to quantitatively  assess the accuracy of a vector potential in reproducing the corresponding magnetic field.
Additional metrics defined in \citet{2006SoPh..235..161S} are either not particularly sensitive, or not providing essential additional information in the cases examined below. 
For the interested reader, they are listed in \appx{tables}.

For the metrics defined here, as well as for the integral in \eq{H}, we use standard numerical prescriptions, as those in Appendix~A of \cite{2013A&A...553A..38V}.
In particular, we compute the curl and divergence operators using a second-order, central-difference discretization scheme for points in the interior of $\vol$. 
Values on the volume-bounding surface, $\surf$, are taken from the input magnetic fields.

\subsection{Quantification of the solenoidal property}\label{s:thomson}
The test cases used in this article must have a value of $\divB$ small enough to be considered numerically solenoidal.
In order to quantify the level of solenoidality, we apply here the decomposition of the energy of the magnetic field in $\vol$ into solenoidal and nonsolenoidal contributions, as in  \cite{2013A&A...553A..38V}.
Using that decomposition, a fraction  $\Ediv$ of the total magnetic energy can be associated to the nonsolenoidal component of the field.
In this article, all energy contributions are  normalized to the total energy, $\E$, of the test case in exam.
More details on the decomposition can be found in \appx{energy}.

For reference, we occasionally include the divergence metric proposed in \cite{2000ApJ...540.1150W} and often used in the literature to test the solenoidal property of discretized fields, defined as the average over all $n$ grid nodes, $\avfi{(\vB)} =(\sum_i |f_i|)/n$, of the fractional flux,
\BE
f_i(\vB)=\frac{\int_v \Nabla\cdot\vB_i \, dv}{\int_{\partial v} |\vB_i| \, dS} \simeq
\frac{\Nabla\cdot\vB_i}{6|\vB|_i /\Delta}\, ,
\label{eq:fi}
\EE
through the surface $\partial v$ of an elementary volume $v$ including the node $i$. 
The rightmost expression in \eq{fi} holds for a grid of uniform and homogeneous resolution $\Delta$.
Therefore, it may be appropriately used as a metric for the methods analysed in this study, since they all are based on uniform Cartesian grids.
The smaller  the value of  $\avfi{}$, the more solenoidal the field. 
However, the actual  value of this metric depends on the number of grid points $n$ and the resolution $\Delta$, therefore it makes most sense to apply it to identical discretized volumes.
In addition, this metric is used when the energy one is not applicable, for instance, when checking the solenoidal property of vector potentials in Coulomb gauge methods, $\avfi{(\vA)}$, as in \sect{divergence}.

\section{Test fields}\label{s:tests}
 \setlength{\imsize}{0.49\textwidth}
 \begin{figure*}
   \centering
   \includegraphics[width=\imsize,clip=true]{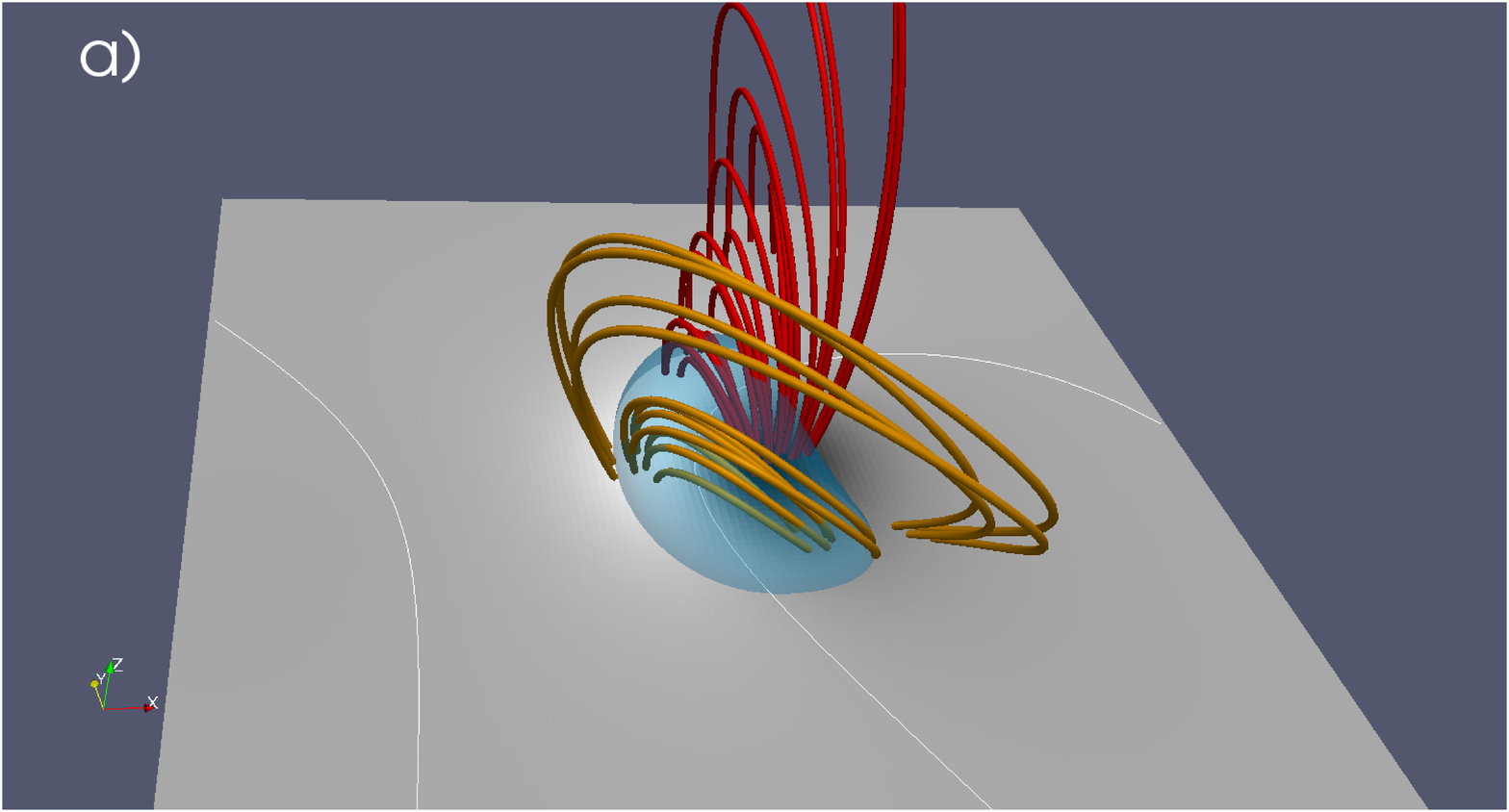}
   \includegraphics[width=\imsize,clip=true]{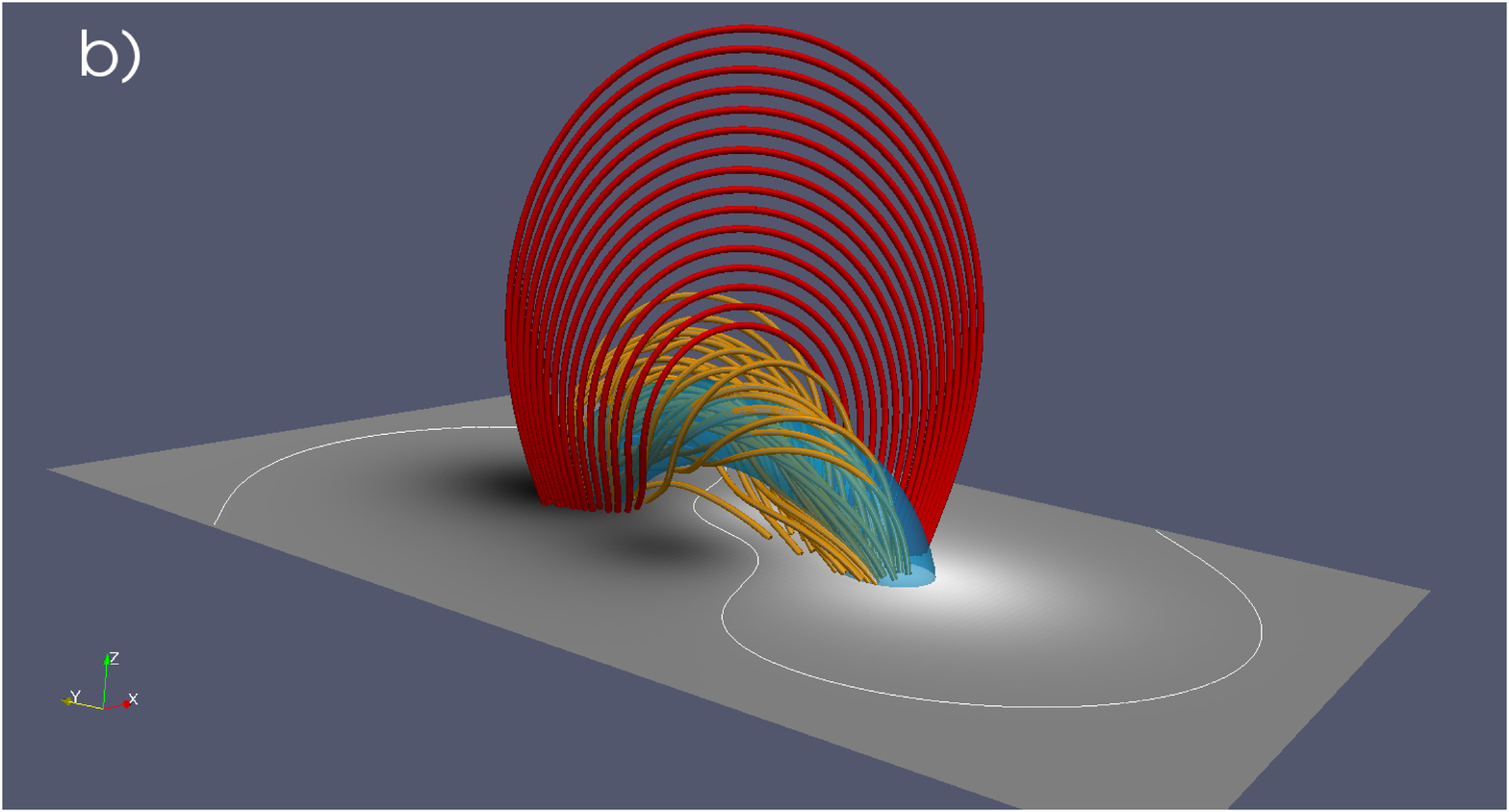}\\
   \includegraphics[width=\imsize,clip=true]{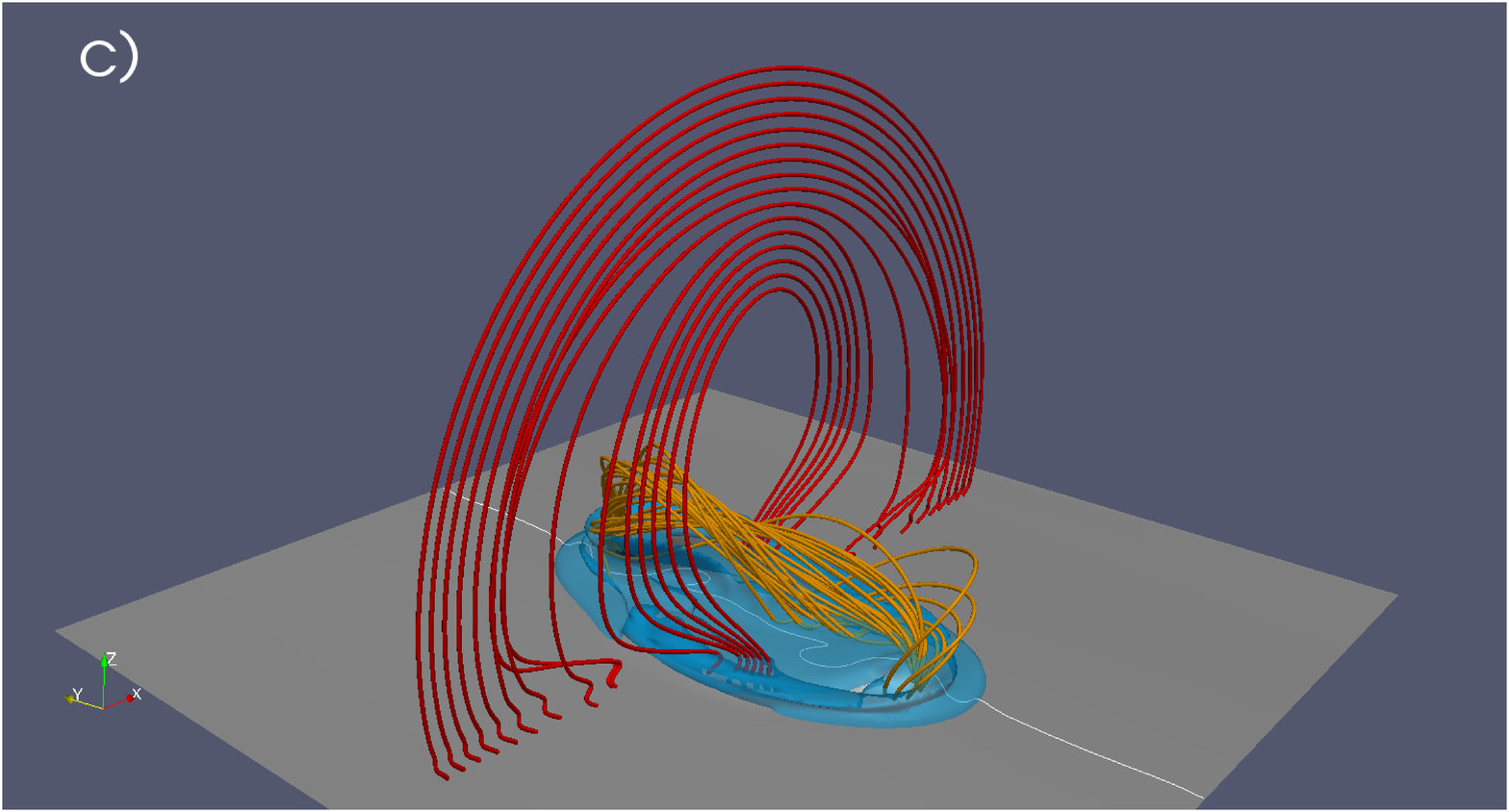}
   \includegraphics[width=\imsize,clip=true]{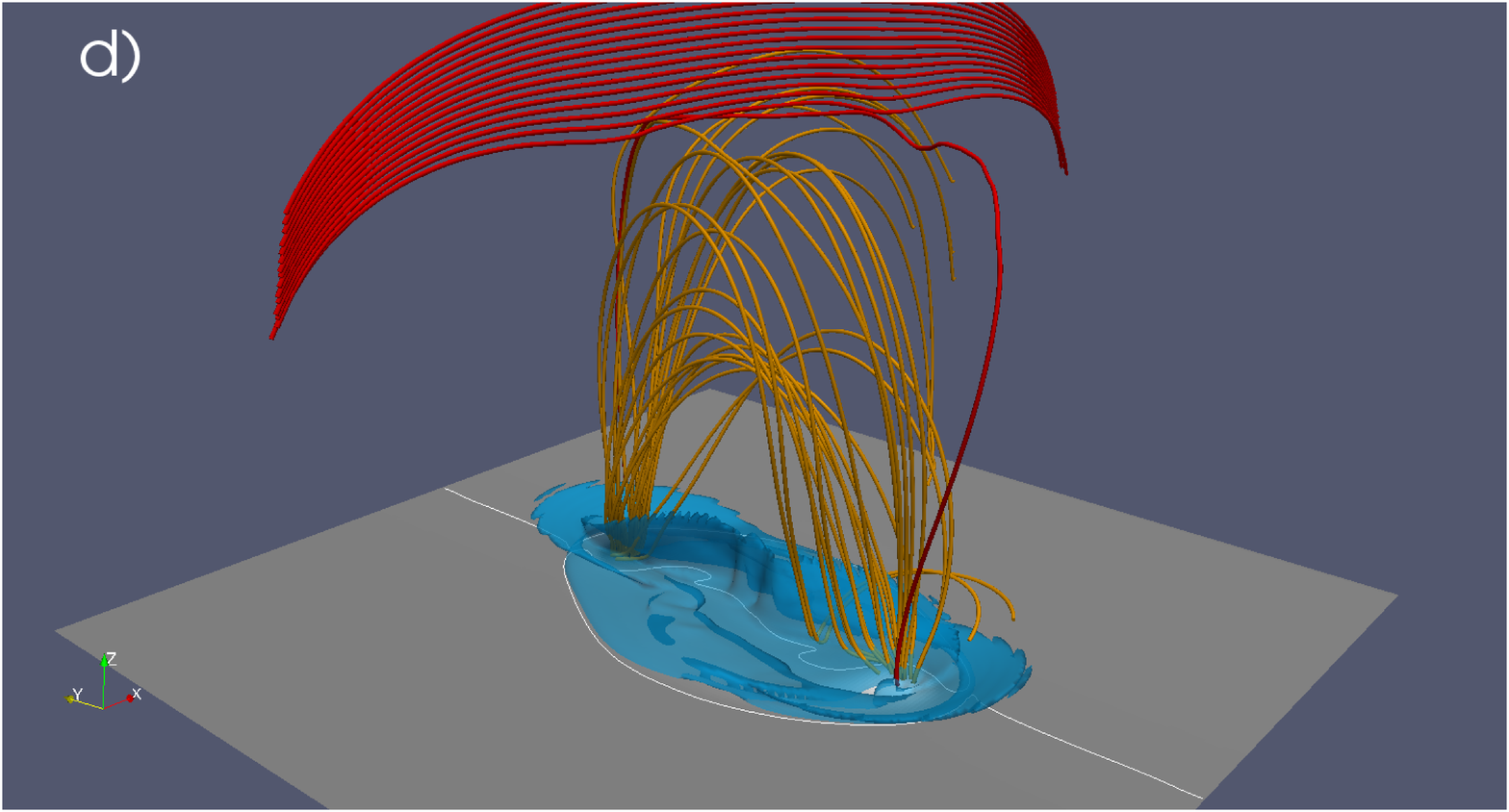}
   \caption{Representative field lines of the four employed test cases:
            \textbf{a)} Low and Lou (\sLL{}) force-free equilibrium, see \sect{LL};  
            \textbf{b)} Titov and D\'emoulin (\sTD{}) force-free equilibrium for $N=1$ and $\Delta=0.06$, see \sect{TD}; 
            \textbf{c)} snapshot at $t=155$ of the stable MHD simulation (\sJst{}), and  
            \textbf{d)} snapshot at $t=140$ of the unstable MHD simulation (\sJun{}), see \sect{JL}.
            Yellow field lines depict the flux rope, red field lines belong to the ambient magnetic field, except for the \sLL{} case where no flux rope is present and the field lines' color scheme does not apply.
            A cyan semi-transparent iso-contour of the current density is shown at 15\%, 35\%, 11\%, and 10\% of its maximum in the four cases a) to d), respectively. 
            }
   \label{f:fl}
 \end{figure*}
In order to be able to critically test the different helicity computation methods of \sect{methods}, the test fields used in this study are chosen such that they represent challenging tests, at the same time including aspects of relevance for solar physics.
From the point of view of the fields' structure, we include both compact structures with concentrated currents as in \fig{fl}b and c, as well as more extended structures with currents threading also the lateral and top  boundaries, as in \fig{fl}a and d.
Similarly, we consider both static (\fig{fl}a and b) and time evolving (\fig{fl}c and d) fields, representing typical magneto-static and magneto-hydrodynamic applications.
In the following, the most relevant properties of the test fields used in this study are discussed in some detail.
In particular, the solenoidal properties of the discretized magnetic fields are quantified using the method described in \sect{fom}.
The results of the analysis of the input fields is given in full in \tab{input}, and a selection thereof is graphically presented in the following sections. 
In particular, \tab{input} also reports the fractional flux defined in \eq{fi} for all test fields considered here.

\subsection{\TLL}\label{s:LL}
The \TLL{} model \cite[][ hereafter \sLL{}]{1990ApJ...352..343L,2006A&A...457.1053W} is a 2.5D solution of the zero-$\beta$ Grad-Shafranov equation, analytical except for the numerical solution of an ordinary differential equation.
In particular, the four \sLL{} cases considered here are different discretizations of the same solution (corresponding to $n=1$, $l=0.3$, $\phi=\pi/4$ in the notation of \cite{1990ApJ...352..343L}), in the same $\vol=$[-1,1]$\times$[-1,1]$\times$[0,2] volume. 
From this solution, four test cases are constructed where the above volume is discretized using 32, 64, 128, and 256 nodes per side.
Since the solution of the ordinary differential equation that defines the \sLL{} field is always the same in the four cases, the only factor changing among the different \sLL{} cases is the resolution.
As a test for helicity methods, \sLL{} represents a large-scale, force-free field with large-scale smooth currents distributed in the entire volume.
This latter aspect is not supported by observations of solar active regions.
However, being almost analytic, the \sLL{} equilibrium  offers a tightly controlled test case.
Here, this test field is used mostly in \sect{LLres} for exploring the dependence of the \tFV{} methods on spatial resolution.

\Fig{Etest}a shows that for the \sLL{} test cases, while  $\Efree\simeq 26$\% and  hardly changes with resolution, the solenoidal error of the input field varies from 0.1\% to 4\% as pixels become coarser (see also \tab{input}).
Hence, the value of $\Hv$ for the \sLL{} cases computed by the different methods will be affected simultaneously by the combined effects of resolution in the computation of the vector potentials on one hand, and by the different degree of violation of the solenoidal property by the test field on the other.

\subsection{\TTD{}}\label{s:TD}
The Titov and D\'emoulin model \cite[][hereafter \sTD{}]{1999A&A...351..707T} is a parametric solution of the 3D force-free equations constituted by a current ring embedded in a confining potential field. 
By considering the portion of the ring above a given (photospheric) plane, the TD equilibrium is possibly the simplest  3D model of a bipolar active region with localized direct currents, see \fig{fl}b.
Differently from the \sLL{} case, the current is tethering only the bottom boundary, while the field is potential on the lateral and top boundaries of the volume considered here.

The \sTD{}  model has significant topological complexity, in the sense that, for different values of its defining parameters, can exhibit a finite twist of $N$ end-to-end turns, bald patches, and an hyperbolic flux tube \citep{2002JGRA..107.1164T}.
In this article we consider six realizations of that solution, in different combinations of twist and spatial resolution.
Unless differently stated, in the following  we refer to  twist of the \sTD{} cases  as the  average twist over the current ring's cross section at the flux rope apex, as computed, \eg in Sect.~4 of \cite{2010A&A...519A..44V}.

Since the equations in \cite{1999A&A...351..707T} are given in implicit form, the twist of each equilibria is obtained by trial and error.
In all six \sTD{} cases considered here, the discretized volume is [-3.18, 3.18]$\times$[-5.10, 5.10]$\times$[0.00, 4.56], the distance between the charges and the current ring centre is $L=0.83$, the depth of the current ring centre is $d=0.83$, and the ring's radius is $R=1.83$. 
We refer to \cite{2010A&A...519A..44V} for definitions and normalizations of the TD parameters, where the Low\_HFT case has the same parameters as the $N=1$ case here.
Based on previous tests  \citep[see \eg][]{Toeroek2004b,2006PhRvL..96y5002K}, all cases presented here are stable equilibria, except for the $N=3$ one, which is almost certainly kink-unstable.

\paragraph{\sTD{}-twist test cases.}
\begin{table}[h]
 \strtable
 \caption{Parameters of the \sTD{} twist test cases, with resolution $\Delta=0.06$; 
          $N$ is the approximate end-to end  number of turns, $a$ and $d$ are the minor radius and the depth of the centre of the current ring, $q$ is the strength of the magnetic 
charges generating the confining field. 
The $N=1$ case here is the same as the $\Delta=0.06$ case of  \tab{TDres}.
We refer to \cite{2010A&A...519A..44V} for definitions and normalizations of the \sTD{} parameters.
 \label{t:TDtw}}
 \begin{tabular}{ c@{\quad}  c@{\quad}  c@{\quad} c@{\quad} }
 \hline
   $N$ &   $a/d$    &$q\times 10^{12}$ & Twist/$\pi$\\
 \hline
     3   &  0.31   &   100     &   -6.054 \\
     1   &  0.80   &   100     &   -2.114 \\ 
    0.5  &  0.80   &    29.5   &   -1.004 \\
    0.1  &  0.80   &     5.5   &   -0.201 \\
 \end{tabular}
\end{table}
We consider equilibria defined by parameters given in \tab{TDtw} resulting into flux ropes with different average twist. 
The resolution (uniform pixel size) is $\Delta=0.06$ for all four cases, \ie given by a grid of 107$\times$171$\times$79 nodes.
The energy decomposition of the \sTD{} twist cases are shown in \fig{Etest}b.  
The solenoidal errors vary between 0.1\% and 2\%, monotonically increasing with twist except for the $N=3$ case, which corresponds to a much thinner flux tube that satisfies the local cylindrical approximation better.
Their contribution in relative energy is always one order of magnitude smaller than the free energy. 
The only exception is the $N\simeq0.05$ case where the twist is lower and the field is almost potential ($\Efree\simeq 0.1\%$).
In this case solenoidal errors are slightly larger than the free energy, but anyway extremely small ($\Ediv=0.2\%$).

\paragraph{\sTD{}-resolution test cases.} 
\begin{table}[h]
 \strtable
 \caption{Parameters of the TD resolution test cases, with $N\simeq1$;
          $\Delta$ is the spatial resolution,  ($n_x$, $n_y$, $n_z$) are the number of nodes in the $x$-, $y$- and $z$- directions, respectively. 
          The case $\Delta=0.06$ here is the same the  $N=1$  case in \tab{TDtw}.
          We refer to \cite{2010A&A...519A..44V} for definitions and normalizations of the \sTD{} parameters.}
 \label{t:TDres}
 \begin{tabular}{c@{\quad} c@{\quad} c@{\quad}  c@{\quad} c@{\quad} }
 \hline
   $\Delta$& $n_x$   & $n_y$& $n_z$ & Twist/$\pi$\\
 \hline
     0.03  & 209  &  337  &153     &   -2.15 \\ 
     0.06  & 107  &  171  & 79     &   -2.14 \\ 
     0.12  &  56  &   88  & 42     &   -2.09 \\ 
 \end{tabular}
\end{table}
The three equilibria for this test are the N1 case in \tab{TDres} at resolution $\Delta=0.06$, and two additional equilibria with exactly the same parameters but with $\Delta=0.03$ and 0.12, respectively.
The energy decomposition of the \sTD{} resolution tests summarized in \fig{Etest}c  shows that the free energy is independent of resolution (at about 2\% of the total energy), and that the solenoidal errors depend only very weakly on it (cf.  \tab{input}).
While the former is expected, the latter confirms that solenoidal errors in the (non-relaxed) numerical implementations of \sTD{} equilibria have a stronger dependence on twist than on resolution that is mostly generated by the match at the interface between flux rope and ambient potential field.
\paragraph{}

In the construction of \sTD{} equilibria, the local matching between current ring and potential environment field at the flux rope's boundary is done in locally cylindrical coordinates, hence some spurious Lorentz forces and solenoidal errors are present at the interface between the two flux systems. 
For such reasons, when employed as an initial state of numerical simulations, an MHD relaxation is normally applied beforehand to remove such  residual forces and errors, see \eg \cite{Toeroek2003e}. 
For the purpose of computing the relative magnetic helicity, however, the relaxation step is unnecessary as long as the errors in the solenoidal property of the field are small enough, which is the case here.

 \setlength{\imsize}{0.49\textwidth}
 \begin{figure*}
   \centering
   \includegraphics[width=\imsize,clip=true]{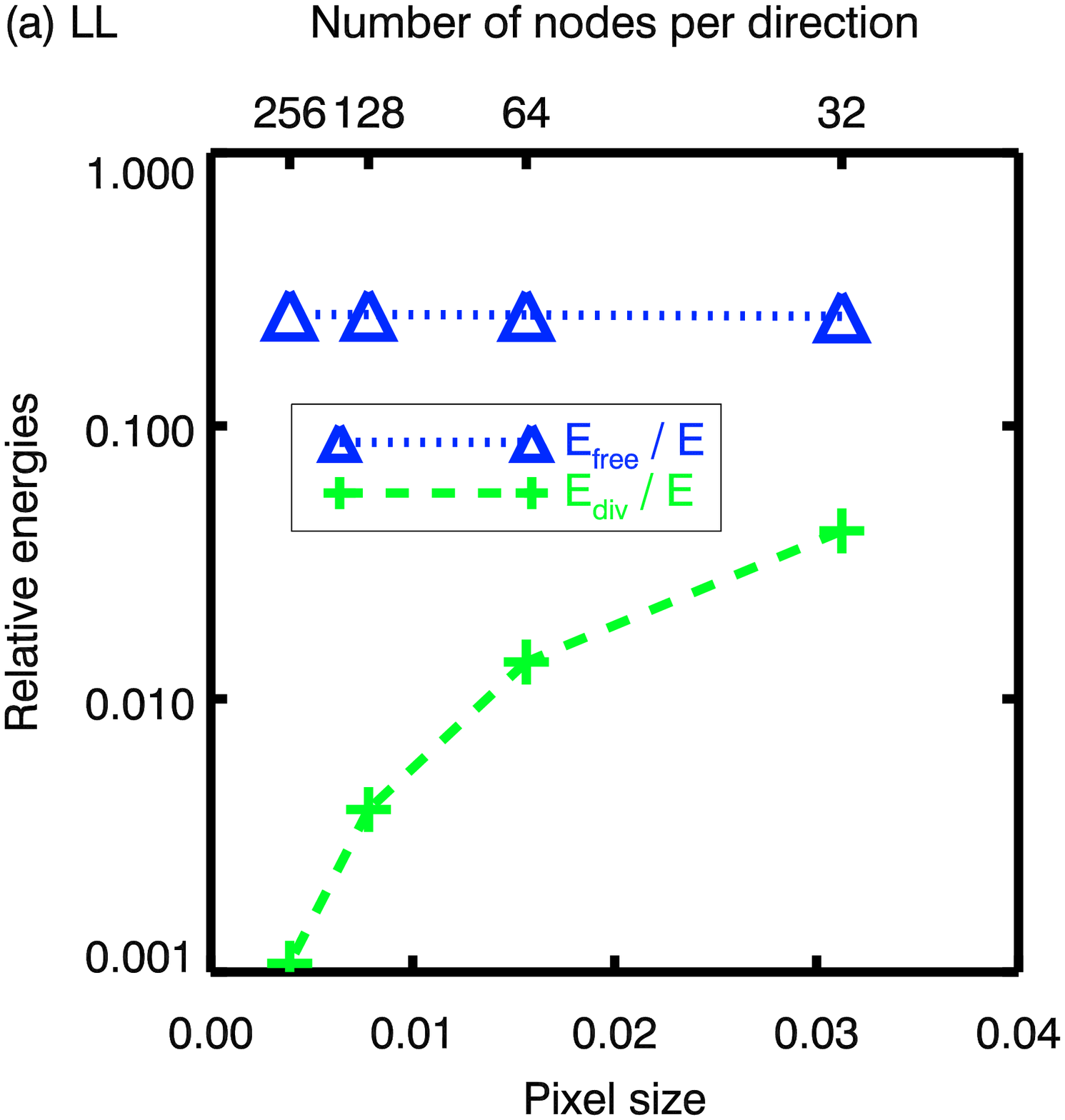}
   \includegraphics[width=\imsize,clip=true]{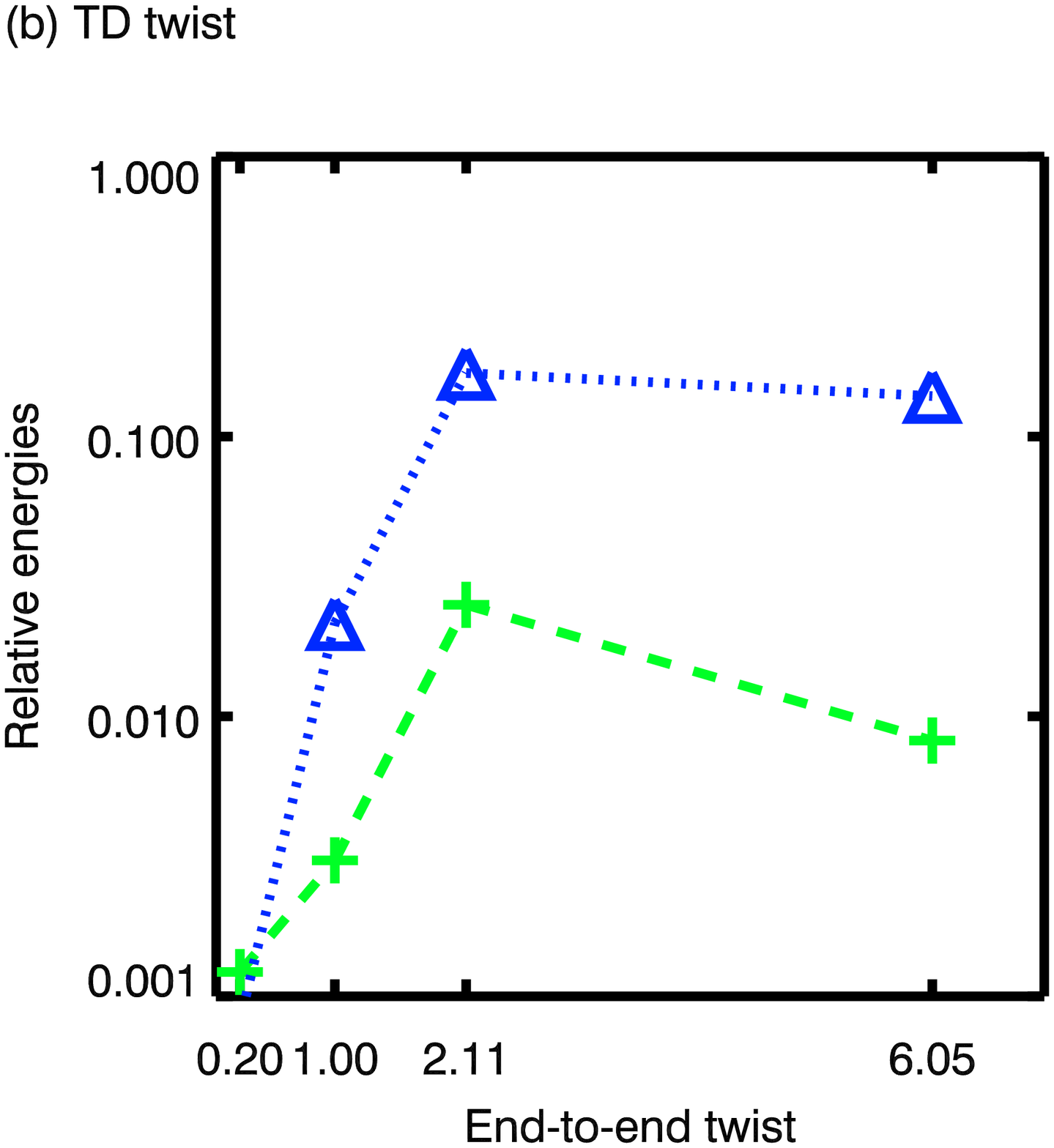}\\
   \includegraphics[width=\imsize,clip=true]{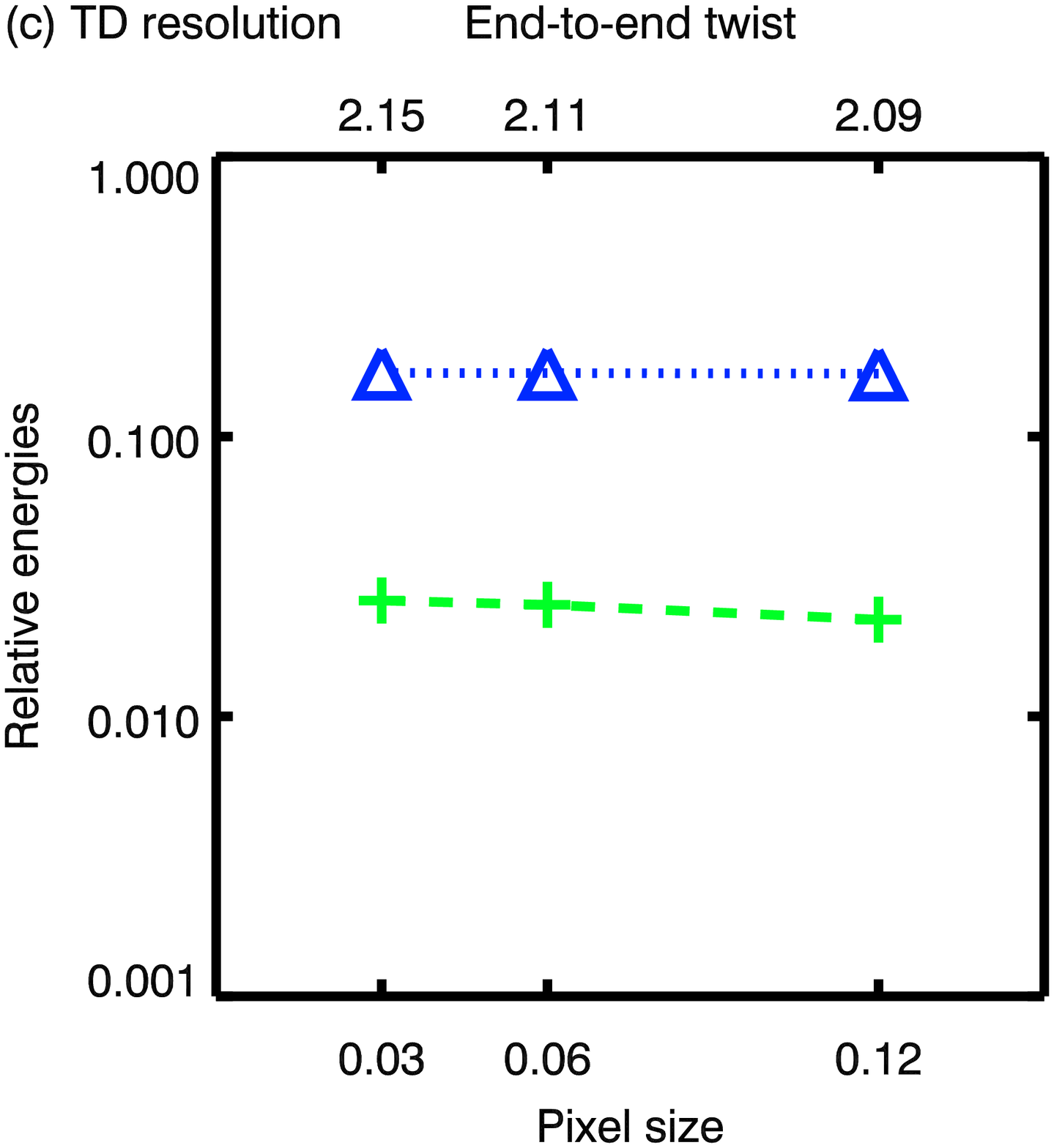}
   \includegraphics[width=\imsize,clip=true]{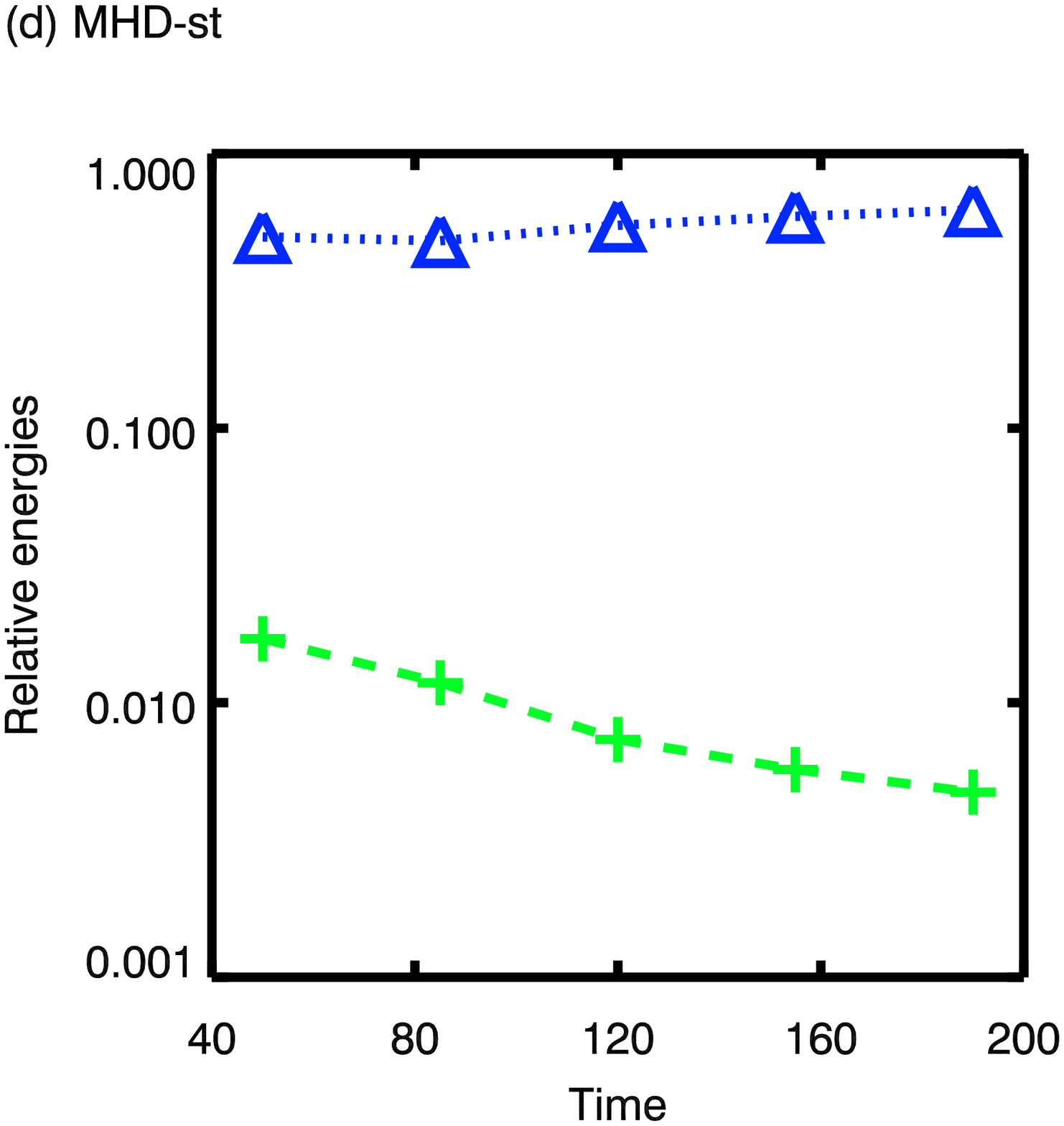}\\
   \includegraphics[width=\imsize,clip=true]{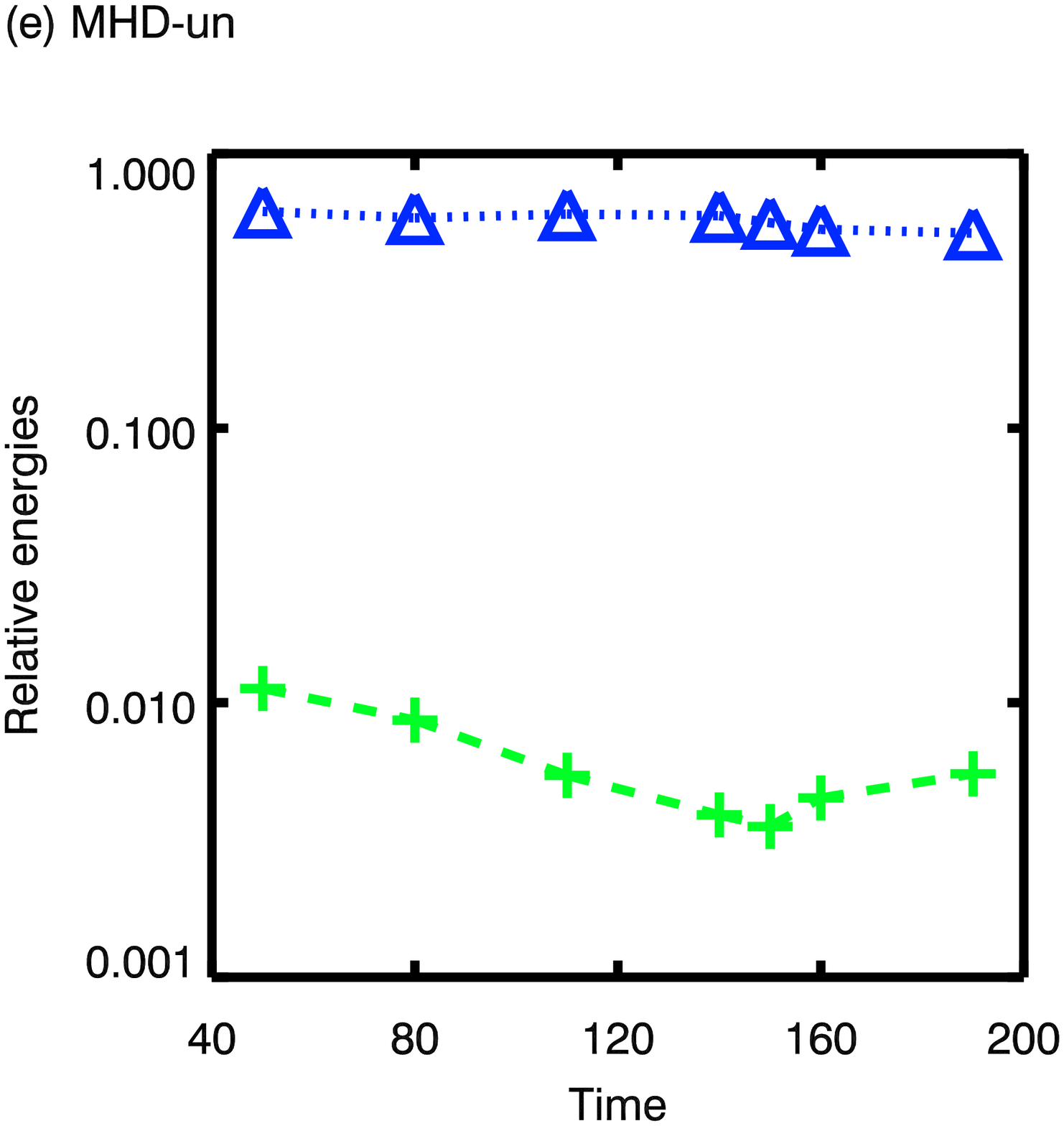}
   \includegraphics[width=\imsize,clip=true]{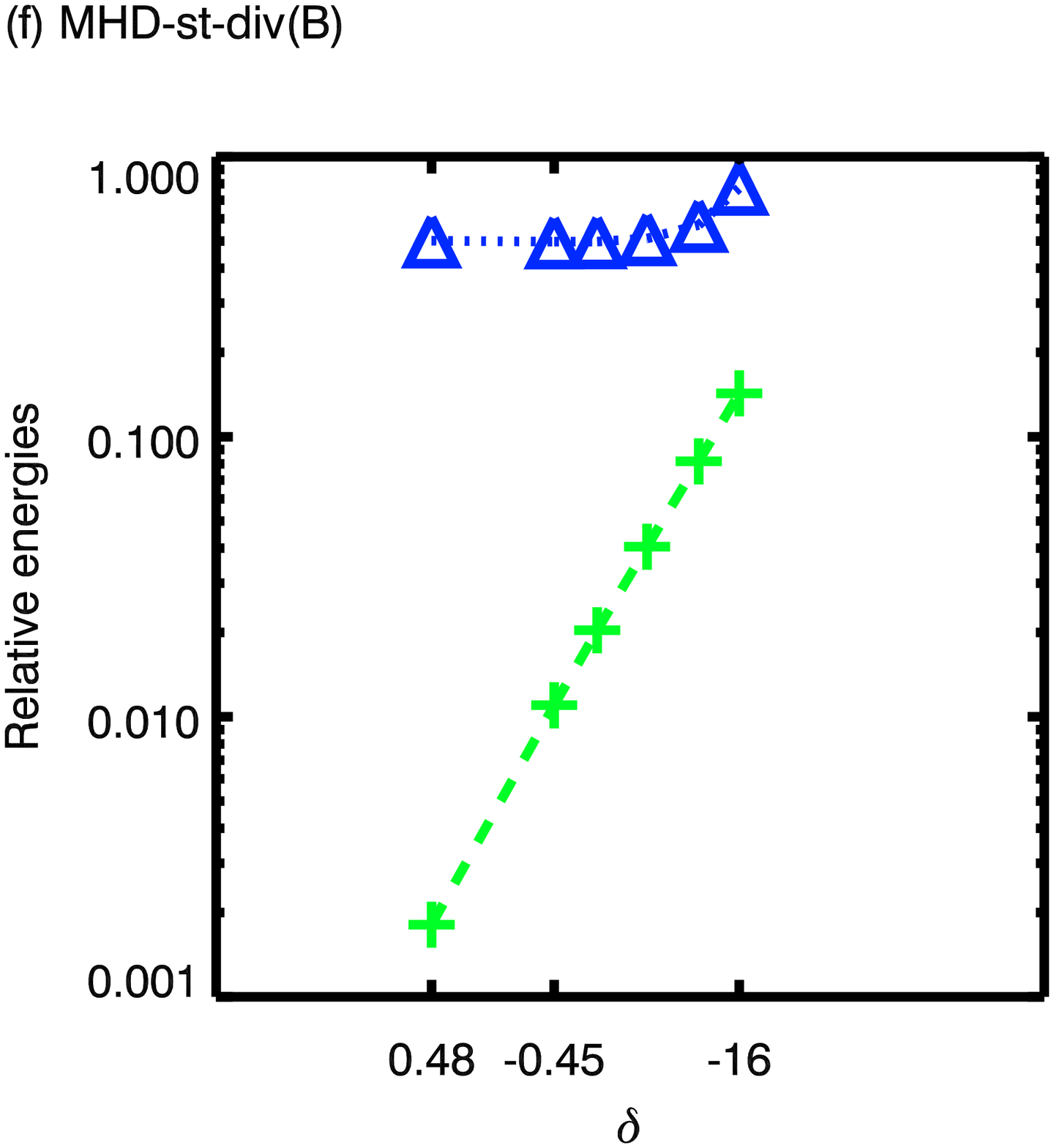}
   \caption{Normalized free (blue connected triangles) and nonsolenoidal (green connected crosses) energies for the test cases
            \textbf{a)} \sLL{};
            \textbf{b)} \sTD{}  as a function of twist;
            \textbf{c)} \sTD{} as a function of resolution;
            \textbf{d)} \sJst{}; 
            \textbf{e)} \sJun{} 
            \textbf{f)} divergence test \sJdiv{}, as a function of $\delta$, see \sect{divergence}.
            }
   \label{f:Etest}
 \end{figure*}

\subsection{\TJst{} and \TJun{}}\label{s:JL}
The MHD cases are simulations of stable (hereafter, \sJst{}) and unstable (hereafter, \sJun{}) evolutions of flux emergence published in by \cite{Leake2013d}  and \cite{Leake2014}, respectively. 
These simulations offer the possibility of studying the time evolution of the helicity as the flux breaks through the photospheric layers, slowly accumulates in the coronal volume, and either reaches stability  or erupts. 

More in detail, the \sJst{} case is a sub-domain  of the Strong Dipole case by \cite{Leake2013d}, which was obtained using a stretched grid. 
Here we consider five snapshots of that simulation, and for each one the magnetic field is interpolated using a grid of 233$\times$233$\times$174 nodes with a uniform mesh size $\Delta=0.86$
discretizing the volume 
[-100,100]$\times$ [-100,100]$\times$ [0, 149] 
(in units of $L_0$, see \cite{Leake2013d} for more details).
The interpolation is expected to introduce spurious solenoidal errors, but is required to accommodate for the requirements of all helicity computation methods.
Compared to the original data, the bottom boundary of the interpolated mesh corresponds to the photospheric level, the top boundary to $z=150$, and the domain is reduced in the lateral extension. 
The time series includes snapshots at $t=$[50, 85, 120, 155, 190]. 
The \sJun{} case, corresponding to the Medium Dipole case of \cite{Leake2014}, is prepared in a similar way, but the time series is $t=$[50, 80, 110, 140, 150, 160, 190].
\cite{Leake2014} identify the starting time of the eruption around $t=120$ and the erupting structure is leaving the domain between $t=150$ and $t=160$, see, \eg their Fig.~11.

In the MHD simulation, the stable and unstable cases are obtained only by changing strength and orientation of the coronal field, but nothing of the emerging flux rope.
In this sense, the two simulations are very similar, albeit resulting in a very different end state.
Note that in the MHD cases,  $\Efree$ , $\Ep$, and hence $\E$  are all dynamically evolving in time. 
In the unstable case, the small variation of the ratio $\Efree(t)/\E(t)$ (of about 0.05 points in \tab{input}) during the eruption, actually corresponds to a variation of 93 units of free energy with the maximum total energy before the eruption being equal to 464 units.

\Fig{Etest}d,e show that, in both \sJst{} and \sJun{} cases, the solenoidal errors slightly decrease during the MHD evolution from at most 1.6\% at the beginning  of the simulations, to 0.5\% at their end.
Due to the interpolation, the actual values of the solenoidal errors are larger than for  the original simulations, but the trend in time is presumably the same.
In addition to the helicity estimation discussed in \sect{JL}, the snapshot at $t=50$ of \sJst{} is also used as the reference case  for the test on the effect of nonzero divergence independently of resolution, as described in \sect{divb_test} and \sect{divergence}.

With respect to \sTD{} and \sLL{} cases, the \sJst{} and \sJun{} tests confront the \tFV{} methods with a higher complexity in the field, with the formation of small scales, and with the coronal re-organization of the connectivity in time, see \fig{fl}c,d.
Moreover, these cases contain large currents and large free energies, of the order of 50\% to 60\% of the total magnetic energy, depending on the type and stage of the evolution, see \tab{input} and \fig{Etest}d,e.
Naturally, such cases are of interest for our discussion since they are supposed to mimic more realistically the difficulties that helicity estimation methods face in solar applications.

\subsection{Parametric nonsolenoidal test case}\label{s:divb_test}
The application of \eq{Hdef} to a field with nonzero divergence makes simply no sense in terms of helicity because \eq{Hdef} becomes gauge-dependent. 
However, in the routine  situation of numerical studies, a finite value of divergence is always present. 
The question arises, what is the value of divergence that is tolerable in computing helicity, \ie that is producing a helicity value close to the one obtained for the solenoidal field?
A complication of the problem is that the value of helicity is in general not known, not even for the test cases presented here.
Hence, we are forced to reformulate our question in terms of variation of $\Hv$ obtained by each method as a function of the increasing divergence.
In practice, we check how each method behaves for increasing violation of the solenoidal property, but we are not in the position of stating which method gives the more ``correct'' value as divergence increases.

To this purpose, we designed a dedicated \TJdiv{} test (hereafter, \sJdiv{}) by considering a numerically solenoidal field to which divergence is added in a controlled way and without changing the resolution, as done in Eqs.~(14,15) of \cite{2013A&A...553A..38V}, to whom we refer for the details. 
In brief, starting from the snapshot of the magnetic field $\vB$ at $t=50$ of the \sJst{} case, a solenoidal field $\vBs$ is produced by removing the divergence part $\vBns$ of $\vB$. 
A parametric, generally nonsolenoidal field,
\BE
 \vBd=\vBs + \delta \vBns \, ,
 \label{eq:Bdiv}
\EE
is then constructed by adding the divergence part back, multiplied by a scalar amplitude $\delta$.
In this way, the original spatial structure of $\vBns$ is kept in $\vBd$ but its amplitude is modified according to the chosen value of $\delta$.
The case $\delta=0$ corresponds the numerically solenoidal field $\vBs$, whereas $\delta=1$ reproduces the original test field $\vB$, \ie the snapshot of \sJst{} at  $t=50$.
The method for building the vector field $\vBns$ is detailed in Sect.~7.1 of \cite{2013A&A...553A..38V}.

By increasing the value of $\delta$, progressively more divergence can be added, yielding a larger amplitude of the nonsolenoidal component.
A trial-and-error tuning of $\delta$ resulted into six \sJdiv{} test fields with nonsolenoidal  contributions as a function of the parameter $\delta$ as summarized in \tab{Bdiv} \referee{and \fig{Etest}f. In the following, different \sJdiv{}}
test cases are identified by the correspondent fraction of  $\Ediv$.
\begin{table}[h]
 \strtable
 \caption{Parameters $\delta$ and corresponding values of relative nonsolenoidal energy for the \sJdiv{} test, see \eq{ediv}.
 \label{t:Bdiv}}
 \begin{tabular}{@{~}l c@{\quad}  c@{\quad}  c@{\quad} c@{\quad}   c@{\quad}  c@{\quad} }
 \hline
   $\delta$     &  0.48 &   -0.45 &   -1.28 &   -2.87  &  -6.04 &   -16.00   \\ 
 \hline
   $\Ediv$(\%)  &  0.2  &     1.1 &     2.0 &    4.0   &    8.2 &    14.4    \\
   $\Ens$ (\%)  & -0.1  &    -1.0 &    -1.9 &    -3.6  &    -6.3&    -8.3 
 \end{tabular}
\end{table}

The dependence  of \sFV{} methods on solenoidal errors as modeled by \eq{Bdiv} is discussed in \sect{divergence}.

\section{Results: \TFV{} methods}\label{s:results}
In this Section we discuss the accuracy of \sFV{} methods presented in \sect{methods} using the test cases introduced in \sect{tests}.
As mentioned above, such methods require the full knowledge of the 3D magnetic field.
Obviously, a non-perfectly solenoidal input field cannot be reproduced by a vector potential, and will affect each method in a different way. 
Therefore, the comparison between methods presented below should be read against the properties of the input field as described in \sect{tests}.
The purpose of these comparisons is not only to assess absolute accuracy in specific tests, but also to  address the sensitivity of the methods towards certain studied parameters, such as resolution, twist, and topological complexity.

As anticipated in \sect{fom}, the vector potentials obtained by \sFV{} methods are judged on their ability of reproducing the test fields and their corresponding potential fields.
Indirectly, this is also a measure of the accuracy of the methods in the computation of $\Hv$.
The full metrics' values are provided in Tables~\ref{t:lltd}-\ref{t:stun}, together with the computed values of helicity, and a partial but representative selection is reproduced in the plots of the next Sections.

\subsection{Dependence on twist in the \TTD{} case}\label{s:TDtw}
 \setlength{\imsize}{0.5\textwidth}
 \begin{figure}[]
   \includegraphics[width=\imsize,clip=true]{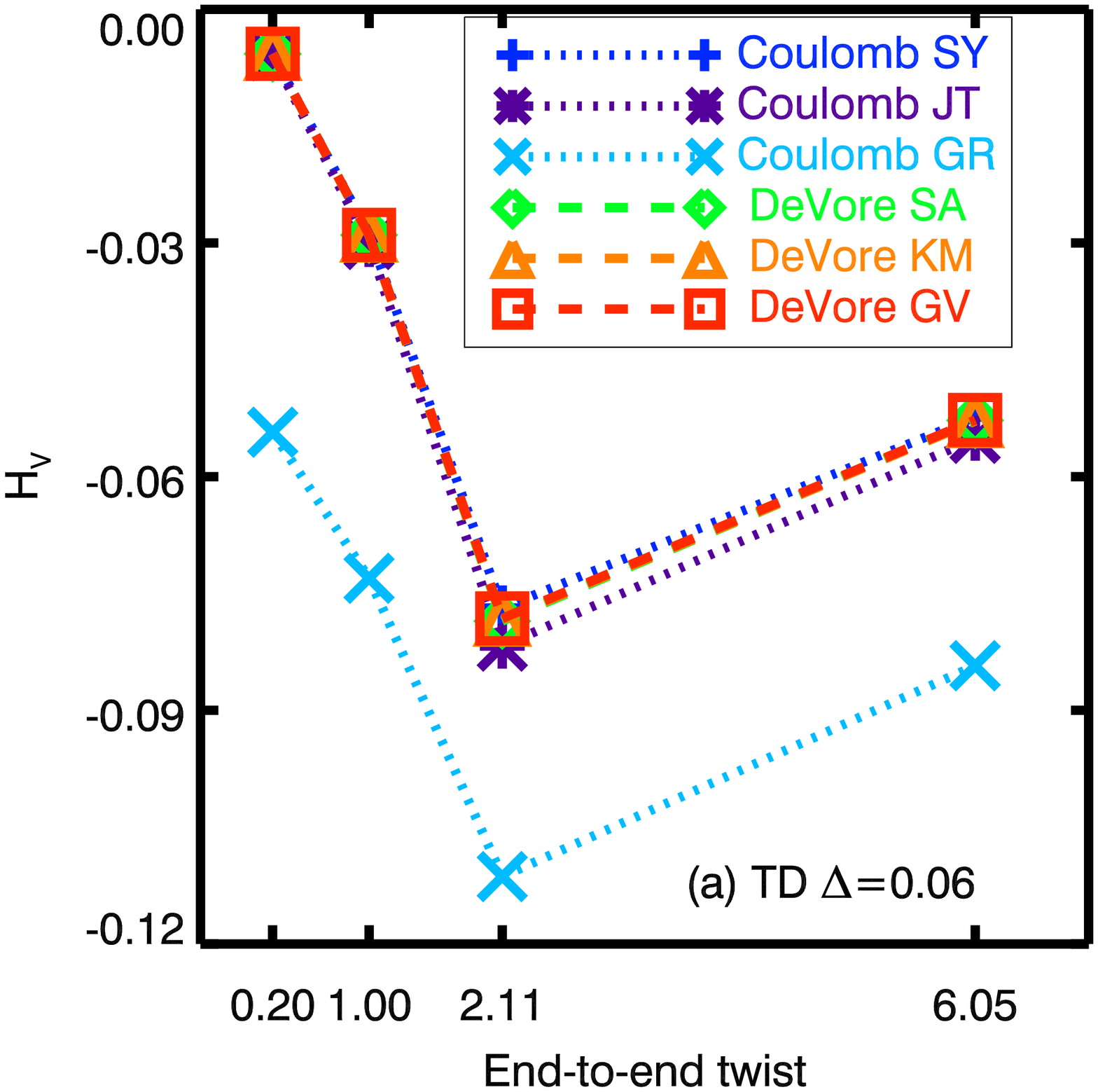}
   \includegraphics[width=\imsize,clip=true]{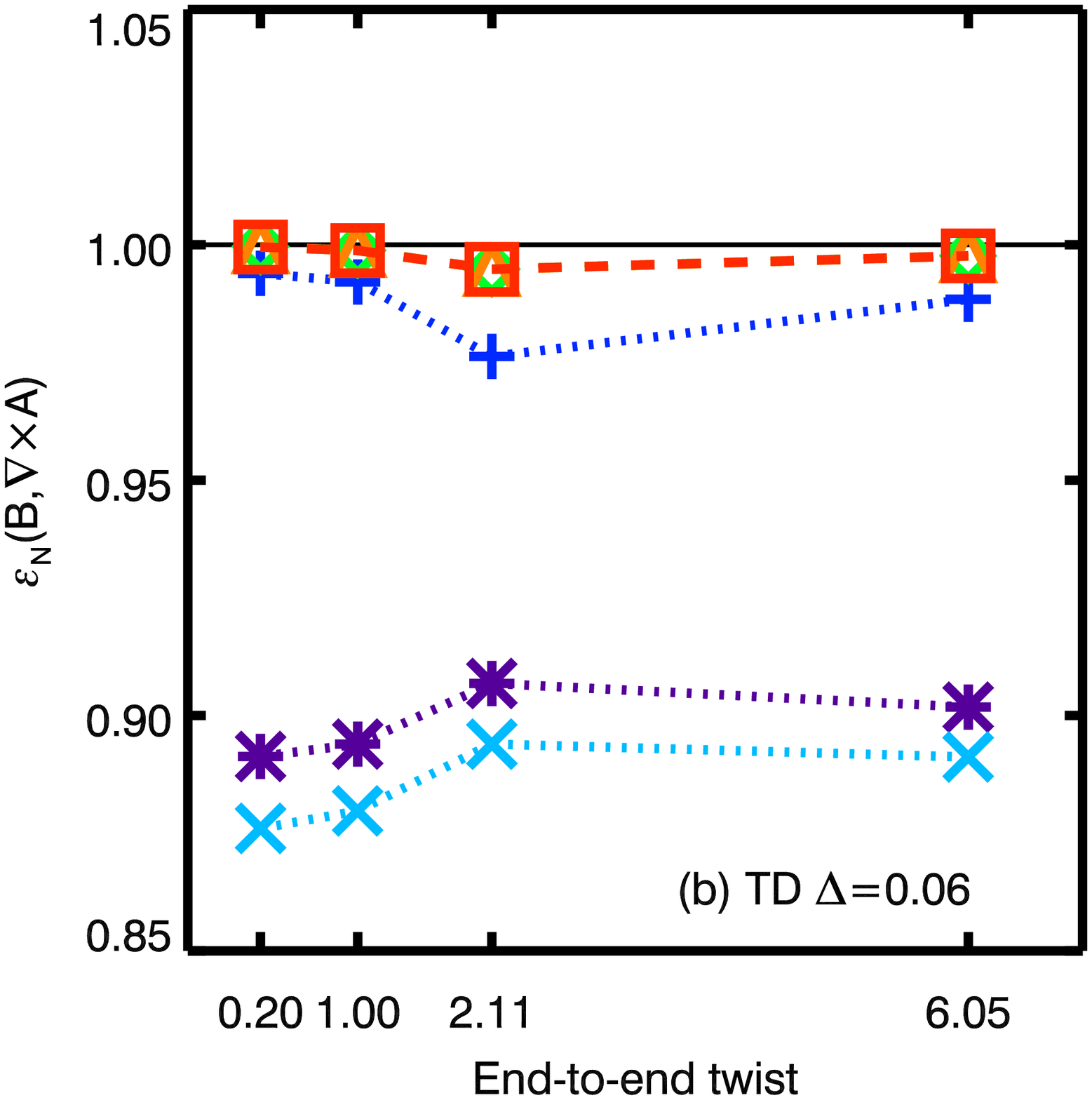}
   \caption{\sTD{}-twist test: a) Normalized helicity $\Hv$; b) Complement of the normalized vector for the test field $\epsN(\vB,\curlA)$.}
   \label{f:TDtw}
 \end{figure}
\Fig{TDtw}a shows the dependence on the twist of $\Hv$ computed by the different \sFV{} methods.
With the exception of the \sGR{} method, all other methods are basically producing the same value of $\Hv$ at all twists.
More quantitatively, the spread in $\Hv$  computed including all twist values and all methods except for the \sGR{} method, is only 2\%.
In fact, DeVore methods are practically indistinguishable from each other. 
For a given twist value, for instance at $N=1$, the spread of $\Hv$ values around the average -0.079 is only 2.3\%, whereas average and spread become -0.084 and 16\%, respectively,  if \sGR{} is included.
The $\Hv$ from the \sGR{} method, on the other hand, follows the same trend as the other methods, but has values about a factor two larger.
In addition, there is no apparent correlation between the spread of $\Hv$ and the value of the twist.
All methods seem to be unaffected by this particular aspect of complexity of the field, at least at the resolution considered here.
On the other hand,  recalling the dependence on twist of the  solenoidal errors discussed in \sect{TD} and \fig{Etest}b,  it is found that larger spreads in $\Hv$ correlates to larger values of divergence.

In order to address specifically the methods' accuracy, we show in \fig{TDtw}b the complement of the normalized vector error, $\epsN(\vB, \Nabla\times\vA)$ (cf. Eq. \ref{eq:En}), between the test field $\vB$ and the rotation of the corresponding vector potential $\vA$ computed by each method.
The DeVore methods sport extremely high accuracy metric  of $\epsN=0.994$ or higher, and are indeed indistinguishable from each other.
The \sSY{} accuracy correlates inversely with the solenoidal errors of that test (cf. \fig{Etest}b), in the sense that the accuracy is lower for proportionally larger values of $\Ediv$, as could be expected.
Given the high accuracy in all twist cases, such a dependence in not very clearly visible in \fig{TDtw}b. 
Among the Coulomb methods, the \sSY{} method has accuracy values (\eg $\epsN=0.976$ at $N=1$) that are only slightly worse than the  DeVore methods.
Similarly to DeVore-gauge methods, the \sSY{} accuracy correlates inversely with the solenoidal errors, but in this case with more pronounced variations.
The vector potential computed by the \sGR{} method has the largest error in reproducing the input field, with values of the complement of the error vector as small as $\epsN=0.88$. 
A trend similar to \sGR{}'s one is found for the \sJT{} method but with a slightly smaller error ($\epsN=0.903$).
Apparently, both  \sGR{} and  \sJT{} method show an accuracy in this test that directly correlates with the solenoidal errors, \ie  the accuracy is lower (smaller values of $\epsN$) for smaller  values of $\Ediv$.
This counterintuitive trend, however, is not confirmed by the tests in \sect{divergence}, and has to be regarded as insignificant.
On the other hand, one could equally conjecture that  \sGR{} and  \sJT{} methods show a direct correlation between $\epsN$ and the free energy $\Efree$, see again \fig{Etest}b. 
In this sense, \sGR{} and  \sJT{} could be said to obtain more accurate results for fields with higher currents, which can be understood in terms of a stronger source term in \eq{C_A}, and is not contradicted by the tests presented in the following sections. 
We notice that, even for similar values of $\epsN$, the helicity values obtained by the \sGR{} and \sJT{} methods are quite different, the latter aligning with the DeVore ones within 2\%-variation. 

As for the average $\Hv$ values, the \sTD{}-twist test cases in \fig{TDtw}a demonstrate that more twist does not necessarily translate into larger helicity values.
Indeed, the $N=3$ case has a smaller helicity than $N=1$.
The higher twist of the $N=3$ case is obtained by reducing the radius of the current channel ($a$ in \tab{TDtw}) without changing the ambient field (\ie the magnetic ``charges'', $q$). 
The dependence of the current on $a$ is weak \citep[see Eq.~(6) of ][]{1999A&A...351..707T}, and the difference in $\Hv$ is mostly due to the difference in the---dominant---mutual helicity between ambient field and current channel.
 
In conclusion, in the \sTD{}-twist test cases all methods, with the exception of the \sGR{} one, provide the same value of helicity within 2\% of variation.
Furthermore, in a  range of different twist ranging from practically zero to 3 full turns, no evidence of direct influence of the amount of twist on the accuracy of the methods is found.
The variation of the accuracy with twist clearly correlates for most of the methods with the solenoidal errors, but an inverse correlation is also found for the \sJT{} and \sGR{} methods.

\subsection{Time evolution in the \TJst{} and \TJun{} cases}\label{s:JLmhd}
 \setlength{\imsize}{0.5\textwidth}
 \begin{figure}
              \includegraphics[width=\imsize,clip=true]{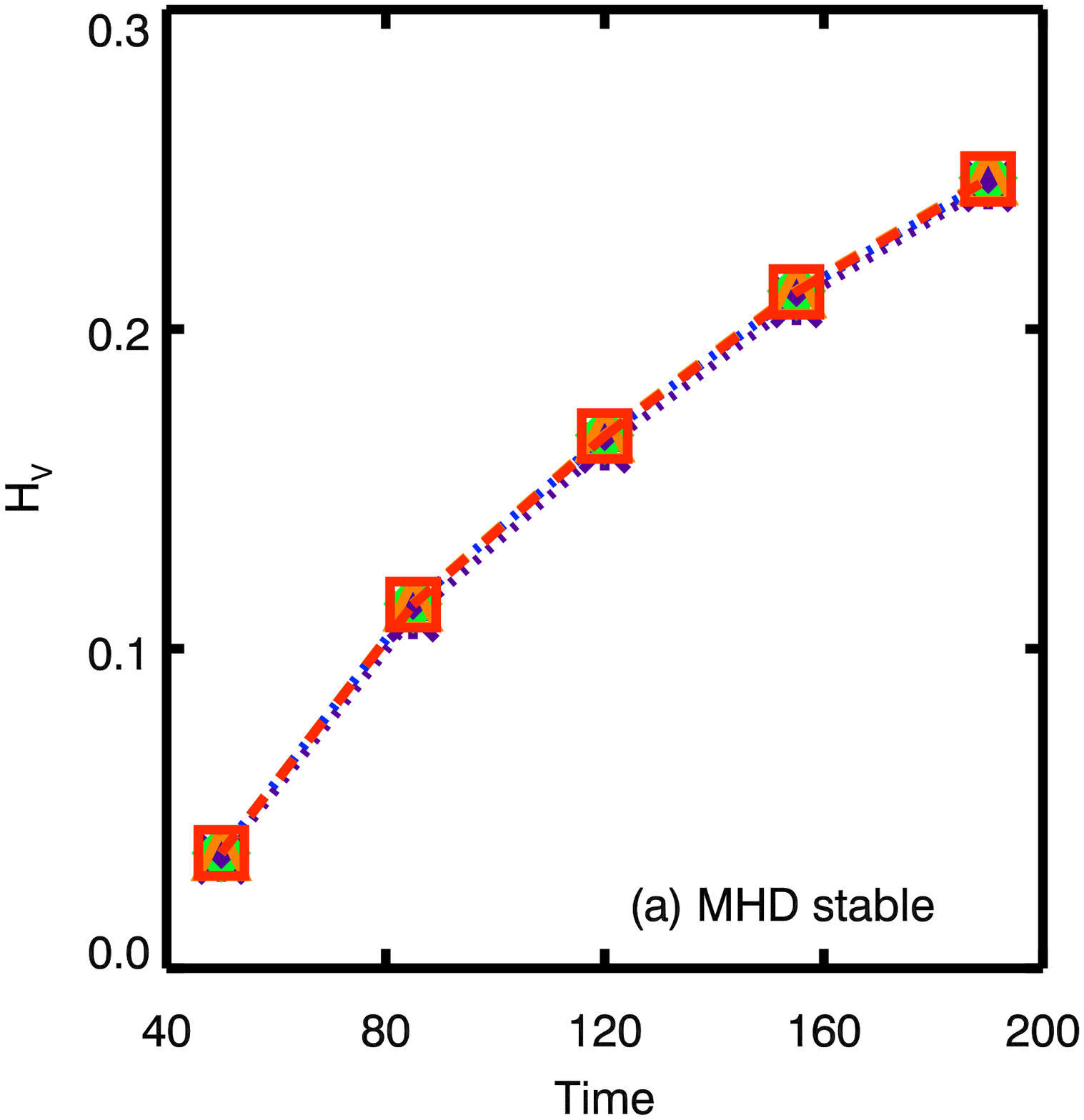}
              \includegraphics[width=\imsize,clip=true]{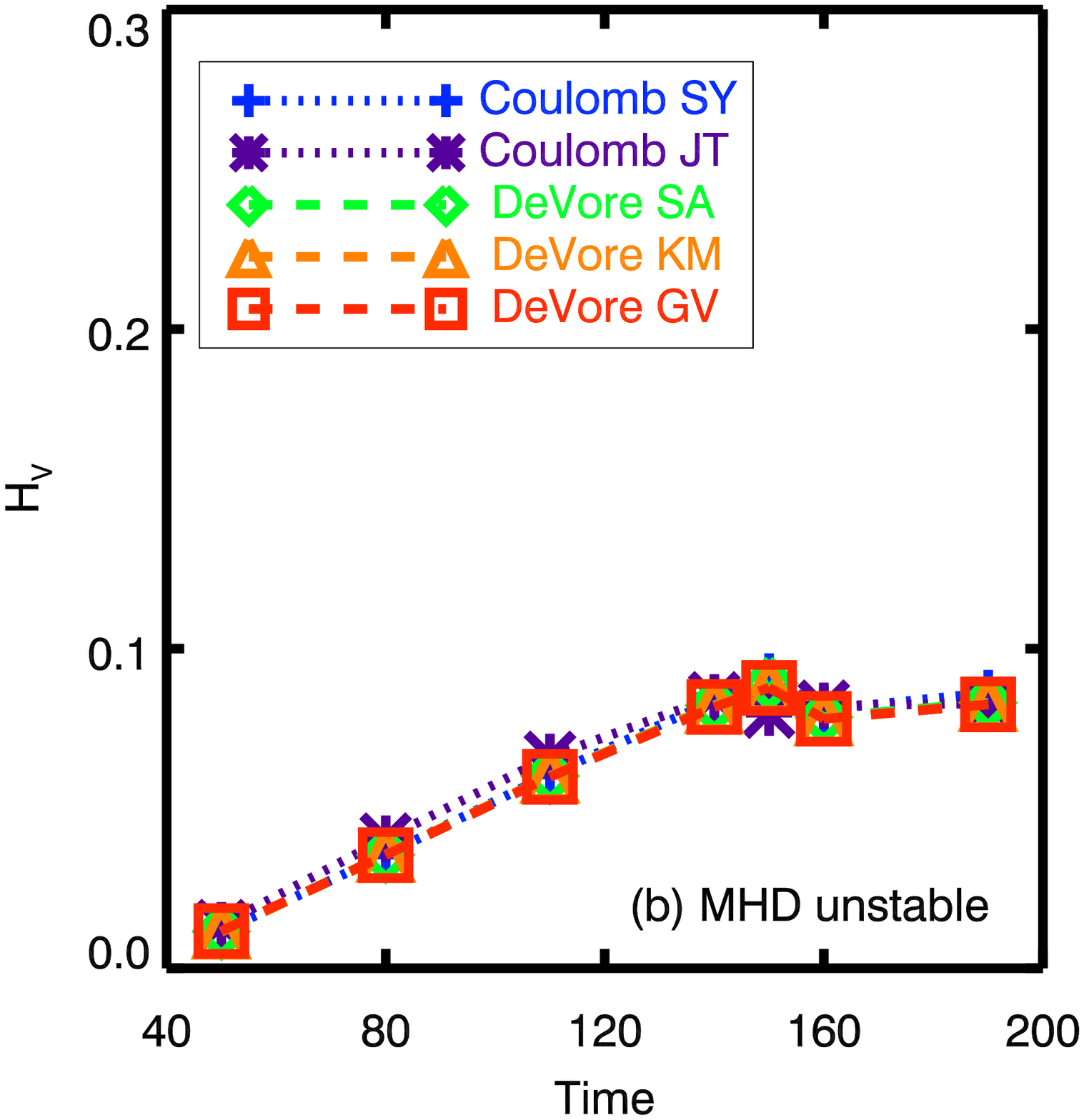}
   \caption{Normalized helicity $\Hv$ for the a) \sJst{} and b) \sJun{}, as a function of time.
            No result from the \sGR{} method is included here due to computing-time limitations.}
   \label{f:JstJunH}
 \end{figure}
\Fig{JstJunH} shows the helicity evolution for the  \sJst{} and \sJun{} cases.
No solution from the \sGR{} method is included for the \sJst{} and \sJun{} cases due to the long computation time that is required to obtain the two time series.
The spread in $\Hv$ computed including all available methods and time snapshots is basically negligible in the \sJst{} case, amounting to just 0.2\%, see \fig{JstJunH}a.
Hence, in the \sJst{} case, even more than  in the \sTD-twist case of \sect{TDtw}, all considered methods yield essentially the same value of $\Hv$, which is a very encouraging result in view of future applications.
The spread in $\Hv$ is very small also in the \sJun{} case, being 3\% at the end of the simulation.
Even though this represents a modestly larger spread in $\Hv$ values, one has to recall that the \sJun{} case corresponds to a time evolution where an eruption rearranges drastically the magnetic field, with ejection of field and currents from the top boundary.
The challenge posed to numerical accuracy in such cases is indeed very high.

 \setlength{\imsize}{0.5\textwidth}
 \begin{figure}
       \includegraphics[width=\imsize,clip=true]{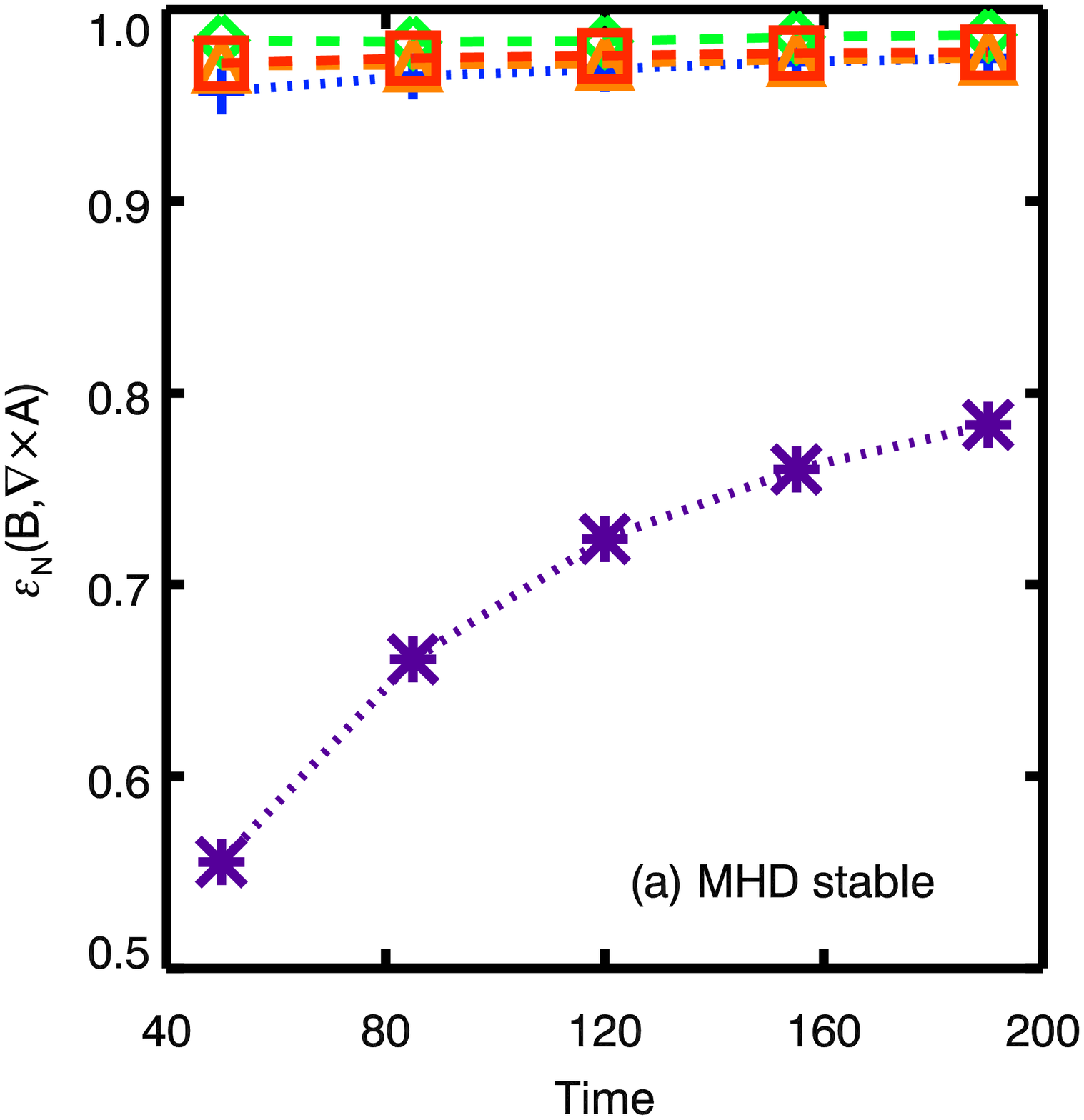} 
       \includegraphics[width=\imsize,clip=true]{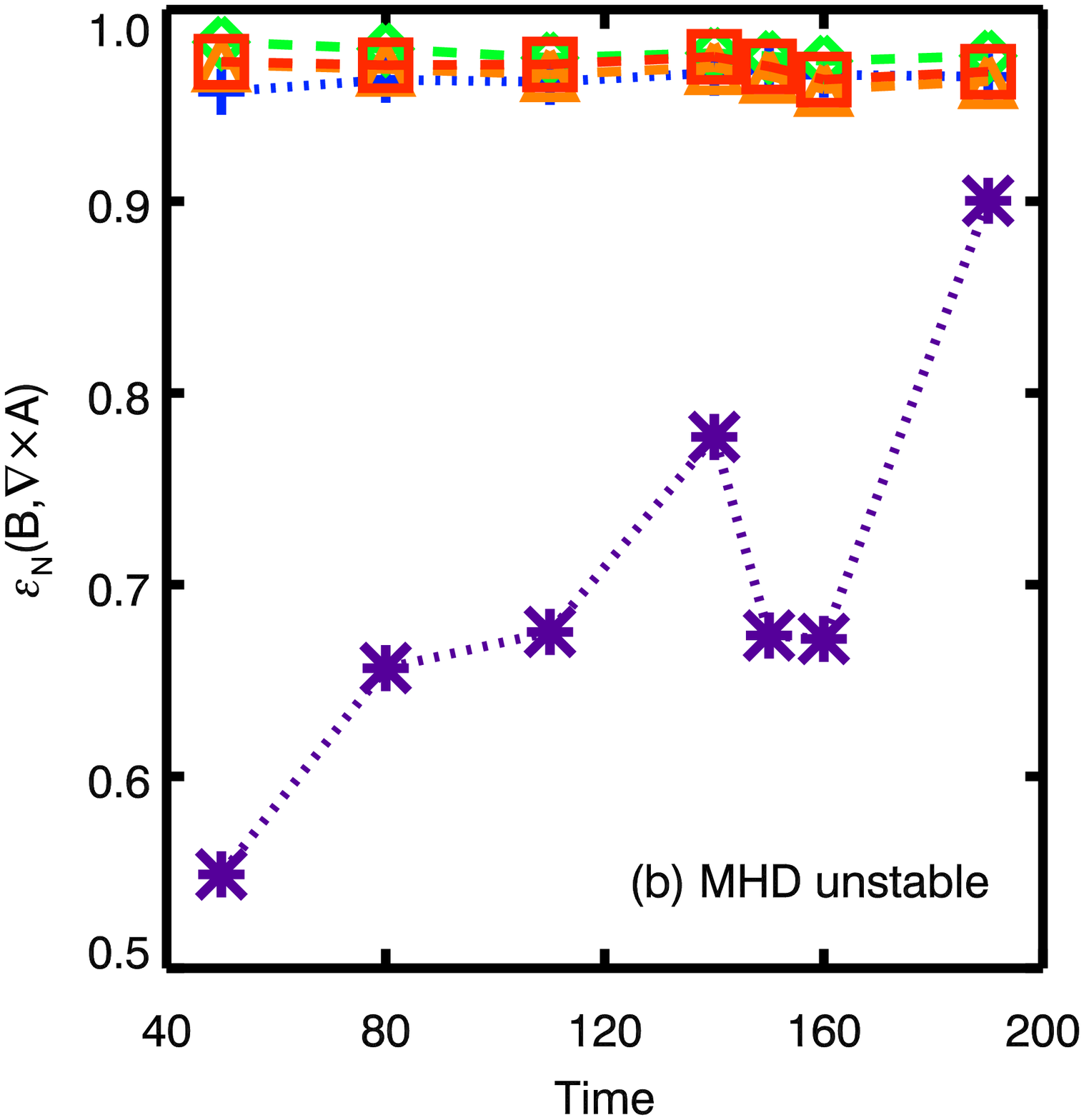}
   \caption{Complement of the normalized vector for the test field, $\epsN(\vB,\curlA)$, for the  \sJst{} (a) and the \sJun{}  (b) cases, as a function of time.}
   \label{f:JstJunEn}
 \end{figure}
Concerning the accuracy metrics for the magnetic field, \fig{JstJunEn} shows the $\epsN(\vB,\curlA)$ (cf. \eq{En}), for the \sJst{}  and \sJun{} cases, respectively, as a function of time.
The accuracy metric shows that \sSY{} and all DeVore methods have comparable accuracy, of about 0.96 and 0.97 on average, respectively.
On the other hand, the \sJT{} method has an accuracy for the complement of the vector error of $\epsN=0.54$ at the beginning of the simulation, increasing up to $\epsN=0.77$ at the end.
Comparing with the evolution in the \sJun{} case, we notice that the metric $\epsN$ drops to worse values in correspondence of the eruption, which correlates in time with the passage of current carrying field through the top boundary (without necessarily contradicting  the direct correlation between $\epsN$ and $\Efree$  noticed for the \sJT{} method in  \sect{TDtw}, cf. \fig{JstJunEn}b with \fig{Etest}e).

Indeed, the simulation \sJst{} and \sJun{} are similar in many respects, but one essential difference from the point of view of the $\Hv$ computation is that the eruption in the \sJun{} case generates the transit of a plasmoid carrying  significant field through the top boundary.
This  seems to have hardly any consequence for DeVore methods,  but may have more severe consequences for Coulomb methods that use the normal component of the field to specify the boundary conditions for $\vA$ (cf. \eq{C_bcAn}).
The reason of the increase in the error might then be due to the fact that the \sJT{} method practically enforces flux balance by concentrating to the top boundary any possible error deriving from a nonsolenoidal input field (see \sect{coulomb}).
As significant field transits through the boundary as a consequence of the eruption, larger solenoidal errors affect the value of the field there and, as a consequence, the solution  of $\vAp$ that is computed form the boundary values.
The correlation between the drop in the metrics of \sJT{} in correspondence with the plasmoid passage through the top boundary supports this speculation.

As for the average $\Hv$ values, it is worth noticing that, for the simulation in exam,  ``larger $\Hv$ value''  does not immediately translate into ``more likely to erupt''.
Indeed,  \fig{JstJunEn} shows that, according to all methods, the \sJun{} case has \textit{less} helicity than the stable one, \sJst{}.

In conclusion, the helicity values $\Hv$  in this most relevant test of MHD evolutions, both stable and unstable, show a very good agreement  between different methods, namely within 3\% in the most challenging \sJun{} case.
It is only using very sensitive metrics such as $\epsN$ that differences between methods of vector potentials' computations can be disclosed.
The results presented in this section are very encouraging for applications of helicity estimation methods to numerical simulations.
The obtained values are largely independent of the specific methods employed, provided that the solenoidal property is sufficiently fulfilled, which allows for accurate and reliable  studies of the properties of helicity as tracer of magnetic field evolution.

\section{Dependence on resolution}\label{s:resolution}
Numerical resolution, taken here to be the voxel size as customary in simulations, affects the solenoidal property of the input field as well as the accuracy of finite volume (\sFV{}) methods in solving for the vector potentials.
Here we address the issue directly, trying to separate each contribution. 

\subsection{\sTD{} resolution test: Major field complexity with little flux through boundaries}\label{s:TDres}
 \begin{figure}
            \includegraphics[width=\imsize,clip=true]{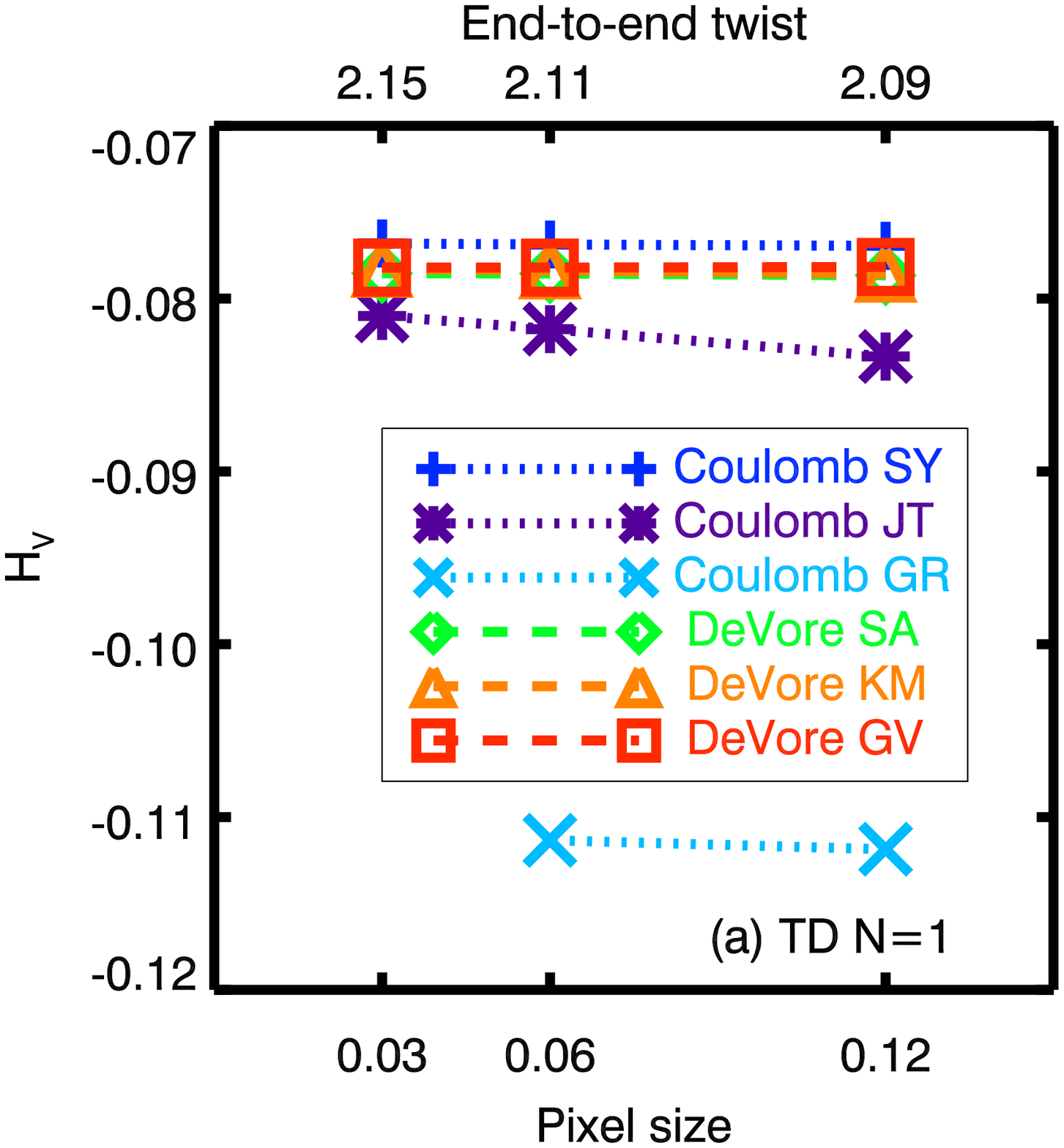} 
            \includegraphics[width=\imsize,clip=true]{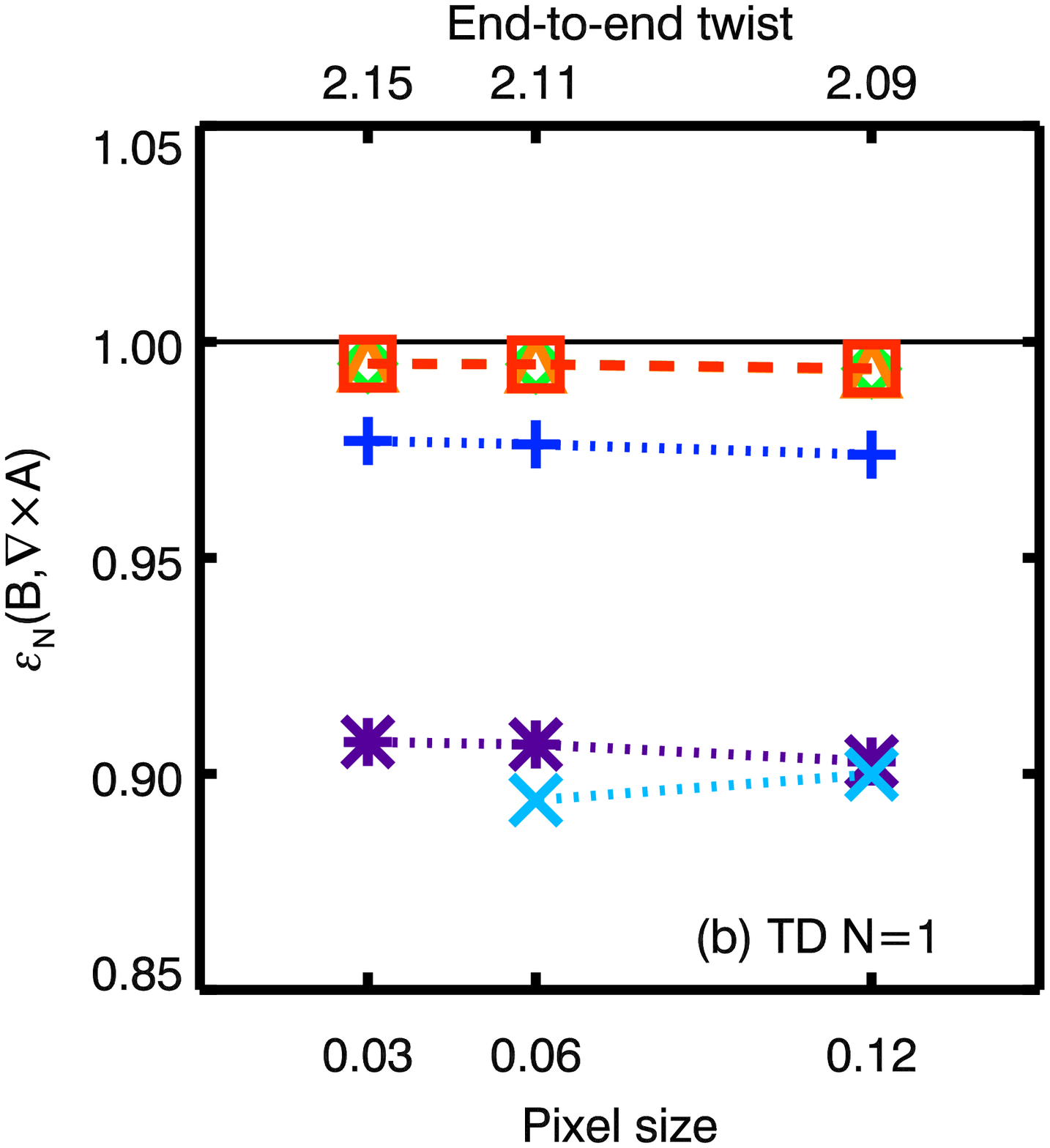}
   \caption{\sTD{} resolution test: (a) Normalized helicity $\Hv$; (b) complement of the normalized vector $\epsN(\vB,\curlA)$.}
   \label{f:TDres}
 \end{figure}
\Fig{Etest}c shows that the solenoidal error in the \sTD{}-resolution cases is practically independent of resolution, and amounts to $ \Ediv\simeq 3\%$  at most in the examined resolution interval, $\Delta=[0.03,0.12]$.
Similarly, the $\Hv$ values for each method separately, see \fig{TDres}a, are all essentially independent of resolution, except for a small variation of about 1\% for the \sJT{} case. 
On the other hand, the spread in values between methods over all resolutions is definitely more significant, being equal to 20\% if all methods are included, and  4\% if \sGR{} is excluded.
Hence, this is a first indication that differences between methods (4\% at least) are more important than resolution (1\% at most) in determining the value of $\Hv$, for similar levels of solenoidal errors.

\Fig{TDres}b shows the complement of the normalized vector for the test field $\epsN(\vB, \curlA)$ (cf. Eq. \ref{eq:En}).
The analysis of the complement of the vector error again separates the methods more markedly.
There are no appreciable differences between the three DeVore methods, all around $\epsN=$0.99.
Errors are larger on average going from \sSY{} ($\epsN=$0.97), to \sJT{} ($\epsN=$0.90), to the \sGR{} ($\epsN=$0.89). 

Since there is no strong dependence on resolution  of the solenoidal error in the \sTD{}-resolution case, then any such dependence that is found in the value of $\Hv$ should be due to specific sensitivity  of the different method to resolution.

\subsection{\sLL{} resolution test: Minor field complexity with significant flux through the boundaries}\label{s:LLres}
 \setlength{\imsize}{0.49\textwidth}
 \begin{figure*}
    \includegraphics[width=\imsize,clip=true]{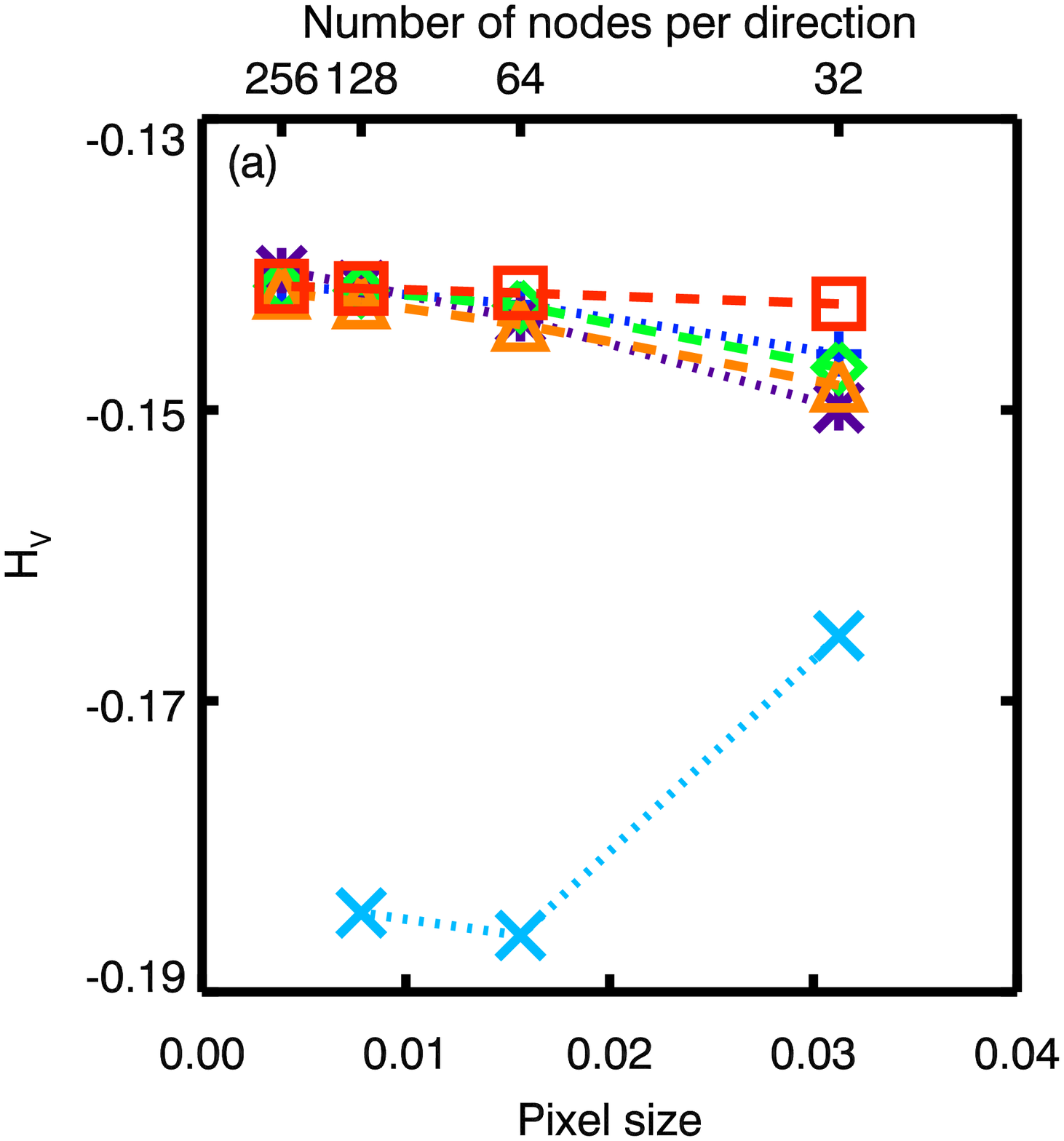}
    \includegraphics[width=\imsize,clip=true]{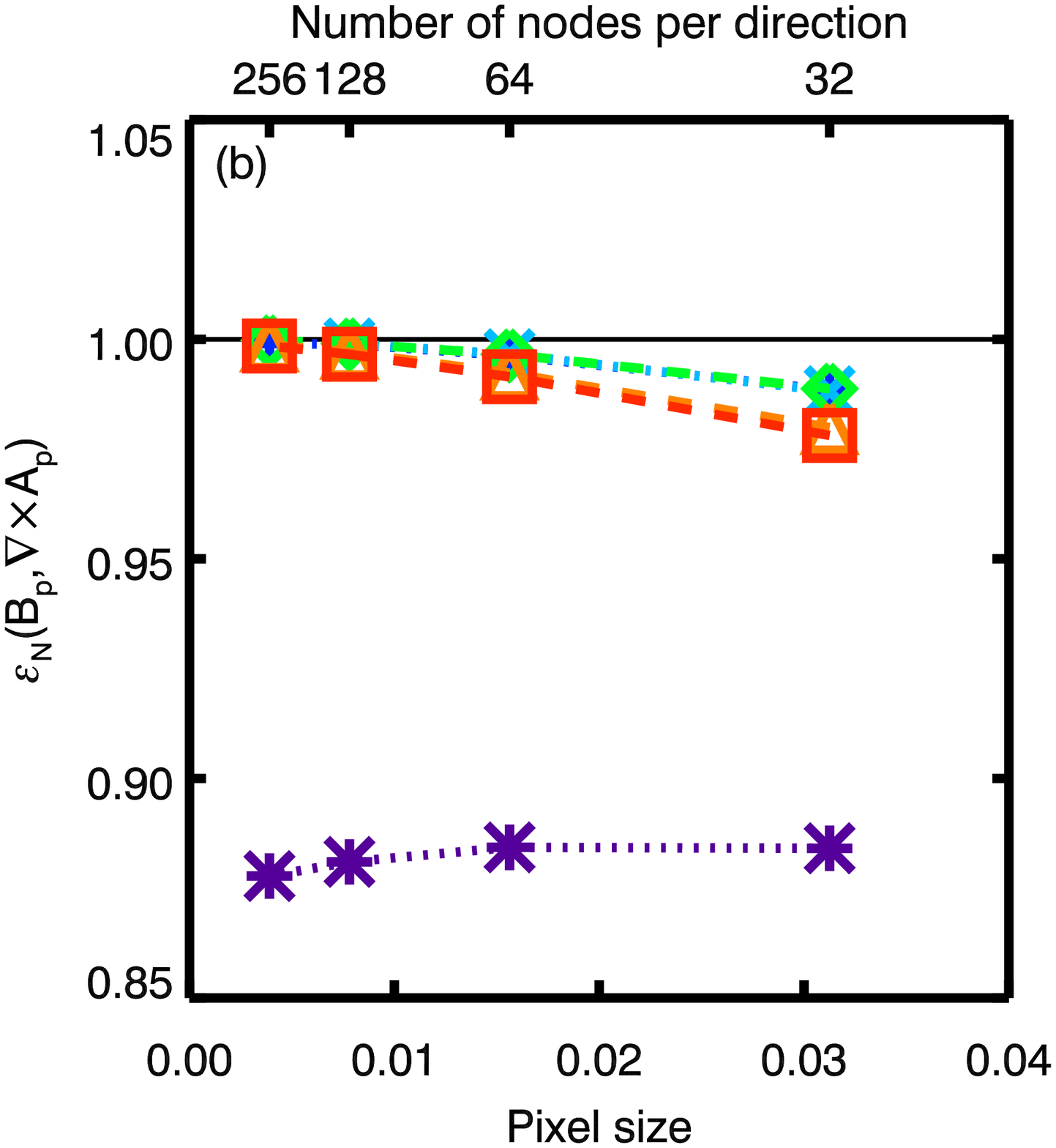}\\
    \includegraphics[width=\imsize,clip=true]{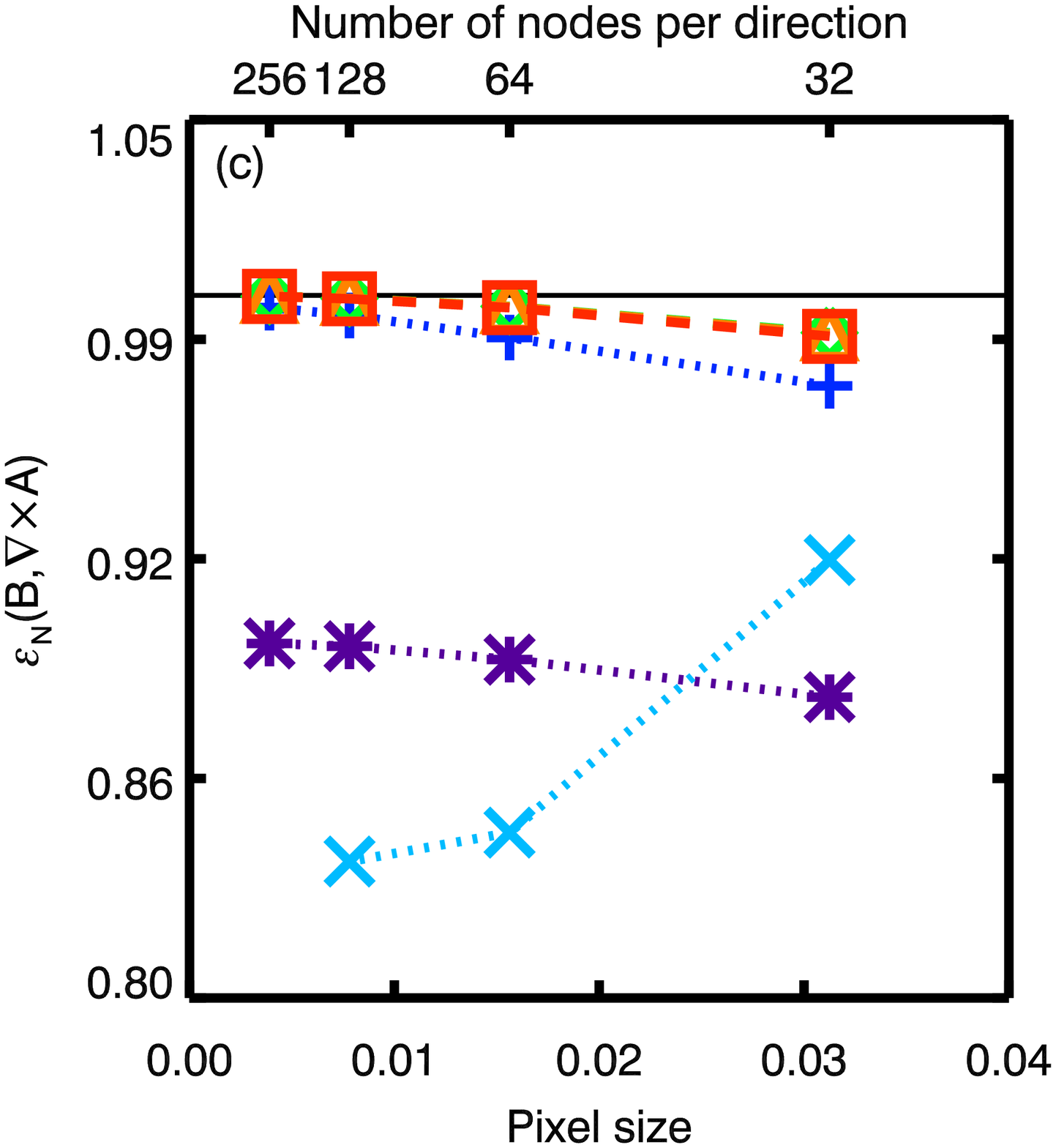}
    \includegraphics[width=\imsize,clip=true]{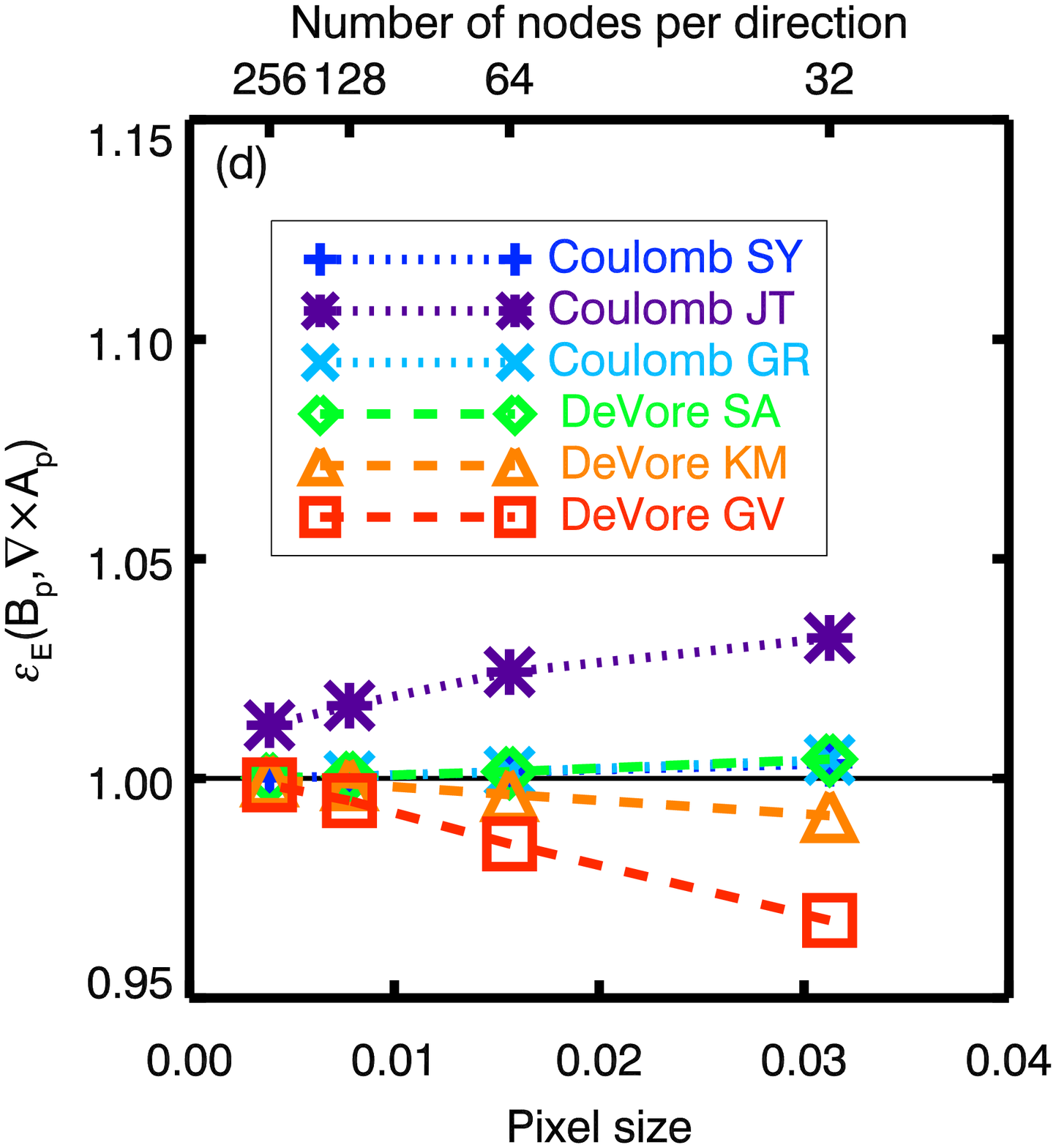}\\
    \includegraphics[width=\imsize,clip=true]{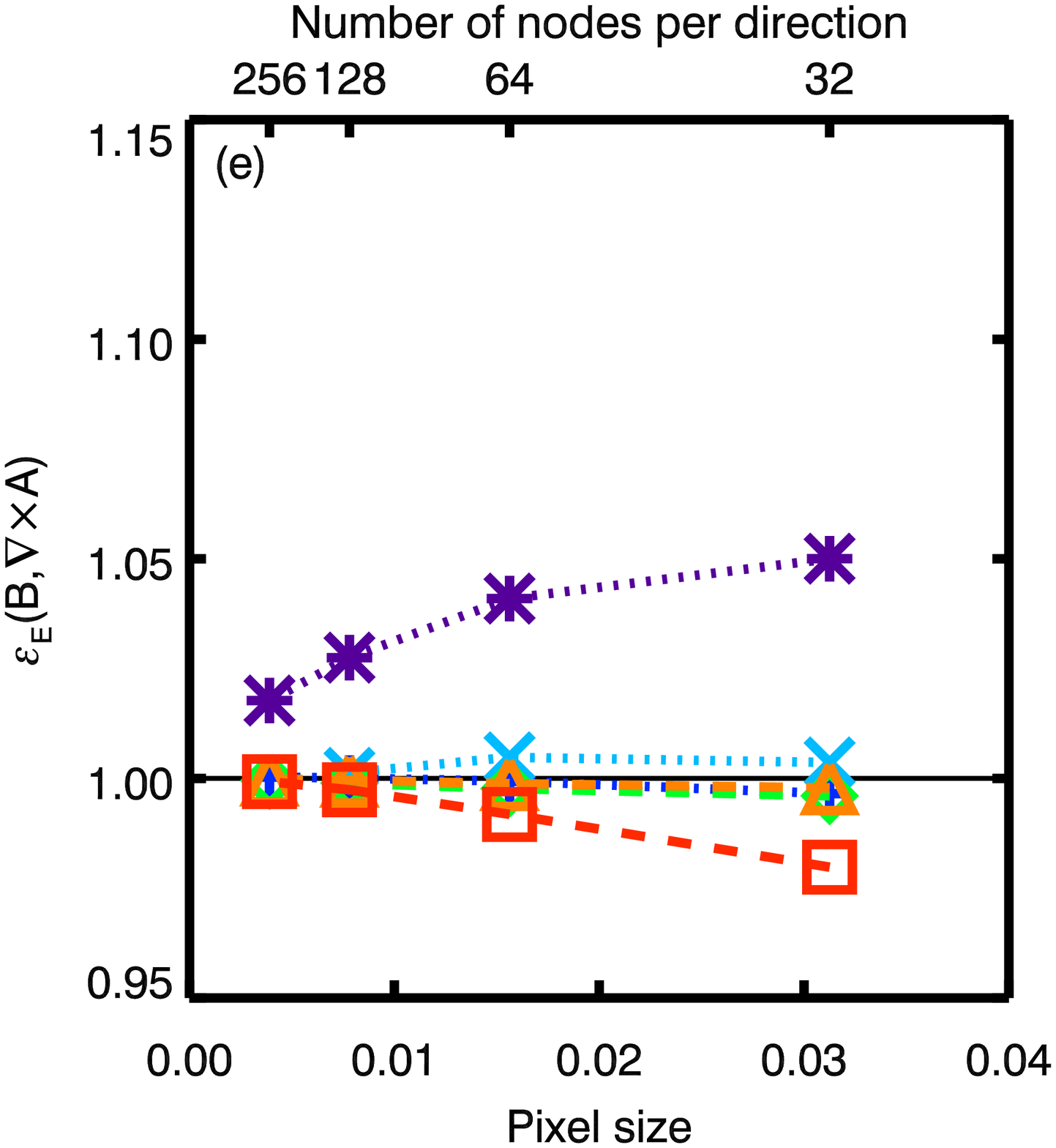}
    \includegraphics[width=\imsize,clip=true]{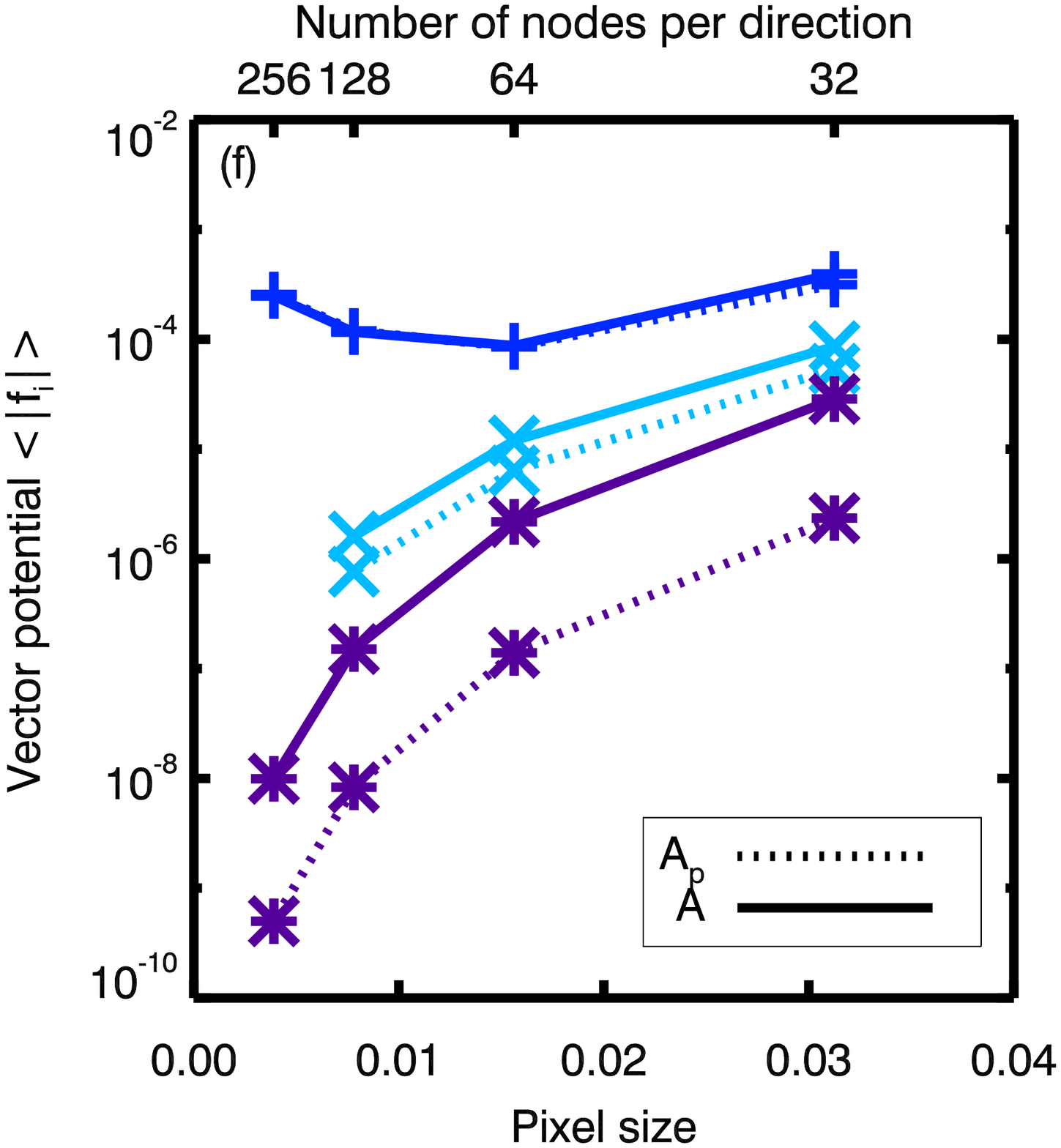}
    \caption{\sLL{} resolution test: 
                 \textbf{a)} Normalized helicity, $\Hv$; 
                  complement of the normalized vector error for the test field for the 
                 \textbf{b)} potential,  $\epsN(\vBp,\curlAp)$, and
                 \textbf{c)} test field, $\epsN(\vB ,\curlA )$; 
                 energy ratios for the 
                 \textbf{d)} potential,  $\epsE(\vBp,\curlAp)$, and
                 \textbf{e)} test field, $\epsE(\vB ,\curlA )$;
                 \textbf{f)} $\avfi{\vAp}$ and $\avfi{\vA}$ for the Coulomb methods only; 
                 as a function of resolution. 
           }
   \label{f:LLres}
 \end{figure*}
In the tests analyzed so far, differences in $\Hv$ are limited, except for the \sGR{} method and, to a slightly lesser extent, the \sJT{} method.
In order to progress further in the analysis we consider the \sLL{} case described in \sect{LL}.

\Fig{LLres} summarizes the detailed analysis of the resolution effects  on the computation of $\Hv$ in that case.
The main result is that  a very clear dependence of $\Hv$ on resolution is found when \sFV{} methods are applied to the \sLL{} case, see in particular \fig{LLres}a.
In more detail, excluding again \sGR{}, the spread in $\Hv$ at $n=64$ is only 0.8\% (11\% if  \sGR{} is included).
This is even smaller than the spread in $\Hv$ for the \sTD{} case of  \sect{TDres}, at a comparable level of $\Ediv$  (cf. the end-to-end twist = 2.11 data point in \fig{Etest}a with the $n=64$ data point  of  \fig{Etest}b).
In the \sLL{} case we can decrease the resolution even further, thereby following the trends over a longer interval.

All methods show an increase of $\Hv$ to more negative values for larger pixel size, with the  exception  of the $n=32$ data point of the \sGR{} method.
This is markedly different from the trend in \sect{TDres} where, for a test field  with basically the same $\Ediv$ across different resolutions, the $\Hv$ values found by different methods were also basically independent of resolution, albeit not the same.
Therefore, the  dependence of $\Hv$ on resolution in the \sLL{} case can be directly related to the presence of larger $\divB$ at lower resolutions, as \fig{Etest}a shows.
On the other hand, for a given resolution, the spread due to different methods ranges from  0.3\% at higher resolution, to  2.3\% at the lower one (5.4\% if  \sGR{} is included).
Among the different methods, \sGV{} give less variable values with $\Delta$, whereas \sJT{} is more sensitive to it (and  \sGR{} even more).

The complement of the normalized vector error and the energy metrics show as usual more differences between the methods. 
The DeVore methods are all very accurate in the computation of $\vAp$, with \sSA{} slightly less sensitive than \sKM{} and \sGV{} (see \fig{LLres}b). 
Differences become entirely negligible in the computation of $\vA$ (cf \fig{LLres}c).
The energy metrics $\epsE$ (cf. Eq \ref{eq:eps}) provide similar information, with more accentuated spreads (see \fig{LLres}d and \fig{LLres}e).
Therefore, we conclude that the most important difference between the three implementations of the DeVore method (\sSA{}, \sKM{}, \sGV{}) is how the potential field is computed.

Among the Coulomb methods, \sSY{} is the more accurate one on average, and is only marginally less accurate than the DeVore methods, having slightly larger inaccuracies in computing $\vAp$ than $\vA$.
The \sGR{} method is even more accurate in this case than most of the DeVore methods in the computation of $\vAp$ (in both $\epsN$ and $\epsE$ metrics), but yields contradictory metrics for $\vA$ (poor $\epsN$ but good $\epsE$).
This results into $\Hv$ values that stand more clearly apart form the trend. 
The  \sGR{}  improvement of $\epsN$ with coarser pixel size in \fig{LLres}c is difficult to interpret without additional testing.
The \sJT{} method lies somewhere in between the other two Coulomb methods, but overall yields  $\Hv$ values  in line with the DeVore methods.
The computation of $\vAp$ (respectively, $\vA$) with the \sJT{} method yields $\epsN\simeq 0.88$ (respectively, $\epsN\simeq 0.90$) but is not particularly sensitive to resolution.
Interestingly, \fig{LLres}d  shows that the spread in energy metric for the potential field has an overall value  including all methods of just 1.2\%, with a maximum value of 2\% at the lowest resolution.
The same metrics for the input fields are 1.6\% and 2.4\%, respectively.
In summary, the energy metric $\epsE$, even though not exactly reproducing the accuracy of the methods as quantified by $\epsN$,  indicates a small spread of energy in the field recomputed from the vector potentials, even at low resolution and for all methods.

DeVore methods can impose the gauge $\Az=0$ and $\Apz=0$ exactly also in numerical implementations.
On the contrary, the gauge conditions $\divA=0$ and $\divAp=0$ in Coulomb methods are numerically fulfilled only up to a finite precision.
Hence, errors in fulfilling the gauge requirements might become a source of inaccuracy by generating spurious, nonsolenoidal components of the vector potentials.
In order to quantify  how well the solenoidal property of the vector potentials is satisfied for each Coulomb-gauge-based FV method, we show in \fig{LLres}f the fractional fluxes, $\avfi{(\vA)}$ and  $\avfi{(\vAp)}$, as defined in \eq{fi}, for the $\vAp$ and $\vA$ solution of the LL resolution test cases, respectively.
All Coulomb methods satisfy the gauge better (\ie have lower $\avfi{}$) at higher resolutions, with similar rates of change, except for the \sSY{} method that has a minimum in correspondence of the $n=64$ data point.
The \sSY{} method satisfies the Coulomb gauge with almost identical accuracy for both $\vAp$ and $\vA$, to a relatively low degree ($\avfi{}\simeq 10^{-4}$). 
The other two Coulomb methods show larger differences in the fulfillment of the solenoidal property of $\vAp$ and $\vA$.
In particular, both the \sGR{} method and the \sJT{} method respect the gauge condition better for $\vAp$ than for $\vA$.
The difference is minimal for \sGR{} ---about  $3\times10^{-5}$--- and about one order of magnitude for \sJT{}.
At the same time, \sJT{} provides the most solenoidal $\vAp$ and $\vA$  of all Coulomb methods, at all resolutions, with values of  $\avfi{}$ below 10$^{-8}$ for both potentials at the highest resolution.

In summary, the proper fulfillment of the Coulomb gauge is attained at different level of accuracy by the different methods, especially as far as  $\vA$ is concerned, but no obvious correlation between violation of the solenoidal property of the vector potentials and accuracy of the vector potential is found.

\section{Dependence on divergence}\label{s:divergence}
 \setlength{\imsize}{0.49\textwidth}
 \begin{figure*}
   \includegraphics[width=\imsize,clip=true]{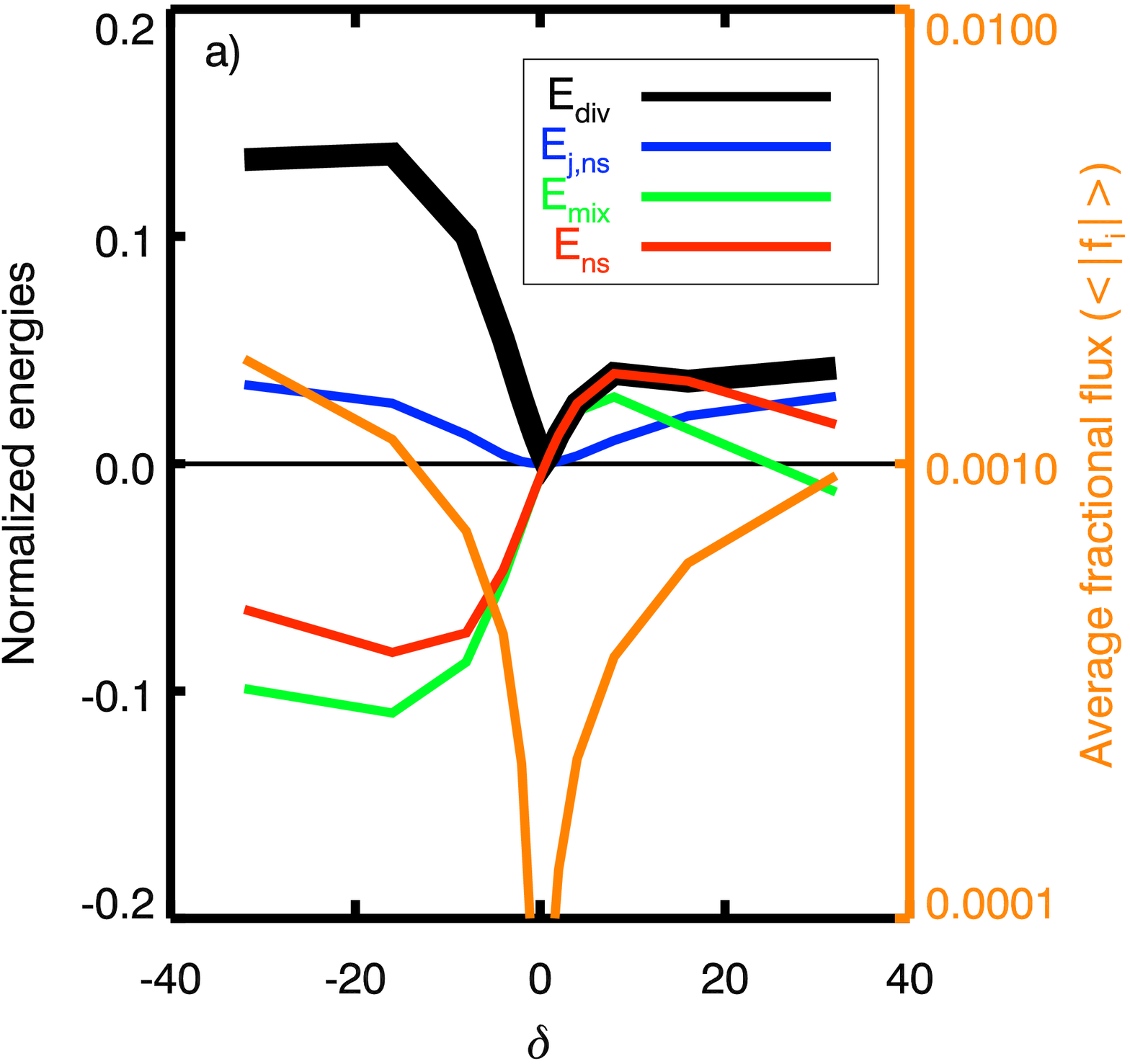} 
   \includegraphics[width=\imsize,clip=true]{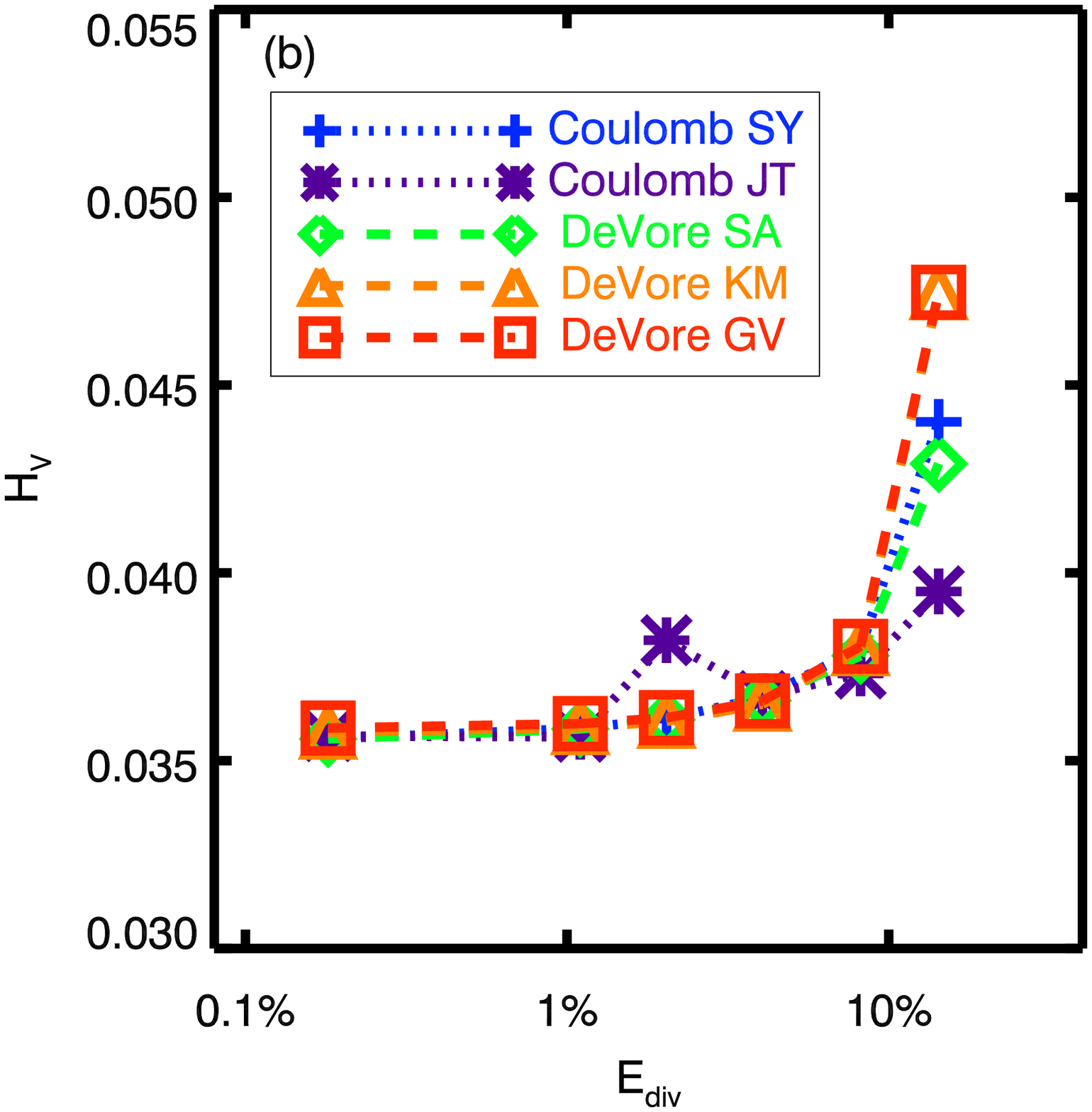}         \\
   \includegraphics[width=\imsize,clip=true]{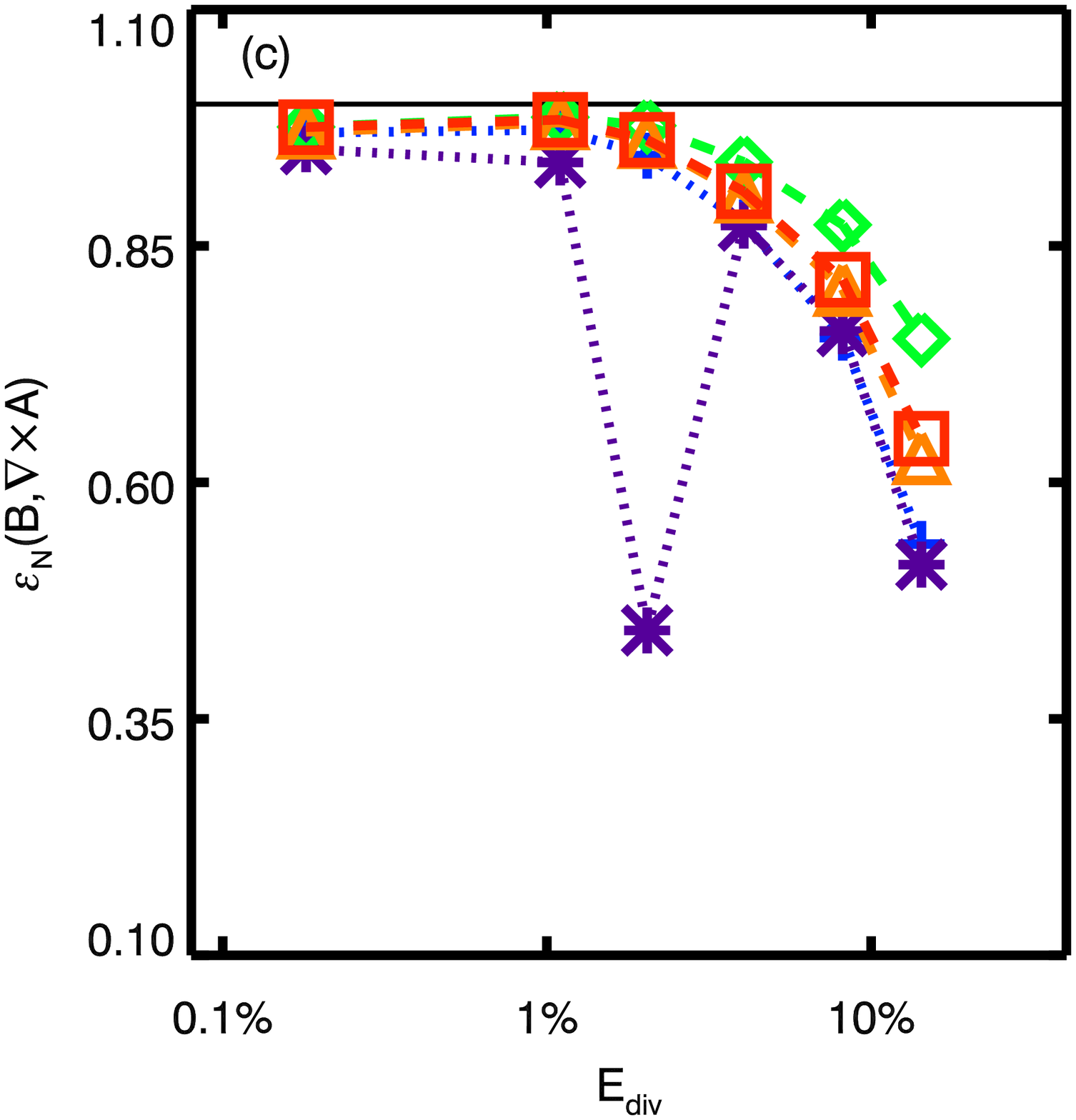}
   \includegraphics[width=\imsize,clip=true]{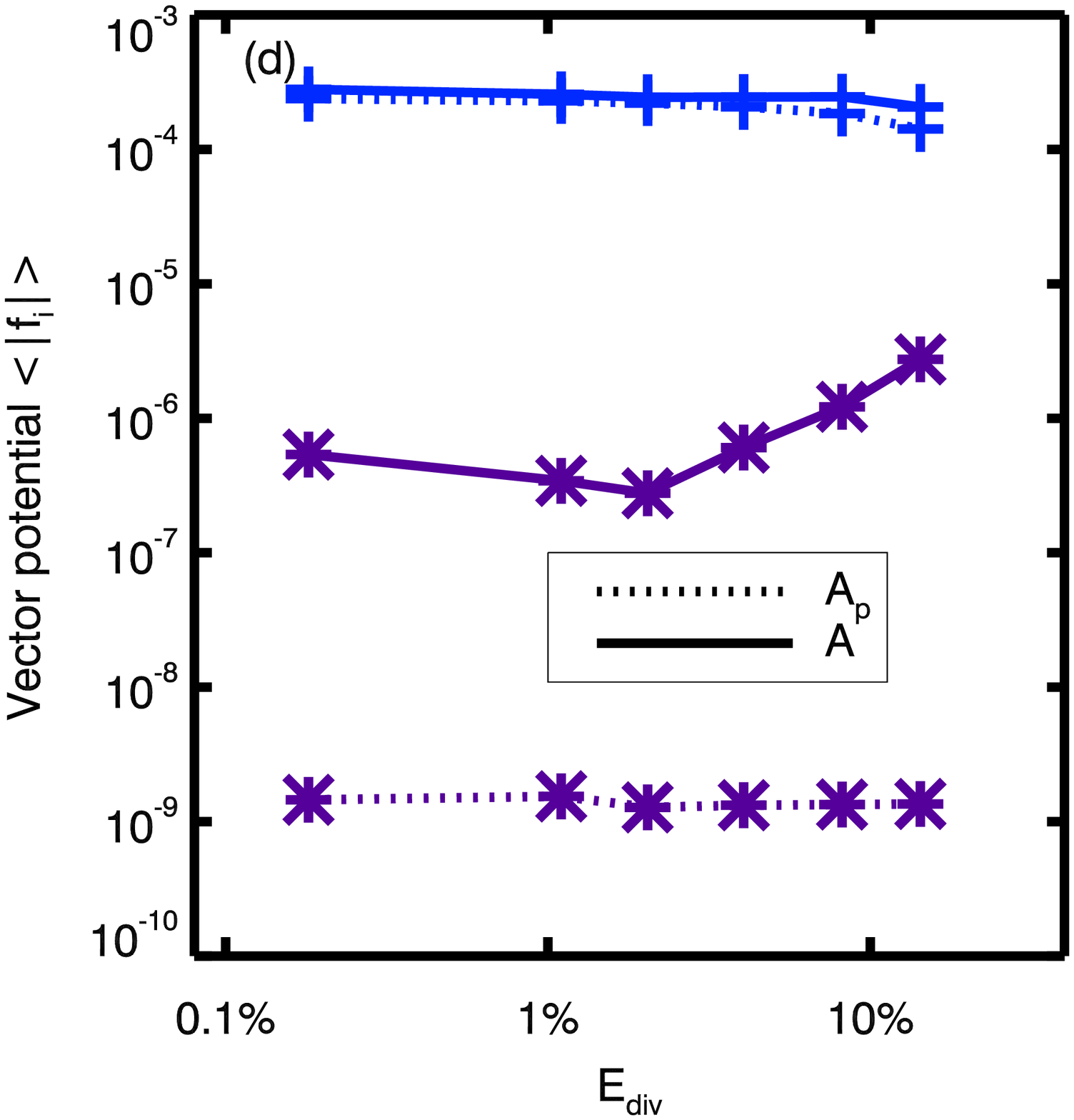}     
   \caption{\sJdiv{} test
          \textbf{ a)} Nonsolenoidal contributions to the energy, as a function of the divergence parameter, $\delta$, see \eqs{Bdiv}{ediv}. 
          As a function of $\Ediv$:                      
          \textbf{ b)} helicity $\Hv$; 
          \textbf{ c)} complement of the normalized vector error for the test field, $\epsN(\vB,\curlA)$;
          \textbf{ d)} $\avfi{\vA}$ for the Coulomb methods only, 
   }
   \label{f:JstDiv}
 \end{figure*}
In \sect{TDres} and \ref{s:LLres} we show that the pixel size affects not only the accuracy of methods in solving for the vector potentials, but also the degree of divergence that is caused by the resolution of the test fields.
More generally, in practical cases, when trying to estimate the helicity of a 3D dataset, a nonzero divergence of $\vB$ is always present. 
It is thus of practical importance to determine the level of confidence of an helicity measurement given its level of nonsolenoidality, here expressed in function of $\Ediv$. 
The present section aims at giving a first test of the impact of the nonsolenoidal effect on the degree of precision of helicity measurements.

Similarly to classical helicity (cf. \eq{variance}), assuming that $\vB$ and $\vBp$ effectively have the same distribution of the normal component on $\surf$, a gauge transformation  $(\vA,\vAp) \longrightarrow (\vA +\Nabla \psi, \vAp +\Nabla \psi_p)$ induce the following difference on the relative magnetic helicity: 
\BE
 \Hdef(\vA +\Nabla \psi,\vAp +\Nabla \psi_p) = \Hdef(\vA,\vAp) - \intv (\psi+\psi_p) \left(\divB-\divBp\right) \dV \, .
 \label{eq:variance_2pot}
\EE
Comparing one \sFV{} method against another is theoretically equivalent to performing a gauge transformation: for a purely solenoidal field they should provide identical values of helicity within the numerical precision of each method. 
While for a finite nonsolenoidal field it is always theoretically possible to find a gauge transformation that will lead to a significantly large difference, in practice most methods are solving for $\vA$ and $\vAp$ in such a way that the amplitude of the vector potentials turns out to be dominant over the contributions from the gauges $\psi$ and $\psi_p$.
Hence, for a given dataset, the difference of amplitude of the vector potentials computed by each method is relatively small. 
Thus, for a given $\Ediv$, the different methods provide helicity values that remain within a certain range of each other. 
As $\Ediv$ is infinitely small, the methods are expected to provide helicity values which are close to each other, while for large $\Ediv$, the methods should provide helicity values presenting a larger spread.    

In the present section, we consider the \sJdiv{} test described in \sect{divb_test}, which allows us to estimate the dispersion of the helicity values obtained with the different methods for different level of nonsolenoidality. 
In the present test the \sGR{} method is not included because of computational limitations.

The amplitude of the nonsolenoidal component in the  \sJdiv{} test field as quantified by  $\Ediv$ for different values of $\delta$ in \eq{Bdiv} is reported in \tab{Bdiv} and shown in \fig{Etest}f, where $\Ediv$ is shown growing from 0.2\% up to 14\% for the considered range of $\delta $ values (the corresponding values of $\avfi{}$ can be find in \tab{input}).  
The contributions $\EdivBJ$ and $\Emix$ to $\Ediv$ in \eq{ediv} and their variation with $\delta$ are shown in  \fig{JstDiv}a.  
The energy $\Ediv$ turns out to be a nonlinear function of $\delta$, with a minimum close, but not at, $\delta=0$.  
This apparent contradiction is due to the definition \eq{ediv} that forbids cancellation between different contributions, and practically shifts the zero of $\Ediv$ to slightly higher values.  
For completeness, the nonsolenoidal error $\Ens$ given by \eq{ediv} without the absolute value, \ie allowing for cancellation between terms, is also plotted in \fig{JstDiv}a.

The values of $\Hv$ obtained in the \sJdiv{} test cases are plotted in \fig{JstDiv}b.
We confirmed that for low values of $\Ediv$, the methods are providing $\Hv$ which are close to each other, while the dispersion of the values obtained are spreading as $\Ediv$ is increasing. 
All method follow a qualitatively similar exponential trend as a function of $\Ediv$, both in $\Hv$ (\fig{JstDiv}b)  as well as in the accuracy metric $\epsN(\vB,\curlA)$  (\fig{JstDiv}c).  
The only exception to a smooth trend is the $\Ediv=$2\%-case of the \sJT{} method, for which no particular reason was identified.

In the most solenoidal case, for $\Ediv=0.2$\%, the relative dispersion of $\Hv$ obtained is of 0.8\%.  
Excluding the $\Ediv=$2\%-case of the \sJT{} method, for each case with $\Ediv \le 4$\%, the relative spread in $\Hv$ remains lower than 1\%.
At $\Ediv=$2\%, taking all the methods into account, the relative spread in the $\Hv$ is of the order of 6\%. 


Excluding the \sGR{} method, we note that in all the previous tests, the maximum spread of $\Hv$ observed was of 4\% (in the \sTD{}-resolution test of \sect{TDres}). 
This incline us to state that within that range of $\Ediv$, the differences between the methods are not related to the nonsolenoidality but rather to the intrinsic numerical error within each method.

For $\Ediv = 8.2$\%, the relative dispersion of $\Hv$ is also relatively low, equal to $1.9$\%. 
However, in the least solenoidal case, for $\Ediv=14.4$\%, the dispersion in helicity estimation obtained for the different methods reaches $18$\% of the average value. 
This is much higher than most of the errors that we have encountered so far. 
We believe that in this range of $\Ediv$, the gauge dependence is directly impacting the estimation of $\Hv$. 

The implementations of DeVore's method (\sKM{}, \sGV{} and \sSA{}) yields very similar results, both in values and in trends.
The trend already noticed in \fig{LLres}b is confirmed by the accuracy of the vector potential quantified by $\epsN$ in  \fig{JstDiv}c.
From the analysis in \sect{LLres}, we tend to attribute this difference to the accuracy in the solution of \eq{dV_phi} being higher for \sSA{} with respect to \sKM{} and  \sGV{}.
Similarly, the \sSY{} and  \sJT{} methods deliver analogous $\Hv$ and  $\epsN$ curves (see \fig{JstDiv}b and c).
However, they strongly differ in the accuracy with which the Coulomb gauge condition is respected, as \fig{JstDiv}d shows.
Confirming the result of the \sLL{} case of \sect{LL}, the \sJT{} method respect the Coulomb-gauge condition extremely well for $\vAp$ ( $\avfi{}\simeq 10^{-9}$), and still excellently for $\vA$ ($\avfi{}\simeq 10^{-6}$ or lower).
Interestingly, $\avfi{(\vAp)}$ is practically independent of $\Ediv$, confirming that the strategy adopted by the \sJT{} method for solving \eq{C_Ap} is very much able to handle flux unbalance resulting from solenoidal errors in $\vol$.
However, even though the solenoidal property of the vector potentials is definitely better verified by  the \sJT{} method than by the  \sSY{}, the accuracy of the vector potential does not reflect this trend.
The \sSY{} vector potentials, indeed, have much higher values  of  $\avfi{(\vAp)}\simeq\avfi{(\vA)}\simeq10^{-4}$ than those by the \sJT{} method.
In this sense, the divergence-cleaning strategy adopted by \sSY{} is less efficient in imposing the Coulomb gauge.
However, such values of $\avfi{}$ seem still low enough to guarantee a relative high accuracy, as testified by  $\epsN$ of \fig{JstDiv}c.

The test discussed here, of course, does not pretend to be general in assessing the influence of the nonsolenoidality on helicity estimations, and further tests are likely to be required. 
However, it represents one well-controlled example that enables an estimate of the degree of confidence of a helicity estimations for given a finite nonsolenoidality level. 

In summary, according to our tests, errors in respecting the solenoidal constraint might be still ignorable as long as $\Ediv$ is below 1\%, but become abruptly more important above that threshold.
For a dataset with $\Ediv$ comprised between 1 and 8\%, using on FV method or the other would lead up to 6\% difference in the estimation of $\Hv$ (excluding \sGR{} method). 
For higher $\Ediv$, the gauge invariance starts to have significant effects.  

According to the above discussion, most of the tests employed in the remaining sections of this work have nonsolenoidal contributions that are mostly ignorable, with few data points where $\Ediv$ (and hence its influence on $\Hv$) is of the order of few percent (cf. \fig{Etest} and \tab{input}).

\section{Discussion of \sFV{} methods}\label{s:discussion}
The previous sections discuss the results of  \sFV{} methods when applied to a variety of test fields that differ for topological complexity, importance and distribution of currents in the volume, stability properties.
Factors that seem to influence the accuracy in the computation of $\Hv$ in \sFV{} methods range from the distribution of strong currents at the top boundary (see \sJT{} in \sect{JLmhd} in particular), to the solver employed for the construction of the reference potential field, as for the DeVore methods in \sect{TDres}.
In this sense, the solar-like separation between a current-carrying and a more potential part of the coronal field seems to favor accuracy in helicity computations, possibly because in this way currents do not affect the boundary conditions for Coulomb methods.
However, as \sTD{} and  \sLL{} tests show, the accuracy of the helicity computed by different methods is not always directly related to the accuracy of the vector potentials in reproducing the corresponding fields. 
Two of the Coulomb methods, \sJT{} and \sSY{}, despite their lesser accuracy in solving for the vector potentials (when compared to the accuracy of the DeVore methods), they still deliver a helicity in line with that obtained by the DeVore methods. 
On the other hand, while, \eg \sGR{} is of similar (in)accuracy as the other two Coulomb methods, it delivers slightly different helicity values.
Likely, given the nonlocal nature of $\Hv$, such details depend on the particular spatial distribution \referee{of the solenoidal errors} on a case by case basis, which may affect how the different distributions of values combine into $\Hv$.
This is in a way confirmed by the remarkable absence of differences in the \sJst{} case, in which $\Hv$ is dominated by the large contribution of the potential field.
Similarly, we find no strong influence on the methods' accuracy by the amount of twist in the field of the \sTD{} case, where the self-helicity $\Hj$ is almost a  factor 100 smaller than the total helicity $\Hv$, \sect{TDtw}.

A strong effect on helicity values is found from errors in the solenoidal property of the input field.
In \sect{divergence} a test field is considered that has increasing value of solenoidal error, as measured by the normalized fraction of the energy associated with magnetic monopoles, $\Ediv$. 

We find a rapid increase of $\Hv$ fluctuations as $\Ediv$ grows above 1\%, see \eg \fig{JstDiv}d. 
We also extensively test the effect of resolution, which is found to affect helicity values basically in two ways: Directly, by  affecting the solution of the Poisson solvers that compute the scalar or vector potentials, and indirectly, by increasing the solenoidal error in the input field and consequently \referee{weakening} the consistency between potentials and boundary conditions.
By increasing the solenoidal error, resolution may influence the helicity value significantly in heavily under-resolved cases, as in \fig{LLres}, for which systematic  quantifications of solenoidal errors must be put in place.
On the other hand, when resolution is not pushed to limits,  differences between methods account for larger spread in $\Hv$ values than resolution, as \fig{TDres}a and \sect{TDres} show. 

From the point of view of the accuracy, vector potentials computed with the DeVore methods reproduce the input field more accurately in most of the cases, see \eg \fig{TDres}.
Among the DeVore methods, accuracy is improved mostly for better solutions of the Laplace equation defining the potential field, \eq{dV_phi}.
Other details of implementation, such as the definition of derivative and integrals, play a minor role.

Coulomb methods, need to impose the solenoidality of the vector potential in the entire volume, and may 
\referee{suffer} more from the inconsistent boundary values for Laplace/Poisson equations that the methods solve.
In this respect, the \sSY{} strategy of divergence cleaning is not as efficient as the parametric tuning of the integration constants (the $c_f$ constants of \sect{methods}) employed by \sJT{} (see \fig{LLres}f and \fig{JstDiv}d). 
On the other hand, the accuracy of the \sSY{} method in the construction of the vector potentials is found in all tests performed here to be always better than that of the \sJT{} method.
Therefore, the accuracy of the vector potential is not directly influenced by the accuracy with which the gauge condition is satisfied.
The computation of $\vA$ poses in general more difficulties for Coulomb methods than that of $\vAp$, in a way contrary to the DeVore methods.
In particular, the \sJT{} method seems to be sensitive to current on the boundaries (as \sJun{} show, see \fig{JstJunEn}), possibly, because they yield inconsistent boundary values for \eqs{C_Ap}{C_A}.
The \sGR{} method, finally, has mostly issues in computing an $\vA$ that reproduces the field accurately enough, which result in the largest departures of $\Hv$ from average values of all methods.

In the tests presented here it is clearly found that the fulfillment of the Coulomb gauge is quite variable among the Coulomb methods.
However, using the \sLL{} test case as a reference, we find that an average value of the fractional flux of $\avfi{(\vA)}$ below $10^{-3}$ yields helicity values comparable with those from  DeVore methods  (\eg within 0.8\% in the moderate-resolution case of $n=64$, see \sect{LL}). 
 
Hence, in a balance between accuracy and applicability, a clear advantage of DeVore methods over Coulomb methods is that the gauge can be imposed exactly in the former, whereas it need to be insured numerically on the latter.
This translates into simpler and more efficient implementations of the method, and to more accurate estimations of $\Hv$.
On the other hand, the Coulomb gauge yields generally simpler analytical expression of more straightforward interpretation by eliminating those terms that depend on the divergence of the vector potentials.
This offers, for instance,  a possibility of a better comparison and integration with \sFI{} methods.
Also, the Coulomb gauge allows a natural interpretation of helicity in terms of Gauss linking number, see \eg \cite{1999PPCF...41B.167B} and references therein, which is an interpretation tool often used in helicity studies.

\section{Comparison with \tTM{} methods}\label{s:resultsothers}
On the grounds of the discussion of the previous sections, and with the limitations there specified, we consider here one of the \sFV{} methods, namely the \sGV{} one, to give the correct value of helicity, and we compare the \sTM{} methods against it.
\referee{As discussed in \sect{othermethods}, the comparison between \sTM{} and \sFV{} methods is by necessity limited to the estimated helicity value, given the different level of approximation and required information of the two groups of methods.}
\subsection{\TTE{} method}\label{s:TE}
 \begin{figure}
   \includegraphics[height=0.4\textwidth,clip=true]{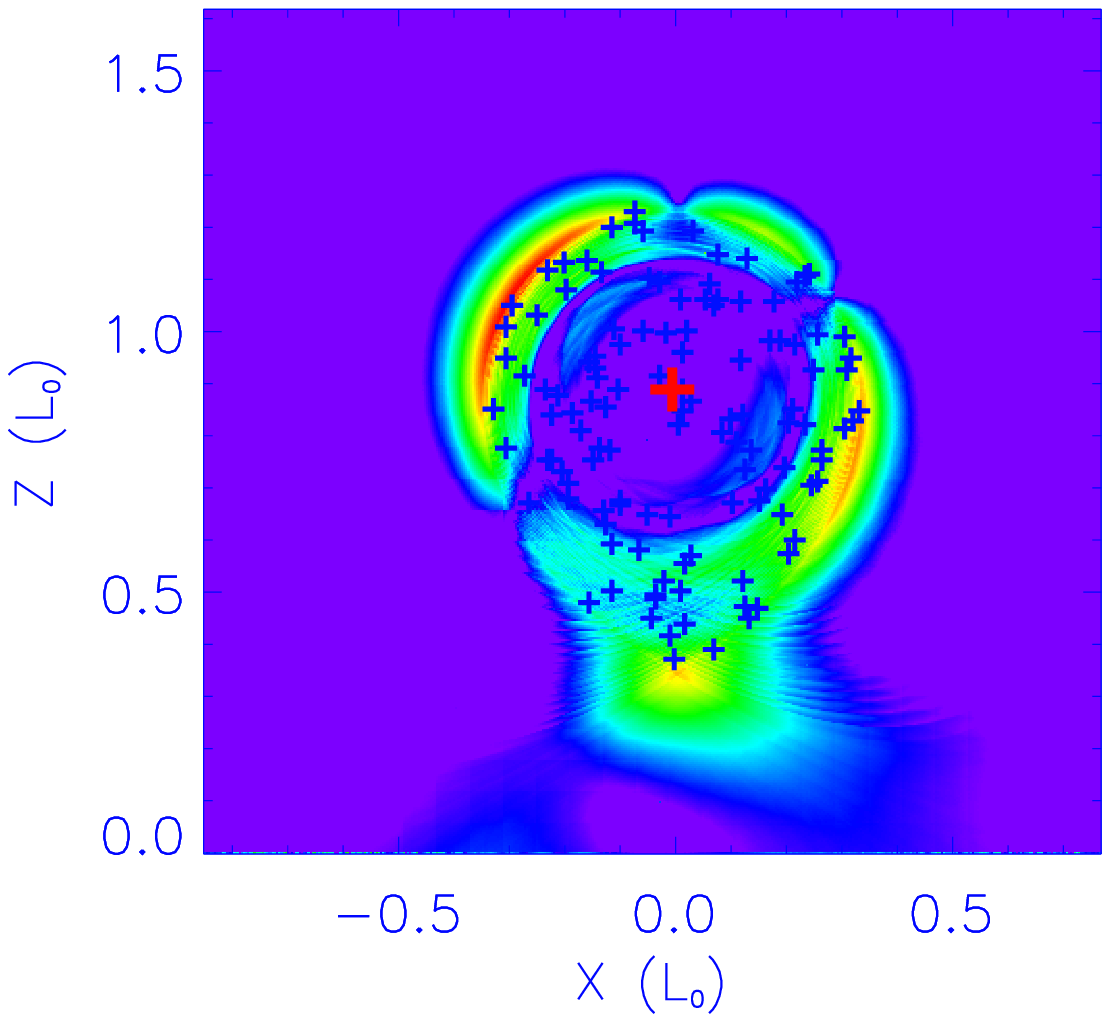}  
   \includegraphics[height=0.4\textwidth,clip=true]{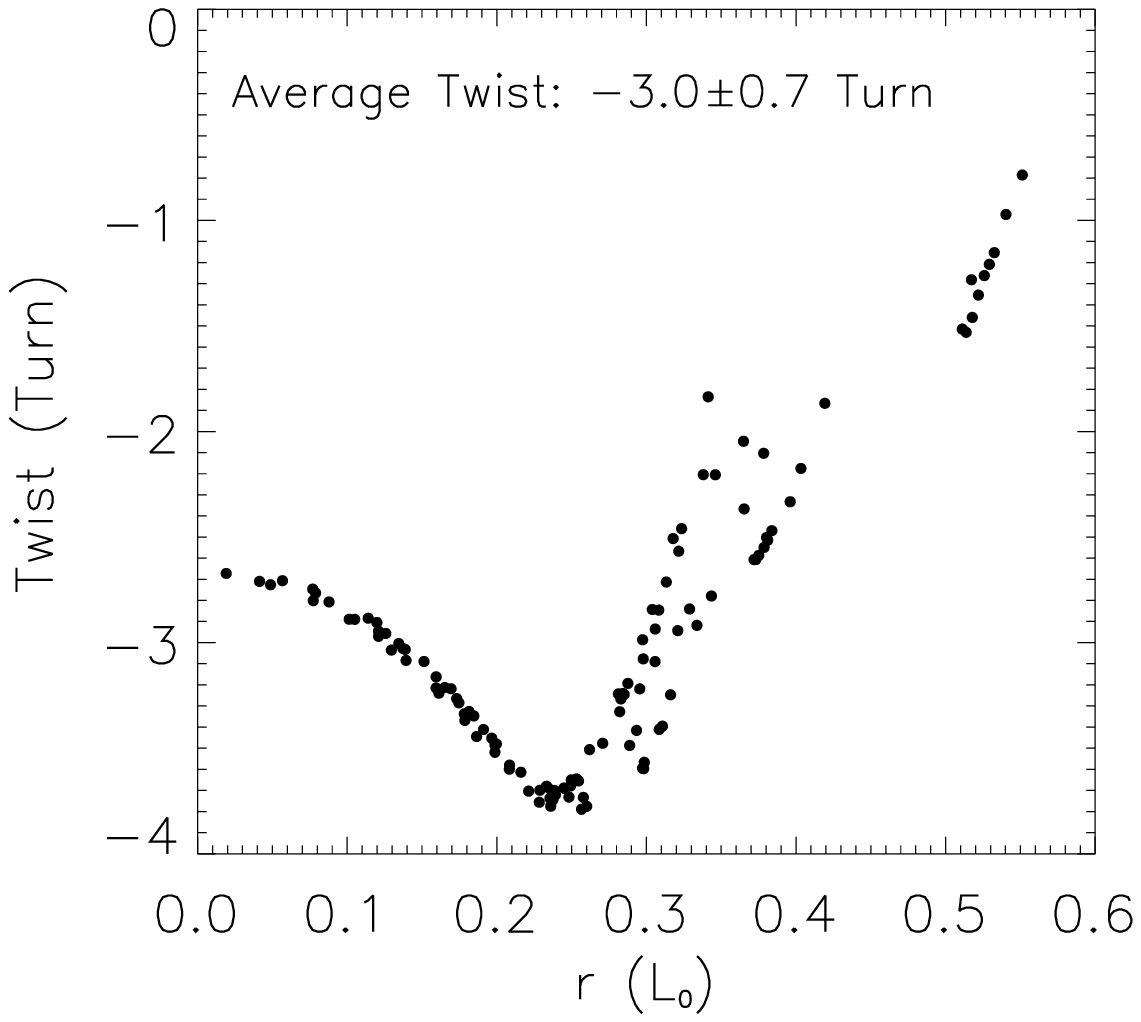}
   \caption{Application of the \sTE{} method to the \sTD{} $N=3$ case. \textbf{Left:} A vertical slice of the Q map in the $xz$-plane at $y = 0$  corresponding to the middle of the flux rope along its axis. Red (blue) plus sign indicates the position of the axis (sample field lines). 
            \textbf{Right:} Twist of the sample magnetic field lines along the distance, $r$, from the origin. 
           }
   \label{f:TD_QSL}
 \end{figure}

The \tTE{} method needs identifiable  flux-rope structures in order to be applied. 
\referee{The dependence of the method on some of its parameters (\eg on the choice of the QSL surface defining the flux rope and  on the location of the axes of the flux rope, see \sect{TE_method}) is specific to the method and requires a testing strategy that is different from the one applicable to the other \sFV{} methods. 
Therefore, we defer that discussion to a separate dedicated paper, in preparation at the time of writing, where  the \sTE{} method is tested using \sTD{}, \sJst{} and  \sJun{} cases, as well as some nonlinear force-free field models \citep{Guo2016tmp}.
Here, we only report the results of the application of the \sTE{} method to  the \sTD{} cases}. 

The 
\referee{Q map \citep[as defined by Eq.~(24) in ][]{2002JGRA..107.1164T} defining the QSL} 
used to identify the flux rope volume is shown in the left panel of \fig{TD_QSL} in the $N=3$ case.
We select 100 sample field lines, which are randomly distributed within the QSL, for the magnetic flux rope. 
We compute the twist of each field line referred to the axis as described in \sect{othermethods}. 
The (nonnormalized) twist of the magnetic flux rope $\Hself$ defined in  \eq{H_TE} is computed as the average of the twist of the sample field lines, and the uncertainties are computed by the standard deviation of their distribution
\referee{(see \tab{td_TE})}.
An example of the twist distribution of the sample magnetic field lines as a function of the distance from the flux rope axis is given in the right panel of \fig{TD_QSL}.

Since the \sTE{} method approximates the total helicity by the self-helicity only, \fig{TD_TE} compares the not-normalized value of $\HdefJ$ of \eq{HdefJ} from the \sGV{} method with the $\Hself$ from the \sTE{} method.
A more extensive view of the comparison is reported in \tab{td_TE}.
\begin{table}[h]
 \caption{Helicity in the twist(upper) and resolution (lower) TD tests, computed with the \sFV{}-\sGV{} method (2$^{nd}$  to 4$^{th}$ column), and  with the \sTE{} method  (5$^{th}$ column). Note that  $\Hdef$,   $\HdefJ$, and $\Hself$ are not normalized.
 \label{t:td_TE}}
 \begin{tabular}{@{~}l | c@{\quad}  c@{\quad}  c@{\quad} | c@{\quad}}
 \hline
  TD case        &  $\Hv$    &     $\Hdef$ &   $\HdefJ$ &     $\Hself$       \\
                 &           & (\sFV{}-\sGV{}) &        & ( \sTE{})          \\
 \hline
   $N=0.1$       & -0.0057 &     -2.23147 &    0.00084 & -0.16  $\pm$ 0.06   \\
   $N=0.5$       & -0.0290 &     -8.33340 &   -0.08374 & -0.66  $\pm$ 0.13   \\
   $N=1  $       & -0.0782 &     -7.21022 &   -0.55240 & -0.50  $\pm$ 0.05  \\
   $N=3  $       & -0.0527 &     -1.78752 &   -0.08976 & -0.087 $\pm$ 0.020  \\
 \hline                                                    
   $\Delta=0.03$ & -0.0782 &     -7.20592 &   -0.54921 & -0.56  $\pm$ 0.06  \\
   $\Delta=0.06$ & -0.0782 &     -7.21022 &   -0.55240 & -0.50  $\pm$ 0.05  \\
   $\Delta=0.12$ & -0.0782 &     -7.21039 &   -0.55444 & -0.49  $\pm$ 0.05  \\
 \end{tabular}
\end{table}
 
We find that the accuracy of the estimation increases for higher resolutions and higher twist.
In particular, in the $N=3$ case, $\Hself$ has the same value of $\HdefJ$ within errors.
In the other cases, larger differences between the \sTE{} method and \sGV{}  method are present.
By contrast, the not-normalized values of the total helicity $\Hdef=-7.2$ for \sGV{} for the case $N=1$, \ie the total helicity is two orders of magnitude larger than the current helicity in the $N=1$ case.

We conclude that the quantity  $\Hself$  provided by the \sTE{} method is an  estimation of $\HdefJ$  rather than of $\HH$ (or equivalently $\Hv$), likely because no modeling of the mutual helicity part is included.
Moreover, the accuracy of the method is higher for highly twisted structures, above $N=1$ in out tests.
For $N=3$, the correct value of $\HdefJ$ is reproduced.

 \setlength{\imsize}{0.49\textwidth}
 \begin{figure}
   \includegraphics[width=\imsize,clip=true]{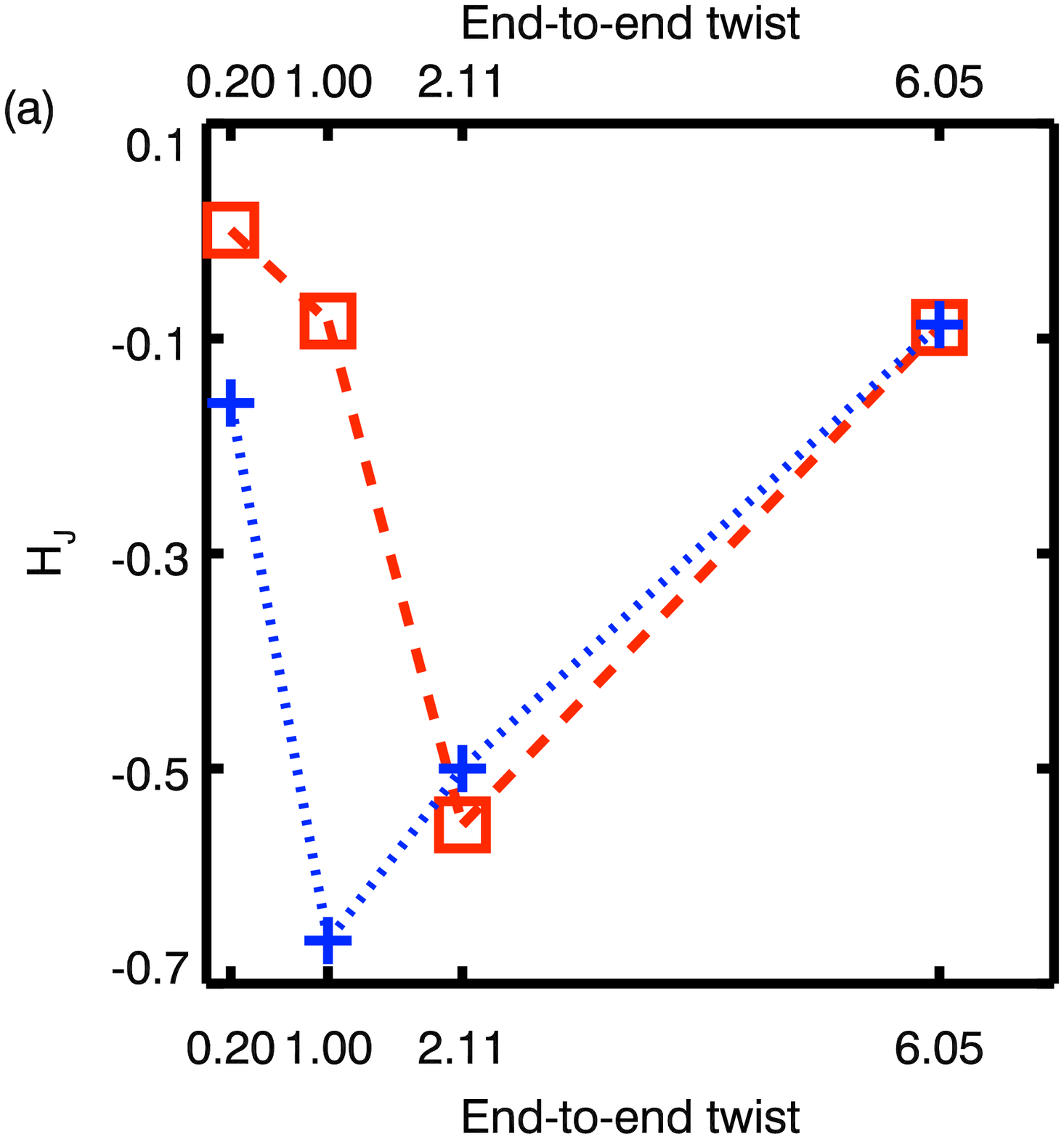}
   \includegraphics[width=\imsize,clip=true]{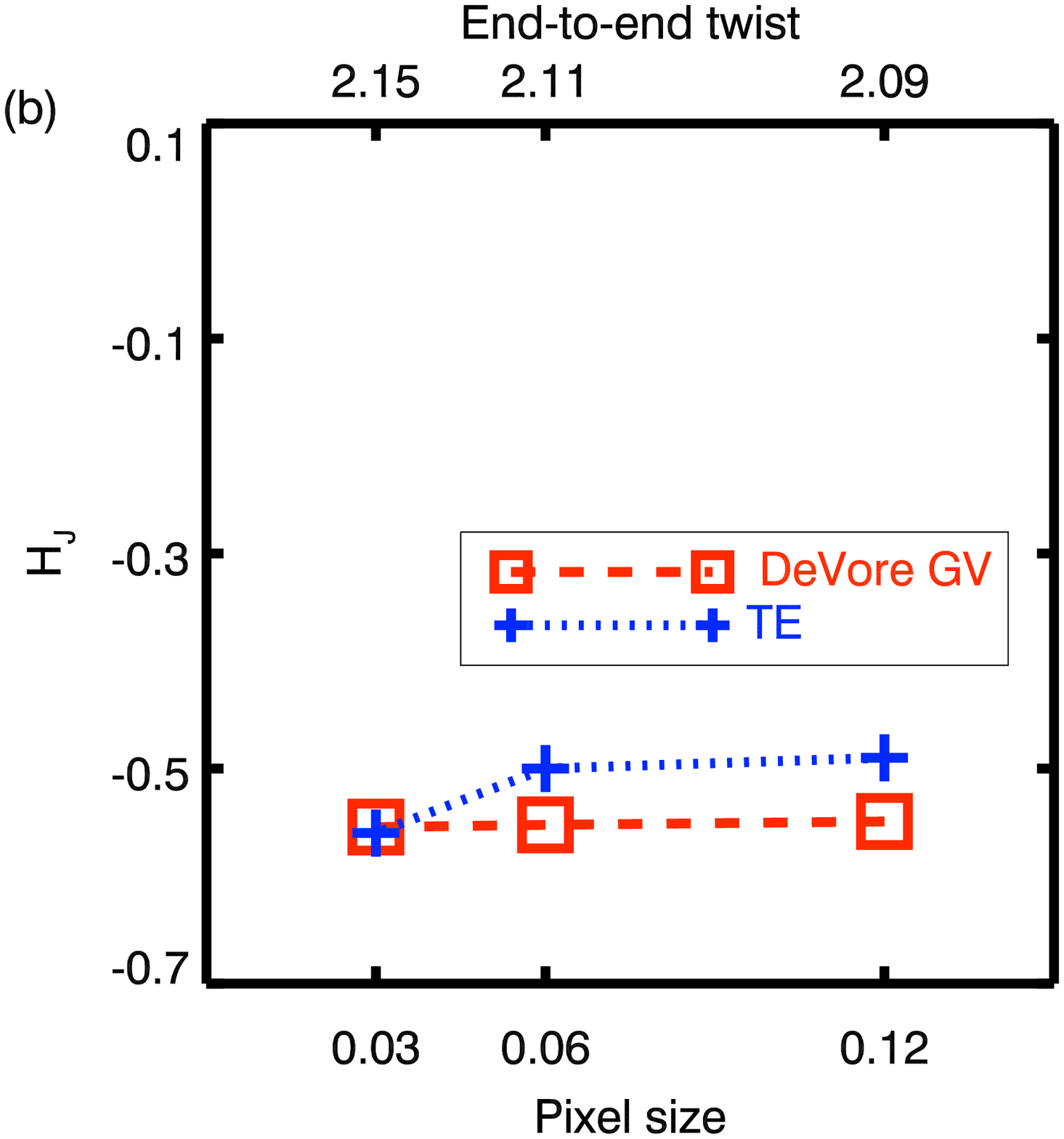}
   \caption{Application of the \sTE{} method to the \sTD{}-twist (\textbf{a}) and -resolution (\textbf{b}) tests.
            The two curves represent the values of $\Hself$ of \eq{H_TE} for the \sTE{} method, and  of the (nonnormalized) value of  $\HdefJ$  for the \sGV{} method.}
   \label{f:TD_TE}
 \end{figure}

\subsection{\TCB{} method}\label{s:CB}
 \begin{figure}
   \includegraphics[width=\imsize,clip=true]{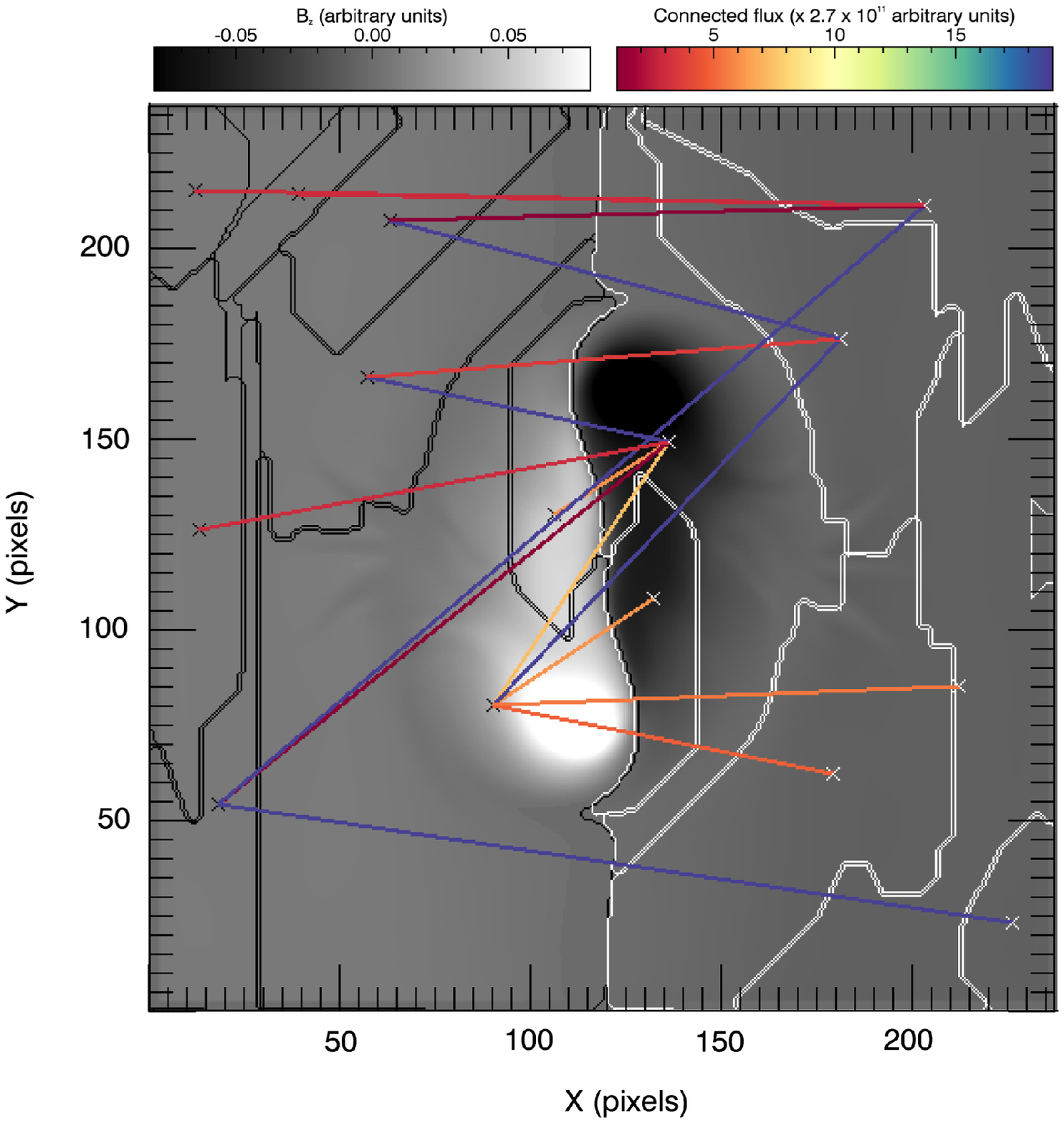}
   \includegraphics[width=\imsize,clip=true]{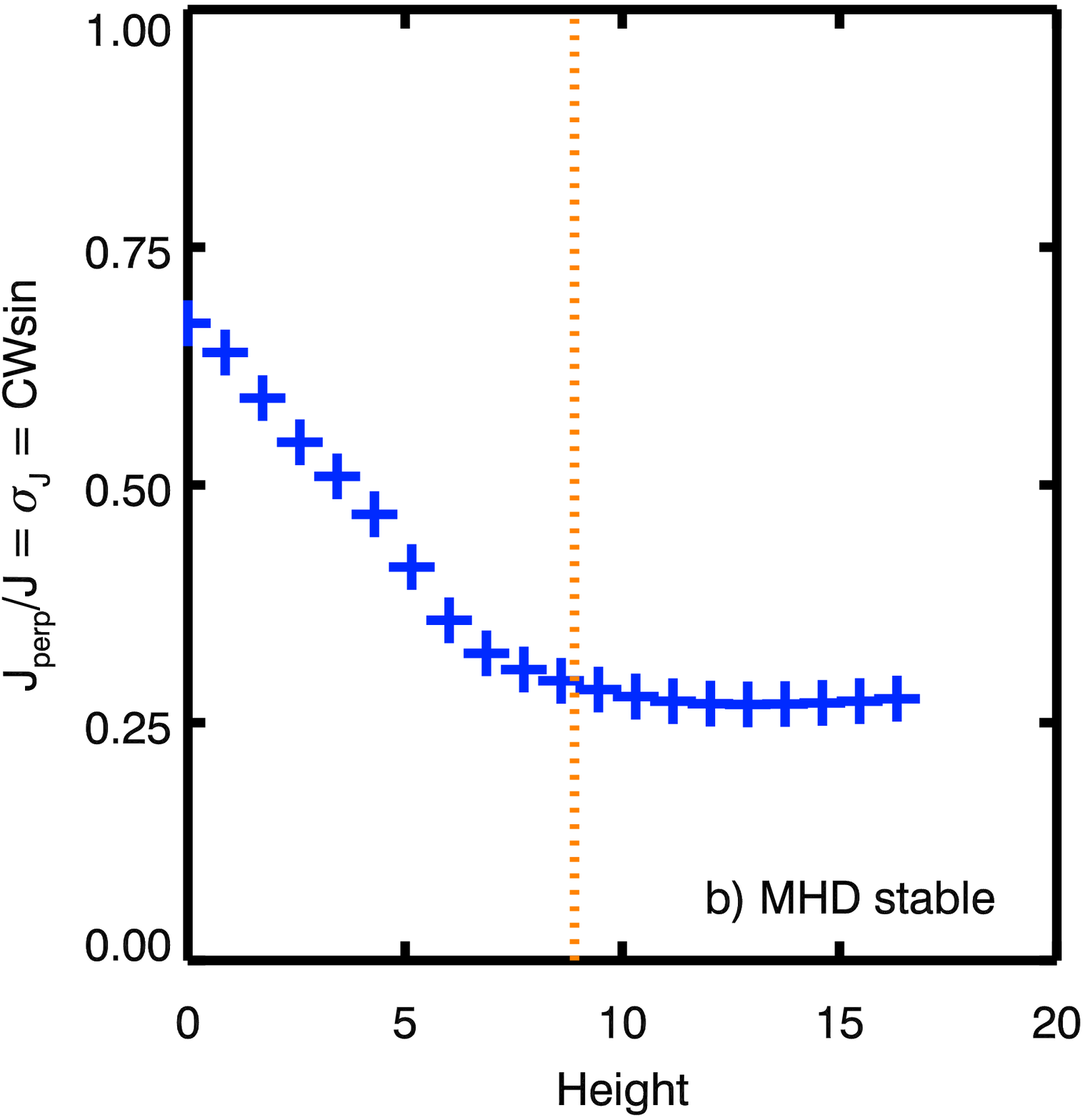}
   \caption{Application of the \sCB{} method to the \sJst{} case. 
(a) Magnetic connectivity of the MHD-emergence stable model at $t=150$ used for the calculation of magnetic helicity in the \sCB{} method. 
The partitioned normal magnetic field component on the lower (assumed photospheric) boundary is shown in gray-scale, with black/white contours outlining positive-/negative-polarity partitions. 
The flux-weighted centroids of the partitions are denoted by crosses. 
Magnetic connections are projected on the lower boundary by colored lines, with different colors denoting different connected-flux contents. 
The connected flux includes approx. 93.6\% of the total flux present in the field of view. 
(b) The force-freeness metric $\sj$ as a function of the height of the bottom boundary for \sJst{}  at $t=155$. 
   The vertical orange dotted line represents the location of the bottom boundary in the tests of \sect{CB}, corresponding to $z=8.9$.
}
   \label{f:JstCB_connectivity}
 \end{figure}
\referee{As for the \sTE{} method, the \sCB{} method can be compared to \sFV{} methods only regarding the estimated helicity values. 
However, we test here for the first time some of the assumptions made in the derivation of the \sCB{} method (force-freeness and minimal connectivity-length principle, see \sect{CB_method}), and their impact on the obtained helicity values. 
These novel comparisons are made possible by the reliability assessment of \sFV{} methods discussed in the previous sections.
Moreover, the comparison of the \sCB{} methods with the \sFI{} methods is discussed in detail by \cite{Pariat2016tmp}.} 

The \tCB{} method is designed for applications to solar magnetograms with a complex flux distribution and an unknown coronal magnetic field. 
If the lower boundary includes only two connected partitions the code automatically uses the linear force-free field approach of \cite{2007ApJ...671.1034G}, in which a single value of the force-free parameter $\alpha$ is used for the entire volume.
When the NLFF mode switches on, the \sCB{} method has the possibility to replaces the total photospheric magnetic flux by the connected magnetic flux, namely the one included in the magnetic connectivity matrix, see \sect{othermethods}.
The \sCB{} code was tuned to use almost the entire flux of the magnetograms in all case presented here, except when differently explicitly stated.

When applied to the \sTD{} cases, where currents are localized, the \sCB{} in LFF-mode tends to pick up mostly the potential field component.
On the other hand, when applied to the \sLL{} cases, where a large-scale $\alpha$ is present, the \sCB{} method yields values of $\Hv$ that are three to four times larger than \sFV{} methods.
This overestimation effect in case of the LFF field approximation was also reported by \cite{Georgoulis_2012}.
For these two sets of tests, where a single dipole appears, the \sCB{} method is hampered by the limitations of the linear force-free theory.
Hence, we do not report further on such applications here.

In the \sJst{} and \sJun{} cases, instead, the magnetogram is complex enough to have more than one connectivity domain.
The \sCB{} method worked in the NLFF mode for all snapshots at all times, in both \sJst{} and \sJun{} cases.
\Fig{JstCB_connectivity}a shows an example of the flux partition and of the resulting connectivity matrix for the \sJst{} case at $t=150$.

 \setlength{\imsize}{0.49\textwidth}
 \begin{figure}
   \includegraphics[width=\imsize]{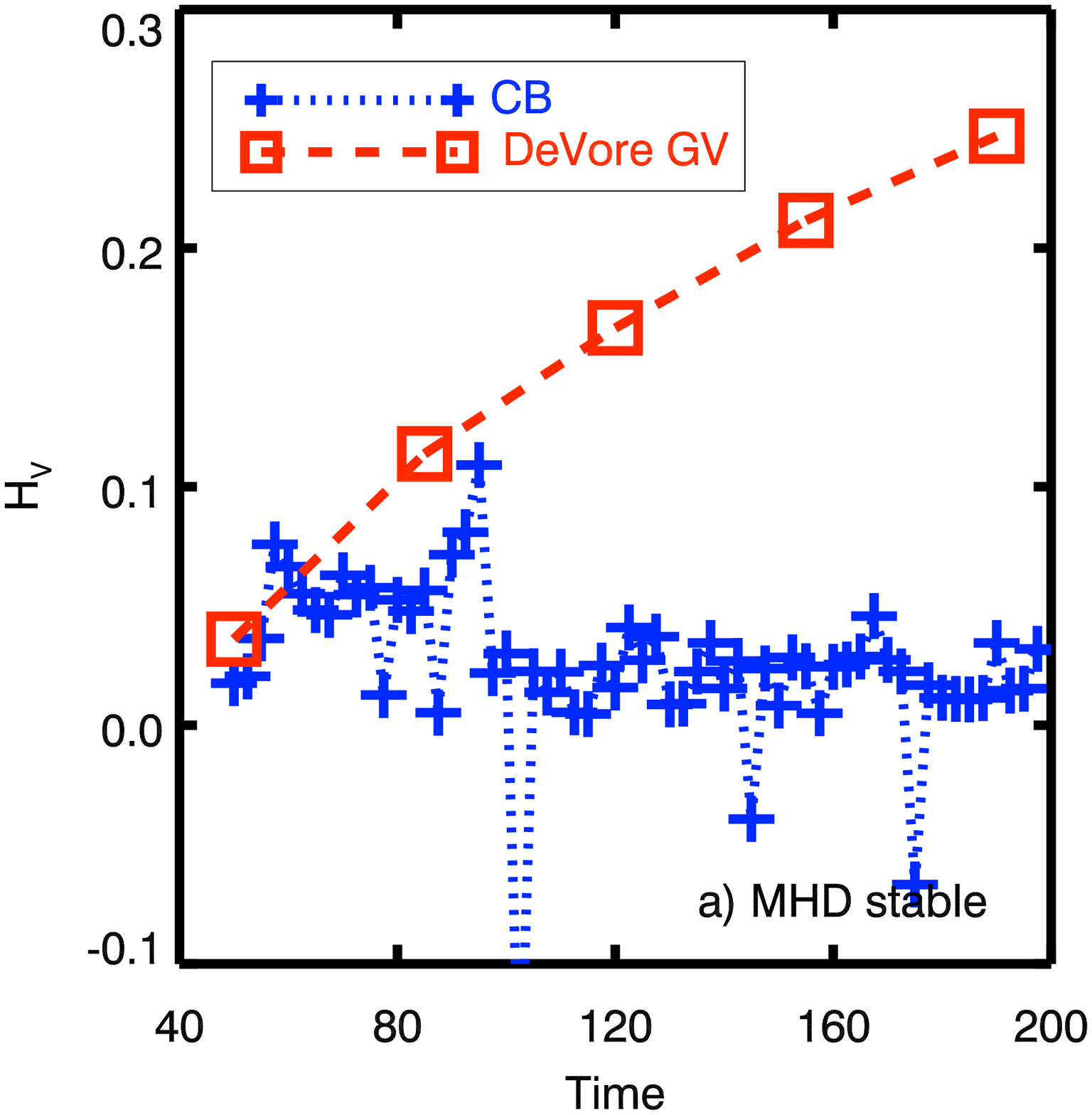}
   \includegraphics[width=\imsize]{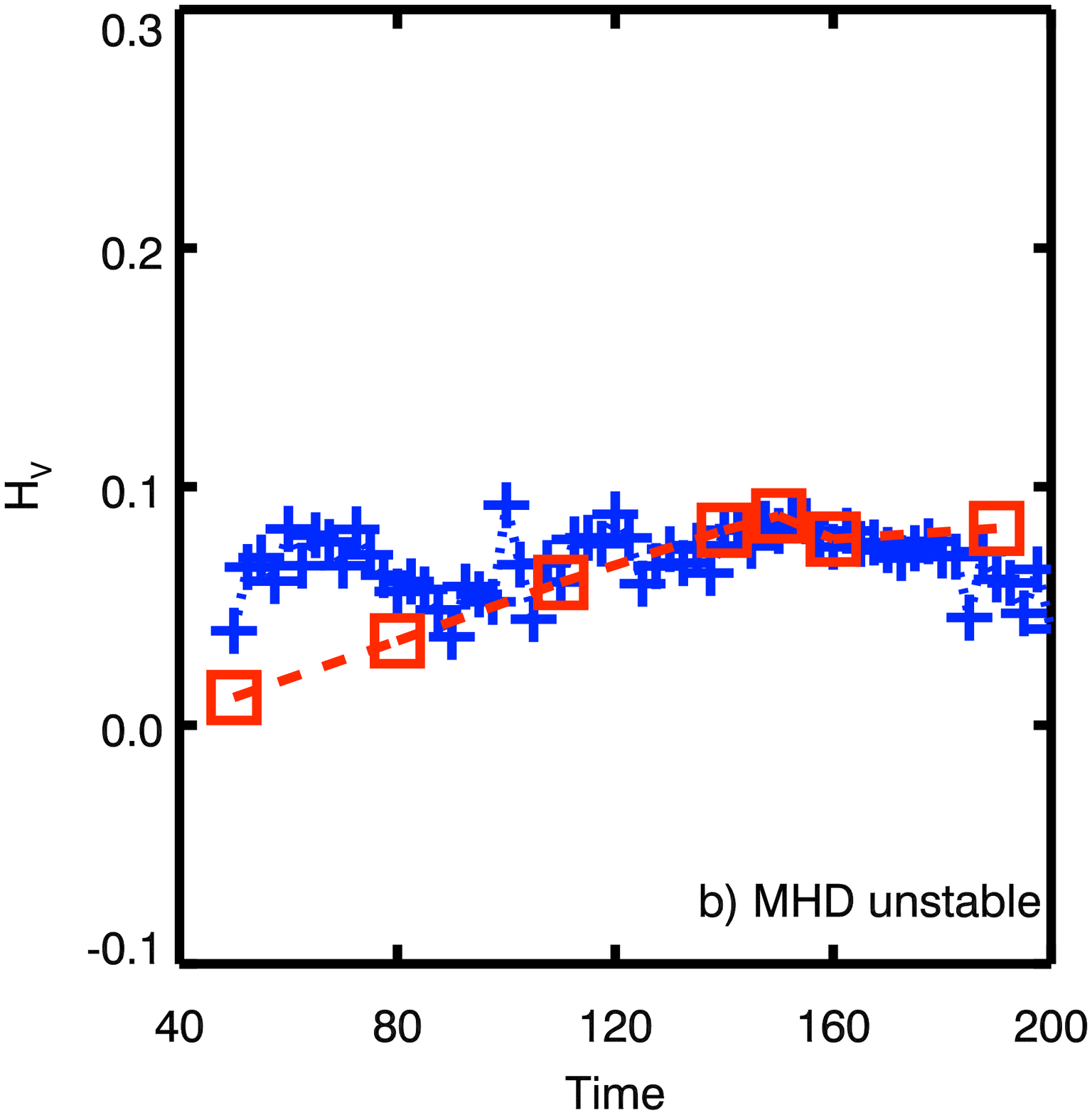}\\
   \includegraphics[width=\imsize]{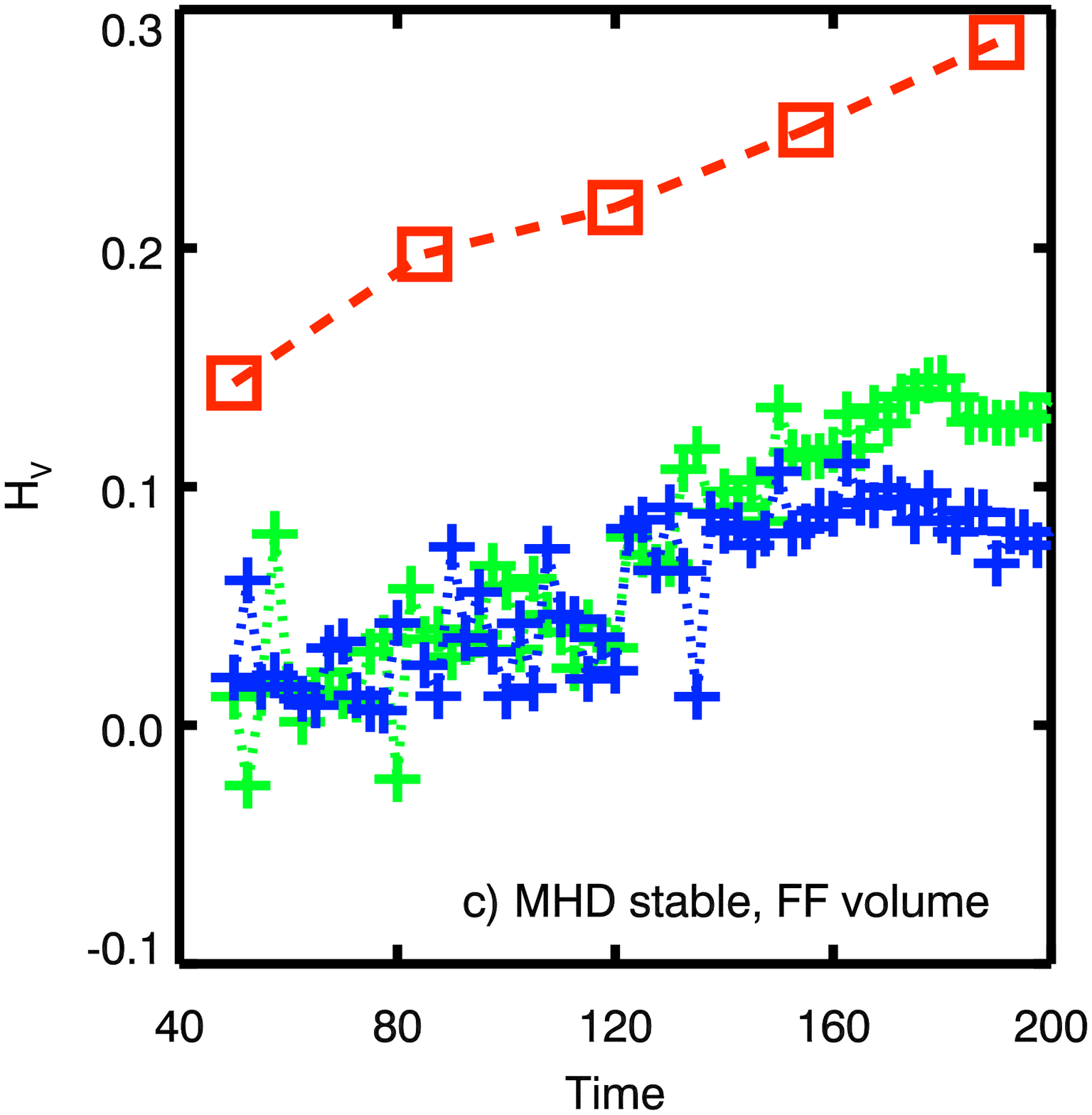}
   \includegraphics[width=\imsize]{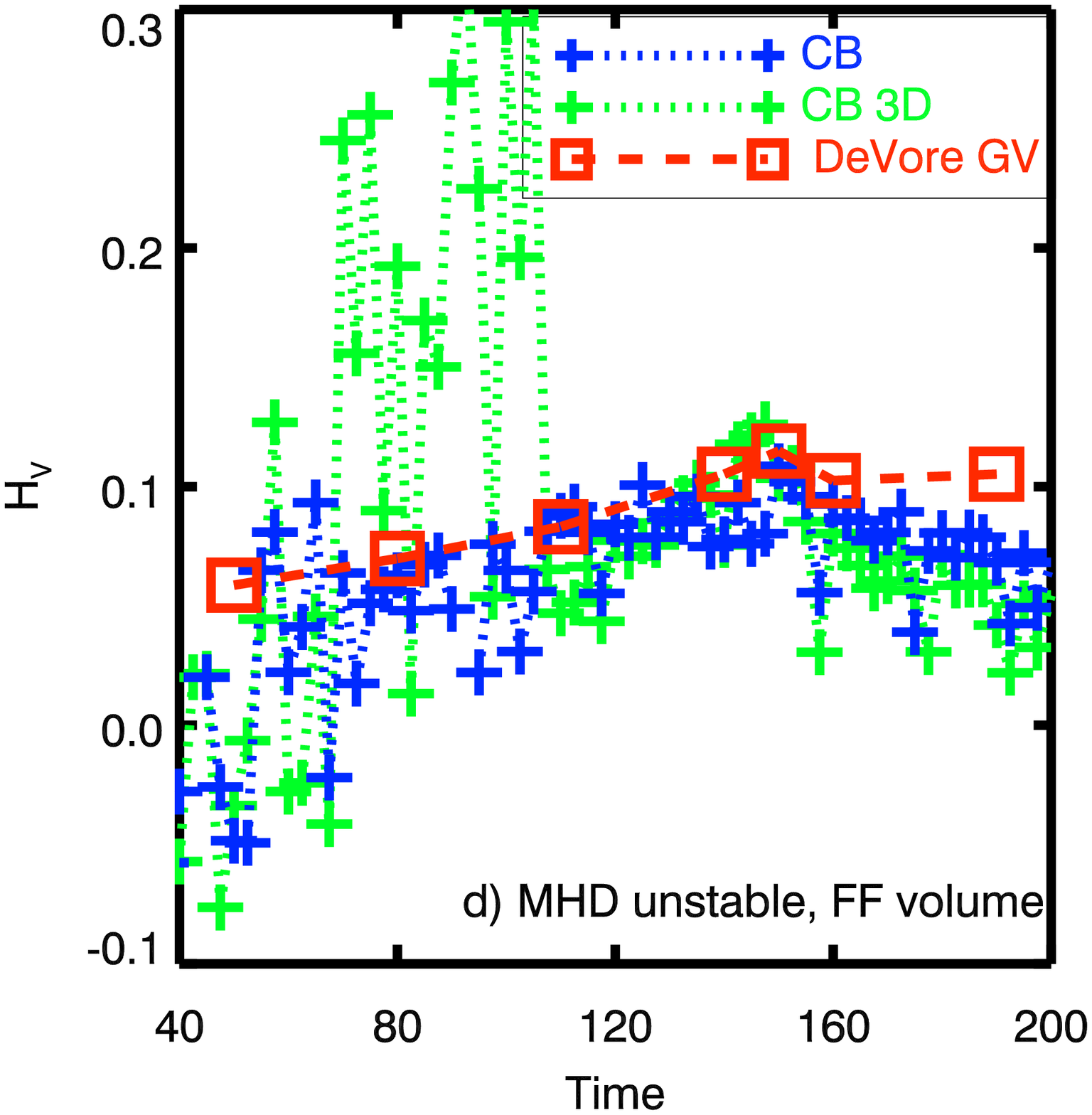}
   \caption{
     Comparison of $\Hv$ between the \sCB{} and the \sGV{} method applied to the \sJst{} (\textbf{a,c}) and \sJun{} cases (\textbf{b,d}) in the full domain (a,b) and in the reduced, more force-free domain (c,d). 
    In panels (c,d) \sCB{} method (blue crosses),  the \sCB{} method  with 3D connectivity information included (green crosses),  and the \sGV{} method (red squares).
         Note that the  \sGV{}  values c,d  are not the same as a,b and in \fig{JstJunH} because the considered volume is different.
   }
   \label{f:JstCB_H}
 \end{figure}

\Fig{JstCB_H} compare the \sCB{} method with the \sGV{} finite volume method.

In the \sJst{} case (\Fig{JstCB_H}a), between  $t=50$ and  $t=95$, the  of $\Hv$ values obtained by the \sCB{} method match reasonably well those from the \sGV{} one, within a factor 2 at most.  
Starting from  $t=95$ onwards, however, the $\Hv$ values obtained by the \sCB{} method settle on a lower, roughly constant value $\Hv=0.016$, on average. 
At the end of the simulations this average is  about eight times lower than the value reached by the \sGV{} method. 

On the other hand, in the \sJun{} case (\Fig{JstCB_H}b), the agreement between the \sGV{} and the \sCB{} methods is very good (within 9\% on average, from $t=95$ onwards), and the two curves overlap for most of that phase. 
A local maxima in the \sCB{} curve is even present at the time of the eruption, very much the same as for the \sGV{} method. 
However, this is not distinguishable from previous, even more pronounced ones, and it would be challenging to identify the time of the eruption only as a decrease of $\Hv$ in the \sCB{} time series.
As a matter of fact,  the examination of the time evolution of the magnetic field at $z=0$ in the \sJst{}  and \sJun{} simulations shows that there is very little differences between the two cases.
Since the \sCB{} method aims at an approximate estimation of helicity that is based only on the flux distribution at the (photospheric) bottom boundary at a given time, the task of distinguishing \sJst{} from \sJun{}  based only on the field on that one plane is a very arduous one, indeed.
This is probably the reason why the \sCB{} method provides similar average values of $\Hv$ for both the \sJst{} (equal to 0.015, from $t=95$ onwards) and the \sJun{} case (equal to 0.070 on the same time interval).


Besides the similarity of the \sJst{} and \sJun{} field distributions at $z=0$, differences with the \sGV{} method may have several origins.
First of all, the \sCB{}-method approach yields a \textit{minimal} value of helicity for a given photospheric configuration. 
In this sense, it is expected that the \sCB{} $\Hv$-curves lay, on average, below the \sGV{} ones, as they do at varying levels in all panels of  \fig{JstCB_H}.

Secondly, the \sCB{} method models the coronal field as a discrete collection of a finite number of constant-$\alpha$ flux tubes.
Hence, it can be expected to have better chances of success if the magnetic field is force-free.
\Fig{JstCB_connectivity}b  shows at a representative time, however, that in large part of the volume of the \sJst{} case the field is not force-free, as quantified by the relative ratio of the current that is perpendicular to the field in the volume, $\sj\equiv(\int_\vol |\vJ_\perp|  \dV)/(\int_\vol |\vJ|  \dV)$. 
In particular, the value of $\sj$ computed for increasing heights of the bottom boundary decreases from 0.7 to 0.25 in the first 20 pixels above the bottom boundary.

In order to test how important is the force-free assumption in this case, we repeat the  $\Hv$ calculation with the \sCB{} and \sGV{} methods in a  reduced volume starting at $z=8.9$, where  $\sj$ at this height has dropped to the value 0.29, see \fig{JstCB_connectivity}b.
The corresponding curves are shown by the blue crosses in \fig{JstCB_H}c and \fig{JstCB_H}d for \sJst{} and \sJun{}, respectively. 
Since the volume is now changed, also the corresponding \sGV{} estimations are recalculated for this reduced,  more force-free volume.
In the \sJst{} case, the average \sCB{} $\Hv$ after $t=95$ is 0.07, which is 3.5 times smaller than the final value obtained by the \sGV{} method (against a factor 8 of the full-volume case).
In the \sJun{} case in  \fig{JstCB_H}d,   the curves obtained by the two methods are slightly closer than in the full-volume case of \fig{JstCB_H}b, although only marginally (the ratio of the end values being 1.5).
Hence, the application to the more force-free, upper part of the volume improves the match between the \sCB{} and the \sGV{} results, markedly so in the \sJst{} case, which is a clear indication that the fulfillment  of the force-free requirement may help to partially compensate for the $\Hv$ underestimation in  \fig{JstCB_H}a. 

Thirdly, the connectivity map between flux partitions is obtained by the \sCB{} method as part of the minimization method discussed in \sect{othermethods}.
However, when the 3D coronal field is available, the true connectivity map can be constructed from the numerical simulation, and the influence of the minimization tested.
The result of such a test are represented by the green symbols in \fig{JstCB_H}c and d,  for the stable and unstable cases, respectively (again in the reduced, more force-free volume).
It must be noticed that using the 3D information implies a decrease of the amount of flux included in the connectivity matrix to 60-80\%, compared to the 95\% or more employed in the two cases above, since tracking of field lines intersecting the lateral and top boundaries cannot be completed in the \sCB{} method, even in case these lines return to the simulation volume by intersecting a different boundary location. 
The contribution of these lines is, then, ignored.

In general, \fig{JstCB_H}c demonstrates that the knowledge of the true connectivity in the \sJst{} case improves the matching between \sCB{} and \sGV{} methods. 
The \sCB{}-average (after $t=95$) $\Hv$ value is slightly above 0.10, which results in just a factor two between the corresponding end values.
In the \sJun{} case, differences with the standard application of the \sCB{} method that does not used the three-dimensional connectivity information are less significant, except for a slight increase of the mismatch with the \sGV{} method (\eg the ratio between end values of $\Hv$ obtained by the two methods is down to 1.9).
Such relatively small variations can be explained in terms of reduced connected flux.
We also recall that, for the \sCB{} method, an error analysis by \cite{Moraitis2014} is available that was not included in the discussion presented in this article. 
It is likely that some of the fluctuations discussed above fall within the error estimation provided by that analysis.

In conclusion, in the \sJun{} case the agreement between \sCB{} and \sGV{} (and, by extension, \sFV{})  methods is within 10\%, which is very good considering the much more  limited amount of information that the \sCB{} method requires.
On the other hand, in the \sJst{} case the helicity is significantly under-estimated, by a factor eight at the end of the simulation. 
The deficit in $\Hv$ that the \sCB{} method shows in the \sJst{} case, can be partially ascribed to the exiguous differences between the \sJst{} and \sJun{} cases in terms of flux distributions at $z=0$.
Additionally, the field at that plane is not quite force-free, and the \sCB{} results are shown to improve for a more force-free test volume. 
Furthermore, the minimization of the connectivity matrix seems to represent connectivity fairly well, in the sense that the implementation of the full three-dimensional information, although improving the \sCB{} estimation in the \sJst{} case, does not entirely remove the under-estimation of $\Hv$.

\section{Conclusions}\label{s:conclusions}
In this work we review, benchmark, and compare the currently available methods for the computation of the relative magnetic helicity, $\Hv$, in finite volumes.
Given that the three-dimensional magnetic-field solutions we use are common for all tested methods, the problem essentially
reduces to computing the vector potential of a given discretized input magnetic field.
The considered methods group into Coulomb ($\divA=0$) and DeVore ($\Az=0$) methods, according to the gauge in which the vector potentials are written. 
A total of six different implementations including three Coulomb methods (\sJT{}, \sSY{}, and \sGR{}) and  three DeVore methods (\sSA{}, \sKM{}, and \sGV{}) are included which differ in the equations they solve and/or in their numerical implementations.
Details of these implementations can be found in 
\cite{2011SoPh..272..243T} (\sJT{}), 
\cite{Yang2013} (\sSY{}),
\cite{Rudenko2014} (\sGR{}),
\cite{2012SoPh..278..347V} (\sGV{}), 
\cite{Moraitis2014} (\sKM{}), and in \sect{devore} (\sSA{}).
Accordingly, a different level of numerical complexity and computational effort is required to solve for the vector potentials, with the Coulomb methods being essentially far more  demanding than DeVore methods (see \sect{methods}). 
As a case in point, the \sGR{} method could not be tested on cases above $128^3$ pixels due to the large running time required by  its current implementation. 

The tested  methods are put under severe strain by choosing a variety of numerical test input fields  that are considered to be relevant for helicity studies in solar-physics applications, from 3D force-free equilibria (\sLL{} and \sTD{} of \sect{LL} and \sect{TD}, respectively), to snapshots of time-dependent non-force-free MHD simulations  of flux emergence (\sect{JL}).
Depending on details of the test field 
\referee{being studied}, the accuracy in the computation of $\Hv$ by different methods is found to vary to some extent, especially for Coulomb methods.
We can, however, definitely conclude that in solar-like cases practically all \sFV{} methods converge to the same helicity value within few percent\footnote{From this statistic, the \sGR{} method is excluded}.
\referee{Such a spread is likely to be overrun by other sources of errors in applications to observed --and reconstructed-- coronal fields, but it is definitely to be considered in helicity estimations of  numerical simulations.}

More in detail, the helicity values $\Hv$  in the most relevant test of time-dependent MHD evolution in a coronal model volume (\ie the  \sJst{} and \sJun{} tests of \sect{JL}) show a very good agreement of few percent between different methods, somewhat independently of the details in the vector potentials computation, see \eg \fig{JstJunH}. 
Such errors are as small as 0.2\% in the case where the field is slowly evolving, and always below 3\% even in the highly dynamical eruptive phase.
Similarly, excluding the \sGR{} case, despite differences in the accuracy of the vector potential computation, helicity values computed by different methods for the \sTD{}-twist case are within 2\%.
In other words, when helicity computations are applied to numerical volumes as in this article, differences in the way $\Hv$ is computed can amount to 3\% at most.

Such an agreement is tested and verified to hold independently of the dynamical evolution of the field and consistently throughout the MHD evolution of an eruption, thereby justifying the application of \sFV{} methods to the study of helicity in numerical simulations, and for benchmarking other helicity computation methods.
For instance, \cite{Pariat2016tmp} employs  \tFV{} methods to benchmark \tFI{} methods applied to \sJst{} and \sJun{} evolution, where the flux of helicity is estimated by the photospheric evolution of the field.

In addition to finite volume methods, we also include other two methods that use (\cite{Guo2010}, \sTE{}) or  optionally make use of (\cite{2012ApJ...759....1G}, \sCB{}) the 3D information of the magnetic field in the volume, see \sect{resultsothers}. 
The \sTE{} method estimates the helicity content of a field by parametrically fitting a flux rope to it.
Therefore, it is applicable to tests that include an identifiable flux rope, namely to \sTD{}, \sJst{}, and \sJun{} cases.
For the \sTD{} cases, we find that the \sTE{} method yields relatively accurate estimations of twist and possibly of $\HdefJ$ in high-twist cases, but not of the total helicity $\Hv$. 
\referee{A report on the application of the \sTE{} method to other cases, including the \sJst{} and \sJun{} ones, is in preparation by \cite{Guo2016tmp}.}

The \sCB{} method is designed to be applied to complex photospheric flux distributions with an unknown coronal magnetic field, where a multi-polar partition of the flux yields a coronal field approximated by a collection of flux tubes of constant $J_z/\Bz$.
In addition, a minimal free energy and the corresponding relative helicity are sought. 
Cases like the \sLL{} and \sTD{} have a too simple connectivity for the \sCB{} method, which then falls back to a single flux-tube, purely linear approximation of the (force-free) field. 
In  the more complex \sJun{} case, the \sCB{} method provides an estimation of the helicity that is, on average, within 10\% of the one obtained by \sFV{} methods.
This is a positive result given that the \sCB{} method employs only the photospheric information, whereas \sFV{} methods use the full 3D information about the coronal field. 
The  \sJst{} case poses more difficulties to the coronal field approximation within the \sCB{} method, which, in this case, underestimate significantly the helicity content of the field. 

\TFV{} methods can be used when the magnetic field is known in the entire volume of interest. 
However, in  applications to solar observations, the magnetic field is typically known only on a surface at photospheric heights, and only with limited accuracy.
Hence, in order to know the magnetic helicity in a given coronal volume, a model need to be computed that approximates the coronal field on the base of its photospheric values, which introduce an additional dependence to the estimated $\Hv$ values.
The impact on helicity values of the employed coronal model, being it from a nolinear force-free extrapolation or from a data-driven simulation, is yet to be tested.
Alternatively, the \sCB{} method can be used, as it is designed for such cases. 
A further alternative would be to compute the flux of helicity passing through the ``photospheric plane'' in time.
Reviewing and benchmarking \sFI{} methods for the estimation of the helicity flux is the subject of \cite{Pariat2016tmp}.

\begin{acknowledgements}
This article results from the work of the ISSI International Team on \textit{Magnetic Helicity estimations in models and observations of the solar magnetic field}.
The authors are pleased to thank  Marc deRosa, James Leake, Bernhard Kliem, and Tibor T\"or\"ok for making their numerical data available.
GV acknowledges the support of the Leverhulme Trust Research Project Grant 2014-051.
EP acknowledges the support of the French Agence Nationale de la Recherche (ANR), under grant ANR-15-CE31-0001 (project HeliSol).
SA acknowledges support of the Russian Foundation of Basic Research under grants 15–02–01089–a, 15–02–03835–a, 16-32-00315-mol-a, and 15–32–20504-mol-a-ved, and the Federal Agency for Scientific Organisations base project II.16.3.2 “Non-stationary and wave processes in the solar atmosphere” and thanks George Rudenko for providing the source code of \sGR{} method and contributing to development of \sSA{} method.
YG is supported by NSFC (11203014 and 11533005), NKBRSF (2011CB811402 and 2014CB744203),
and the mobility grant from the Belgian Federal Science Policy Office (BELSPO).
KM and MKG were partially supported by the European Union (European Social Fund – ESF) and Greek national funds through the Operational Program “Education and Lifelong Learning” of the National Strategic Reference Framework (NSRF) – Research Funding Program: Thales. Investing in knowledge society through the European Social Fund.
JKT acknowledges support from the Austrian Science Fund (FWF): P25383-N27.
FC acknowledge the support by the International Max-Planck Research School (IMPRS) for Solar System Science at the University of G\"ottingen.
SY is supported by grants KJCX2-EWT07 and XDB09040200 from the Chinese Academy of Sciences; grants 11221063, 11078012, 11178016, 11173033, 11125314, 10733020, 10921303, 11303053, 11573037, and 10673016 of National Natural Science Foundation of China; grant 2011CB811400 of National Basic Research Program of China; and KLSA2015 of the Collaborating Research Program of National Astronomical Observatories.
Part of the calculations presented here were done on the quadri-core bi-Xeon computers of the Cluster of the Division Informatique de l'Observatoire de Paris (DIO).
\end{acknowledgements}

\begin{appendix}
\section{Decomposition of the magnetic energy}\label{s:energy}
The method that we employ to quantify the error in the solenoidal property \citep{2013A&A...553A..38V} is basically a numerical verification of Thomson's theorem, and allows to quantify the effect of a (numerical) finite divergence of the magnetic field in terms of associated energies.
In order to obtain such a decomposition, the magnetic field is firstly separated into a potential and a current carrying part.
Secondly, each part is additionally split into a solenoidal and a nonsolenoidal contribution, using a Helmholtz decomposition.
As a result, the magnetic energy $\E$ is split into
\BE
\E= \Eps +\EJs +\EdivBp +\EdivBJ +\Emix \,,
  \label{eq:thomson}
\EE
where $\Eps$ and $\EJs$  are the energies associated to the potential and current-carrying solenoidal contributions,
$\EdivBp$ and $\EdivBJ$  are those of the nonsolenoidal contributions, and   $\Emix$ is a nonsolenoidal mixed term, see Eqs.~(7,8) in \cite{2013A&A...553A..38V} for the corresponding expressions.
All terms in Eq.~(\ref{eq:thomson}) are positively defined, except for $\Emix$.
For a perfectly solenoidal field, it is $\Eps=\Ep$, $\EJs=E-\Ep$, and $\EdivBp=\EdivBJ=\Emix=0$, \ie Thomson's theorem is recovered.

In most of the analysis in this article we consider a single number for characterizing the energy associated to nonsolenoidal components of the field, given by 
\BE
  \Ediv=\EdivBp +\EdivBJ +|\Emix| \,,
  \label{eq:ediv}
\EE
which generally overestimates such errors by hindering possible cancellations.
However, since we consider numerically accurate models, we neglect such overestimations, unless explicitly stated.
In that case, also the sum with sign $\Ens$ is considered, corresponding to \eq{ediv} with $\Emix$ instead of $|\Emix|$.
The full decomposition for each test case considered in the article can be found in \tab{input}.
Also, in the article we associate the free energy of the field with the solenoidal component of its current-carrying part, \ie $\Efree=\EJs$.

For a given discretized field, the accuracy of the decomposition can be easily quantified by checking to what precision the sum of the right hand side of Eq.~(\ref{eq:thomson}) reproduces the total energy $\E$, normalised to $\E$.
In relative terms, such an error is smaller than $10^{-7}$ in all cases discussed in this article.
In this article, all energy contributions are  normalized to the total energy, $\E$, of the test case in exam.

\begin{table}
 \caption{
 Energy metrics of  test cases, see \app{energy}. the first two columns indicate the test field identification label and the total magnetic energy in model units [$\E$]. The following columns indicate  the potential/solenoidal [$\Eps$], current-carrying/solenoidal  [$\EJs$],   potential/nonsolenoidal [$\EdivBp$],  current-carrying/nonsolenoidal [$\EdivBJ$], nonsolenoidal mixed [$\Emix$] contributions normalised to the corresponding total magnetic energy, $\E$. The last column contains the  average fractional flux [$ \avfi{(\vB)}$]. 
 In the text, any contribution $E_{\rm xx}$ is intended to be normalized to $\E$, \ie it is here indicated as $E_{\rm xx}/E$, for each test case separately. 
 }
 \label{t:input}
 \centering
 \scriptsize
 \strtable
 \begin{tabular}{@{~}l  c@{\quad}  c@{\quad}  c@{\quad}  c@{\quad} c@{\quad} c@{\quad} c@{\quad} }
 \hline
     Label    & $\E$ & $\Eps/\E$  &$\EJs/\E$  &$\EdivBp/\E$ & $\EdivBJ/\E$& $\Emix/\E$ & $\avfi{(\vB)} \times 10^{5}$\\
 \hline
\sLL{} n= 32             & 5.48e+02  &  0.78  &  0.25 & 2.34e-03 & 6.52e-04 &-3.82e-02 & 26.43  \\  
\sLL{} n= 64             & 5.30e+02  &  0.76  &  0.25 & 2.60e-04 & 8.22e-05 &-1.33e-02 &  3.34  \\  
\sLL{} n=128             & 5.25e+02  &  0.75  &  0.26 & 2.37e-05 & 9.15e-06 &-3.90e-03 &  0.62  \\  
\sLL{} n=256             & 5.24e+02  &  0.75  &  0.26 & 2.38e-06 & 1.32e-06 &-1.07e-03 &  0.27  \\ 
 \hline                                                                                       
\sTD{} N=0.1$\Delta=0.06$& 4.69e+01  &  1.00  &  0.00 & 1.18e-05 & 1.02e-05 &-1.20e-03 &  0.39  \\  
\sTD{} N=0.5$\Delta=0.06$& 3.35e+01  &  0.98  &  0.02 & 1.18e-05 & 2.68e-05 & 3.02e-03 &  0.71  \\ 
\sTD{} N=1  $\Delta=0.03$& 1.08e+01  &  0.81  &  0.17 & 3.08e-06 & 5.94e-04 & 2.53e-02 &  0.75  \\ 
\sTD{} N=1  $\Delta=0.06$& 1.09e+01  &  0.81  &  0.17 & 1.21e-05 & 6.12e-04 & 2.44e-02 &  1.89  \\ 
\sTD{} N=1  $\Delta=0.12$& 1.09e+01  &  0.81  &  0.17 & 6.33e-05 & 6.78e-04 & 2.14e-02 &  6.86  \\ 
\sTD{} N=3  $\Delta=0.06$& 3.80e+00  &  0.85  &  0.14 & 1.35e-05 & 1.16e-04 & 8.07e-03 &  1.35  \\ 
 \hline                                                                                       
\sJst{} t= 50            & 4.04e+02  &  0.52  &  0.50 & 3.86e-05 & 3.34e-04 &-1.67e-02 &  9.99  \\  
\sJst{} t= 85            & 5.31e+02  &  0.53  &  0.48 & 3.49e-05 & 3.00e-04 &-1.15e-02 & 13.08  \\ 
\sJst{} t=120            & 6.03e+02  &  0.46  &  0.55 & 2.08e-05 & 2.62e-04 &-7.06e-03 & 13.66  \\   
\sJst{} t=155            & 6.42e+02  &  0.41  &  0.59 & 1.49e-05 & 2.31e-04 &-5.46e-03 & 12.27  \\ 
\sJst{} t=190            & 6.67e+02  &  0.38  &  0.62 & 1.34e-05 & 2.22e-04 &-4.49e-03 & 12.10  \\ 
 \hline                                                                                       
\sJun{} t= 50            & 3.67e+02  &  0.39  &  0.62 & 3.92e-05 & 2.40e-04 &-1.10e-02 & 11.25  \\ 
\sJun{} t= 80            & 4.43e+02  &  0.42  &  0.58 & 4.36e-05 & 2.55e-04 &-8.34e-03 & 22.83  \\
\sJun{} t=110            & 4.64e+02  &  0.40  &  0.60 & 3.09e-05 & 2.93e-04 &-5.11e-03 & 27.10  \\ 
\sJun{} t=140            & 4.46e+02  &  0.41  &  0.59 & 2.16e-05 & 2.50e-04 &-3.63e-03 & 32.29  \\
\sJun{} t=150            & 4.07e+02  &  0.44  &  0.56 & 2.23e-05 & 2.57e-04 &-3.26e-03 & 46.16  \\
\sJun{} t=160            & 3.81e+02  &  0.47  &  0.53 & 2.32e-05 & 3.02e-04 &-4.17e-03 & 43.88  \\
\sJun{} t=190            & 3.62e+02  &  0.49  &  0.51 & 2.33e-05 & 3.35e-04 &-5.14e-03 & 32.42  \\
 \hline                                                                                       
\sJdiv{} 0.2\%           & 4.14e+02  &  0.50  &  0.50 & 3.92e-05 & 3.14e-04 &-1.45e-03 & 14.95  \\
\sJdiv{} 1  \%           & 4.06e+02  &  0.51  &  0.50 & 3.85e-05 & 1.70e-04 &-1.08e-02 &  9.77  \\
\sJdiv{} 2  \%           & 4.03e+02  &  0.52  &  0.50 & 3.88e-05 & 4.90e-04 &-1.98e-02 & 16.36  \\
\sJdiv{} 4  \%           & 4.09e+02  &  0.53  &  0.51 & 4.16e-05 & 2.25e-03 &-3.82e-02 & 47.11  \\
\sJdiv{} 8  \%           & 4.64e+02  &  0.49  &  0.57 & 5.43e-05 & 9.19e-03 &-7.25e-02 & 85.67  \\
\sJdiv{} 14 \%           & 1.01e+03  &  0.31  &  0.77 & 9.64e-05 & 3.06e-02 &-1.12e-01 &165.15  \\
 \hline
 \end{tabular}
\end{table}
\section{Accuracy of vector potentials}\label{s:tables}
In addition to the metrics \eqs{En}{eps}, we include here also the remaining 
\BA
C_{\rm Vec} =\frac{\sum_i\vX_i\cdot\vY_i}{(\sum_i|\vX_i|^2\sum_i|\vY_i|^2)^{1/2}} \label{eq:Cvec}\\
C_{\rm CV}  =\frac{1}{N}\sum_i\frac{\vX_i\cdot\vY_i}{|\vX_i||\vY_i|}  \\
\epsM =1-\frac{1}{N}\sum_i\frac{|\vY_i-\vX_i|}{|\vX_i|}                           \label{eq:epsM}
\EA
respectively the vector correlation, the Cauchy-Schwartz metric, and the complement of the mean vector error, from \cite{2006SoPh..235..161S}.
We provide the full set of values for each case considered in the paper in the following tables.
In particular, \tab{lltd} reports the metrics for the \sLL{}  and \sTD{}, \tab{stun} those for \sJst{} and \sJun{}, and \tab{stun_div} those for the divergence and gauge tests of \sect{divergence}.

\begin{table*}
 \caption{Metrics of accuracy of vector potentials as defined in \eqs{En}{eps} and  \eqss{Cvec}{epsM}, and helicity $\Hv$ as defined in \eq{H}, for  the \sLL{} (upper part) and \sTD{} (lower part) test cases.
 }
 \label{t:lltd}
 \centering
 \tiny
 \strtable
 \begin{tabular}{
 @{~}l  |
 c@{\quad}  c@{\quad}  c@{\quad} c@{\quad} c@{\quad}  |
 c@{\quad}  c@{\quad}  c@{\quad} c@{\quad} c@{\quad}  |
 c@{\quad}
       }
\firsthline
 Label  & \multicolumn{5}{c}{$\vBp$ vs $\Nabla \times\vAp$}   &
          \multicolumn{5}{c}{$\vB$  vs $\Nabla \times\vA$}    & $\Hv$ \\
          \cline{2-11}
        & $C_{\rm Vec}$  &$C_{\rm CS}$  &$\epsN$ & $\epsM$& $\epsilon_{\rm E}$
        & $C_{\rm Vec}$  &$C_{\rm CS}$  &$\epsN$ & $\epsM$& $\epsilon_{\rm E}$  \\
  \firsthline
      \sSY{} 32  &   0.9997   &  1.0000    & 0.9885   &  0.9938   &  1.0033   &  0.9996   &  0.9924   &  0.9742   &  0.9130   &  0.9965  &  -0.1462\\
      \sSY{} 64  &   1.0000   &  1.0000    & 0.9963   &  0.9981   &  1.0014   &  0.9999   &  0.9970   &  0.9880   &  0.9496   &  0.9994  &  -0.1427\\ 
     \sSY{} 128  &   1.0000   &  1.0000    & 0.9989   &  0.9992   &  1.0005   &  1.0000   &  0.9990   &  0.9943   &  0.9726   &  1.0001  &  -0.1418\\
     \sSY{} 256  &   1.0000   &  1.0000    & 0.9990   &  0.9978   &  1.0001   &  1.0000   &  0.9996   &  0.9965   &  0.9836   &  1.0003  &  -0.1415\\
      \sKM{} 32  &   0.9992   &  0.9998    & 0.9799   &  0.9854   &  0.9915   &  0.9997   &  1.0000   &  0.9891   &  0.9958   &  0.9978  &  -0.1483\\
      \sKM{} 64  &   0.9999   &  0.9999    & 0.9923   &  0.9929   &  0.9963   &  1.0000   &  1.0000   &  0.9968   &  0.9989   &  0.9988  &  -0.1441\\
     \sKM{} 128  &   1.0000   &  0.9999    & 0.9970   &  0.9965   &  0.9988   &  1.0000   &  1.0000   &  0.9991   &  0.9997   &  0.9996  &  -0.1426\\
     \sKM{} 256  &   1.0000   &  1.0000    & 0.9988   &  0.9982   &  0.9996   &  1.0000   &  1.0000   &  0.9997   &  0.9999   &  0.9999  &  -0.1419\\
      \sJT{} 32  &   0.9974   &  0.8772    & 0.8841   &  0.6307   &  1.0320   &  0.9978   &  0.8583   &  0.8856   &  0.5295   &  1.0500  &  -0.1499\\
      \sJT{} 64  &   0.9974   &  0.8658    & 0.8843   &  0.5771   &  1.0242   &  0.9980   &  0.8409   &  0.8963   &  0.4688   &  1.0410  &  -0.1437\\
     \sJT{} 128  &   0.9967   &  0.8595    & 0.8810   &  0.5420   &  1.0166   &  0.9975   &  0.8313   &  0.9001   &  0.4272   &  1.0275  &  -0.1414\\
     \sJT{} 256  &   0.9958   &  0.8562    & 0.8778   &  0.5207   &  1.0121   &  0.9968   &  0.8262   &  0.9009   &  0.4011   &  1.0177  &  -0.1404\\
      \sGV{} 32  &   0.9988   &  0.9997    & 0.9781   &  0.9797   &  0.9677   &  0.9995   &  1.0000   &  0.9884   &  0.9958   &  0.9798  &  -0.1427\\
      \sGV{} 64  &   0.9998   &  0.9999    & 0.9915   &  0.9892   &  0.9851   &  0.9999   &  1.0000   &  0.9965   &  0.9989   &  0.9918  &  -0.1420\\
     \sGV{} 128  &   1.0000   &  1.0000    & 0.9966   &  0.9942   &  0.9949   &  1.0000   &  1.0000   &  0.9990   &  0.9997   &  0.9974  &  -0.1416\\
     \sGV{} 256  &   1.0000   &  1.0000    & 0.9985   &  0.9968   &  0.9985   &  1.0000   &  1.0000   &  0.9997   &  0.9999   &  0.9993  &  -0.1415\\
      \sSA{} 32  &   0.9996   &  1.0000    & 0.9887   &  0.9959   &  1.0044   &  0.9997   &  1.0000   &  0.9894   &  0.9959   &  0.9960  &  -0.1471\\
      \sSA{} 64  &   1.0000   &  1.0000    & 0.9966   &  0.9989   &  1.0015   &  1.0000   &  1.0000   &  0.9969   &  0.9989   &  0.9977  &  -0.1428\\
     \sSA{} 128  &   1.0000   &  1.0000    & 0.9991   &  0.9997   &  1.0005   &  1.0000   &  1.0000   &  0.9991   &  0.9997   &  0.9989  &  -0.1418\\
     \sSA{} 256  &   1.0000   &  1.0000    & 0.9998   &  0.9999   &  1.0001   &  1.0000   &  1.0000   &  0.9997   &  0.9999   &  0.9995  &  -0.1415\\
      \sGR{} 32  &   0.9997   &  1.0000    & 0.9882   &  0.9950   &  1.0042   &  0.9992   &  0.9525   &  0.9247   &  0.7829   &  1.0036  &  -0.1655\\
      \sGR{} 64  &   1.0000   &  1.0000    & 0.9966   &  0.9989   &  1.0015   &  0.9977   &  0.8661   &  0.8471   &  0.5270   &  1.0048  &  -0.1861\\
     \sGR{} 128  &   1.0000   &  1.0000    & 0.9991   &  0.9997   &  1.0005   &  0.9974   &  0.8573   &  0.8388   &  0.4970   &  1.0013  &  -0.1846\\
\hline
  \sSY{} N01 0.06&   1.0000    & 1.0000    & 0.9994   &  0.9991   &  1.0002   &  1.0000   &  0.9997   &  0.9939   &  0.9816   &  0.9999  &  -0.0057  \\
  \sSY{} N05 0.06&   1.0000    & 1.0000    & 0.9993   &  0.9991   &  1.0002   &  1.0000   &  0.9997   &  0.9921   &  0.9792   &  0.9959  &  -0.0289  \\
  \sSY{} N1 0.03 &   1.0000    & 1.0000    & 0.9987   &  0.9960   &  1.0001   &  0.9998   &  0.9995   &  0.9771   &  0.9683   &  0.9745  &  -0.0768  \\
  \sSY{} N1 0.06 &   1.0000    & 1.0000    & 0.9992   &  0.9989   &  1.0003   &  0.9998   &  0.9991   &  0.9763   &  0.9628   &  0.9751  &  -0.0769  \\
  \sSY{} N1 0.12 &   1.0000    & 1.0000    & 0.9977   &  0.9977   &  1.0009   &  0.9998   &  0.9982   &  0.9739   &  0.9510   &  0.9755  &  -0.0769  \\
  \sSY{} N3 0.06 &   1.0000    & 1.0000    & 0.9992   &  0.9989   &  1.0003   &  0.9999   &  0.9995   &  0.9884   &  0.9759   &  0.9889  &  -0.0524  \\
\sKM{} N01 0.06  &   1.0000    & 0.9999    & 0.9920   &  0.9910   &  0.9889   &  1.0000   &  1.0000   &  0.9995   &  0.9997   &  1.0003  &  -0.0057  \\
\sKM{} N05 0.06  &   1.0000    & 0.9999    & 0.9919   &  0.9910   &  0.9890   &  1.0000   &  1.0000   &  0.9988   &  0.9996   &  1.0025  &  -0.0290  \\
 \sKM{} N1 0.03  &   1.0000    & 1.0000    & 0.9959   &  0.9954   &  0.9943   &  0.9996   &  1.0000   &  0.9949   &  0.9992   &  1.0166  &  -0.0783  \\
 \sKM{} N1 0.06  &   1.0000    & 0.9999    & 0.9920   &  0.9911   &  0.9892   &  0.9996   &  1.0000   &  0.9948   &  0.9990   &  1.0151  &  -0.0784  \\
 \sKM{} N1 0.12  &   0.9999    & 0.9998    & 0.9844   &  0.9832   &  0.9806   &  0.9996   &  1.0000   &  0.9936   &  0.9981   &  1.0118  &  -0.0786  \\
 \sKM{} N3 0.06  &   1.0000    & 0.9999    & 0.9919   &  0.9911   &  0.9892   &  0.9998   &  1.0000   &  0.9976   &  0.9994   &  1.0032  &  -0.0529  \\
\sJT{} N01 0.06  &   0.9973    & 0.9446    & 0.8899   &  0.7118   &  1.0238   &  0.9975   &  0.9461   &  0.8913   &  0.7198   &  1.0361  &  -0.0060  \\
\sJT{} N05 0.06  &   0.9973    & 0.9444    & 0.8901   &  0.7114   &  1.0237   &  0.9976   &  0.9470   &  0.8939   &  0.7240   &  1.0319  &  -0.0302  \\
 \sJT{} N1 0.03  &   0.9975    & 0.9444    & 0.8943   &  0.7087   &  1.0172   &  0.9981   &  0.9548   &  0.9074   &  0.7491   &  0.9971  &  -0.0810  \\
 \sJT{} N1 0.06  &   0.9974    & 0.9429    & 0.8918   &  0.7069   &  1.0231   &  0.9980   &  0.9543   &  0.9068   &  0.7506   &  1.0097  &  -0.0818  \\
 \sJT{} N1 0.12  &   0.9972    & 0.9403    & 0.8864   &  0.7043   &  1.0330   &  0.9979   &  0.9534   &  0.9028   &  0.7529   &  1.0290  &  -0.0833  \\
 \sJT{} N3 0.06  &   0.9975    & 0.9443    & 0.8924   &  0.7108   &  1.0232   &  0.9979   &  0.9518   &  0.9018   &  0.7391   &  1.0231  &  -0.0551  \\
  \sGV{} N01 0.06&   1.0000    & 1.0000    & 0.9990   &  0.9988   &  0.9985   &  1.0000   &  1.0000   &  0.9995   &  0.9997   &  0.9991  &  -0.0057  \\
  \sGV{} N05 0.06&   1.0000    & 1.0000    & 0.9988   &  0.9981   &  0.9985   &  1.0000   &  1.0000   &  0.9988   &  0.9996   &  1.0014  &  -0.0290  \\
  \sGV{} N1 0.03 &   1.0000    & 1.0000    & 0.9991   &  0.9979   &  0.9995   &  0.9996   &  1.0000   &  0.9950   &  0.9992   &  1.0164  &  -0.0782  \\
  \sGV{} N1 0.06 &   1.0000    & 1.0000    & 0.9980   &  0.9960   &  0.9983   &  0.9996   &  1.0000   &  0.9948   &  0.9990   &  1.0146  &  -0.0782  \\
  \sGV{} N1 0.12 &   1.0000    & 0.9999    & 0.9950   &  0.9924   &  0.9943   &  0.9996   &  1.0000   &  0.9939   &  0.9983   &  1.0099  &  -0.0782  \\
  \sGV{} N3 0.06 &   1.0000    & 1.0000    & 0.9981   &  0.9965   &  0.9983   &  0.9998   &  1.0000   &  0.9977   &  0.9995   &  1.0035  &  -0.0527  \\
\sSA{} N01 0.06  &   1.0000    & 1.0000    & 0.9995   &  0.9997   &  1.0001   &  1.0000   &  1.0000   &  0.9996   &  0.9997   &  0.9998  &  -0.0057  \\
\sSA{} N05 0.06  &   1.0000    & 1.0000    & 0.9995   &  0.9997   &  1.0001   &  1.0000   &  1.0000   &  0.9988   &  0.9996   &  1.0021  &  -0.0290  \\
 \sSA{} N1 0.03  &   1.0000    & 1.0000    & 0.9999   &  0.9999   &  1.0000   &  0.9996   &  1.0000   &  0.9950   &  0.9992   &  1.0164  &  -0.0785  \\
 \sSA{} N1 0.06  &   1.0000    & 1.0000    & 0.9994   &  0.9997   &  1.0001   &  0.9996   &  1.0000   &  0.9948   &  0.9990   &  1.0147  &  -0.0785  \\
 \sSA{} N1 0.12  &   1.0000    & 1.0000    & 0.9980   &  0.9987   &  1.0005   &  0.9996   &  1.0000   &  0.9938   &  0.9982   &  1.0111  &  -0.0787  \\
 \sSA{} N3 0.06  &   1.0000    & 1.0000    & 0.9994   &  0.9997   &  1.0001   &  0.9998   &  1.0000   &  0.9977   &  0.9995   &  1.0027  &  -0.0528  \\
 \sGR{} N01 0.0  &   1.0000    & 1.0000    & 0.9996   &  0.9997   &  1.0001   &  0.9967   &  0.9523   &  0.8760   &  0.6852   &  1.0088  &  -0.0543  \\
 \sGR{} N05 0.0  &   1.0000    & 1.0000    & 0.9996   &  0.9997   &  1.0001   &  0.9969   &  0.9578   &  0.8797   &  0.6924   &  1.0071  &  -0.0731  \\
 \sGR{} N1 0.06  &   1.0000    & 1.0000    & 0.9995   &  0.9997   &  1.0001   &  0.9976   &  0.9706   &  0.8939   &  0.7218   &  0.9970  &  -0.1113  \\
 \sGR{} N1 0.12  &   1.0000    & 1.0000    & 0.9982   &  0.9988   &  1.0004   &  0.9978   &  0.9733   &  0.8999   &  0.7438   &  0.9974  &  -0.1119  \\
 \sGR{} N3 0.06  &   1.0000    & 1.0000    & 0.9995   &  0.9997   &  1.0001   &  0.9974   &  0.9682   &  0.8912   &  0.7141   &  1.0015  &  -0.0843  \\
  \lasthline
 \end{tabular}
\end{table*}
\begin{table*}
 \caption{Same as \tab{lltd} for the \sJst{} (upper part) and \sJun{} (lower part) test cases.
 }
 \label{t:stun}
 \centering
 \tiny
 \strtable
 \begin{tabular}{
 @{~}l  |
 c@{\quad}  c@{\quad}  c@{\quad} c@{\quad} c@{\quad}  |
 c@{\quad}  c@{\quad}  c@{\quad} c@{\quad} c@{\quad}  |
 c@{\quad}
       }
\firsthline
 Label  & \multicolumn{5}{c}{$\vBp$ vs $\Nabla \times\vAp$}   &
          \multicolumn{5}{c}{$\vB$  vs $\Nabla \times\vA$}    & $\Hv$ \\
          \cline{2-11}
        & $C_{\rm Vec}$  &$C_{\rm CS}$  &$\epsN$ & $\epsM$& $\epsilon_{\rm E}$  
        & $C_{\rm Vec}$  &$C_{\rm CS}$  &$\epsN$ & $\epsM$& $\epsilon_{\rm E}$  \\ 
  \firsthline
      \sSY{} 50  &   1.0000   &  1.0000   &  0.9986   &  0.9982   &  1.0005   &  0.9996   &  0.9994   &  0.9573  &   0.9432   &  1.0026  &   0.0360\\
      \sSY{} 85  &   1.0000   &  1.0000   &  0.9987   &  0.9983   &  1.0006   &  0.9995   &  0.9994   &  0.9652  &   0.9502   &  1.0061  &   0.1138\\
     \sSY{} 120  &   1.0000   &  1.0000   &  0.9987   &  0.9983   &  1.0005   &  0.9994   &  0.9994   &  0.9690  &   0.9545   &  1.0047  &   0.1664\\
     \sSY{} 155  &   1.0000   &  1.0000   &  0.9988   &  0.9983   &  1.0004   &  0.9995   &  0.9995   &  0.9726  &   0.9585   &  1.0035  &   0.2112\\
     \sSY{} 190  &   1.0000   &  1.0000   &  0.9988   &  0.9983   &  1.0004   &  0.9996   &  0.9995   &  0.9746  &   0.9612   &  1.0022  &   0.2463\\
      \sKM{} 50  &   1.0000   &  1.0000   &  0.9972   &  0.9974   &  0.9969   &  0.9995   &  0.9996   &  0.9707  &   0.9694   &  0.9838  &   0.0360\\
      \sKM{} 85  &   1.0000   &  1.0000   &  0.9973   &  0.9974   &  0.9969   &  0.9993   &  0.9996   &  0.9712  &   0.9678   &  0.9894  &   0.1141\\
     \sKM{} 120  &   1.0000   &  1.0000   &  0.9972   &  0.9973   &  0.9970   &  0.9992   &  0.9996   &  0.9720  &   0.9673   &  0.9896  &   0.1670\\
     \sKM{} 155  &   1.0000   &  1.0000   &  0.9972   &  0.9972   &  0.9970   &  0.9993   &  0.9996   &  0.9740  &   0.9674   &  0.9892  &   0.2121\\
     \sKM{} 190  &   1.0000   &  1.0000   &  0.9972   &  0.9971   &  0.9970   &  0.9993   &  0.9996   &  0.9748  &   0.9674   &  0.9880  &   0.2475\\
      \sJT{} 50  &   0.9498   &  0.7208   &  0.5794   &  0.2706   &  1.0077   &  0.9769   &  0.6899   &  0.5553  &   0.1634   &  1.0097  &   0.0343\\
      \sJT{} 85  &   0.9635   &  0.7396   &  0.6200   &  0.3131   &  1.0058   &  0.9824   &  0.7419   &  0.6611  &   0.2918   &  1.0127  &   0.1102\\
     \sJT{} 120  &   0.9657   &  0.7541   &  0.6389   &  0.3434   &  1.0068   &  0.9851   &  0.7921   &  0.7239  &   0.4082   &  1.0132  &   0.1630\\
     \sJT{} 155  &   0.9660   &  0.7616   &  0.6490   &  0.3566   &  1.0095   &  0.9866   &  0.8288   &  0.7601  &   0.4850   &  1.0146  &   0.2085\\
     \sJT{} 190  &   0.9652   &  0.7641   &  0.6513   &  0.3565   &  1.0139   &  0.9874   &  0.8575   &  0.7834  &   0.5426   &  1.0159  &   0.2447\\
      \sGV{} 50  &   1.0000   &  1.0000   &  0.9963   &  0.9956   &  0.9951   &  0.9997   &  0.9996   &  0.9721  &   0.9696   &  0.9927  &   0.0361\\
      \sGV{} 85  &   1.0000   &  1.0000   &  0.9966   &  0.9959   &  0.9955   &  0.9997   &  0.9996   &  0.9747  &   0.9687   &  0.9930  &   0.1138\\
     \sGV{} 120  &   1.0000   &  1.0000   &  0.9967   &  0.9959   &  0.9963   &  0.9997   &  0.9996   &  0.9760  &   0.9685   &  0.9920  &   0.1665\\
     \sGV{} 155  &   1.0000   &  1.0000   &  0.9967   &  0.9957   &  0.9968   &  0.9997   &  0.9997   &  0.9774  &   0.9685   &  0.9915  &   0.2116\\
     \sGV{} 190  &   1.0000   &  1.0000   &  0.9966   &  0.9954   &  0.9971   &  0.9997   &  0.9997   &  0.9777  &   0.9684   &  0.9908  &   0.2470\\
      \sSA{} 50  &   1.0000   &  1.0000   &  0.9997   &  0.9999   &  1.0004   &  0.9997   &  0.9996   &  0.9839  &   0.9788   &  0.9878  &   0.0360\\
      \sSA{} 85  &   1.0000   &  1.0000   &  0.9996   &  0.9999   &  1.0005   &  0.9995   &  0.9995   &  0.9830  &   0.9774   &  0.9949  &   0.1140\\
     \sSA{} 120  &   1.0000   &  1.0000   &  0.9996   &  0.9999   &  1.0004   &  0.9994   &  0.9995   &  0.9835  &   0.9770   &  0.9964  &   0.1668\\
     \sSA{} 155  &   1.0000   &  1.0000   &  0.9997   &  0.9999   &  1.0003   &  0.9995   &  0.9996   &  0.9857  &   0.9778   &  0.9969  &   0.2119\\
     \sSA{} 190  &   1.0000   &  1.0000   &  0.9997   &  0.9999   &  1.0003   &  0.9996   &  0.9996   &  0.9868  &   0.9787   &  0.9966  &   0.2473\\
\hline
      \sSY{} 50  &   1.0000   &  1.0000   &  0.9983   &  0.9979   &  1.0006   &  0.9996   &  0.9994   &  0.9571  &   0.9415   &  0.9954   &  0.0114 \\
      \sSY{} 80  &   1.0000   &  1.0000   &  0.9983   &  0.9978   &  1.0008   &  0.9994   &  0.9992   &  0.9634  &   0.9471   &  0.9998   &  0.0356 \\
     \sSY{} 110  &   1.0000   &  1.0000   &  0.9983   &  0.9979   &  1.0006   &  0.9992   &  0.9992   &  0.9623  &   0.9410   &  0.9998   &  0.0606 \\
     \sSY{} 140  &   1.0000   &  1.0000   &  0.9985   &  0.9980   &  1.0005   &  0.9992   &  0.9992   &  0.9670  &   0.9414   &  0.9975   &  0.0837 \\
     \sSY{} 150  &   1.0000   &  1.0000   &  0.9985   &  0.9981   &  1.0005   &  0.9991   &  0.9990   &  0.9690  &   0.9542   &  0.9963   &  0.0913 \\
     \sSY{} 160  &   1.0000   &  1.0000   &  0.9987   &  0.9983   &  1.0005   &  0.9991   &  0.9991   &  0.9660  &   0.9532   &  0.9970   &  0.0809 \\
     \sSY{} 190  &   1.0000   &  1.0000   &  0.9988   &  0.9985   &  1.0005   &  0.9991   &  0.9992   &  0.9650  &   0.9517   &  0.9967   &  0.0862 \\
      \sKM{} 50  &   1.0000   &  1.0000   &  0.9969   &  0.9973   &  0.9964   &  0.9995   &  0.9995   &  0.9713  &   0.9697   &  0.9840   &  0.0114 \\
      \sKM{} 80  &   1.0000   &  1.0000   &  0.9968   &  0.9971   &  0.9964   &  0.9992   &  0.9994   &  0.9690  &   0.9661   &  0.9887   &  0.0355 \\
     \sKM{} 110  &   1.0000   &  1.0000   &  0.9968   &  0.9970   &  0.9966   &  0.9989   &  0.9992   &  0.9662  &   0.9654   &  0.9885   &  0.0601 \\
     \sKM{} 140  &   1.0000   &  1.0000   &  0.9969   &  0.9969   &  0.9968   &  0.9988   &  0.9992   &  0.9699  &   0.9671   &  0.9877   &  0.0820 \\
     \sKM{} 150  &   1.0000   &  1.0000   &  0.9965   &  0.9964   &  0.9968   &  0.9987   &  0.9990   &  0.9654  &   0.9612   &  0.9832   &  0.0880 \\
     \sKM{} 160  &   1.0000   &  1.0000   &  0.9963   &  0.9959   &  0.9968   &  0.9985   &  0.9986   &  0.9585  &   0.9533   &  0.9817   &  0.0782 \\
     \sKM{} 190  &   1.0000   &  1.0000   &  0.9966   &  0.9963   &  0.9969   &  0.9985   &  0.9989   &  0.9626  &   0.9582   &  0.9811   &  0.0826 \\
      \sJT{} 50  &   0.9577   &  0.6606   &  0.5260   &  0.1438   &  1.0201   &  0.9844   &  0.6659   &  0.5489  &   0.0916   &  1.0049   &  0.0135 \\
      \sJT{} 80  &   0.9694   &  0.6525   &  0.5468   &  0.1176   &  1.0225   &  0.9873   &  0.7210   &  0.6565  &   0.2344   &  1.0111   &  0.0406 \\
     \sJT{} 110  &   0.9666   &  0.6500   &  0.5440   &  0.1262   &  1.0229   &  0.9863   &  0.6805   &  0.6754  &   0.1751   &  1.0113   &  0.0661 \\
     \sJT{} 140  &   0.9794   &  0.7475   &  0.6665   &  0.3743   &  1.0027   &  0.9908   &  0.7064   &  0.7771  &   0.3502   &  1.0017   &  0.0848 \\
     \sJT{} 150  &   0.9486   &  0.6605   &  0.4911   &  0.0177   &  1.0546   &  0.9768   &  0.7915   &  0.6735  &   0.2263   &  1.0220   &  0.0799 \\
     \sJT{} 160  &   0.9513   &  0.7466   &  0.5273   &  0.1055   &  1.0873   &  0.9760   &  0.8024   &  0.6717  &   0.3748   &  1.0391   &  0.0821 \\
     \sJT{} 190  &   0.9955   &  0.9477   &  0.8599   &  0.7440   &  1.0001   &  0.9964   &  0.9748   &  0.9002  &   0.8177   &  0.9985   &  0.0829 \\
      \sGV{} 50  &   0.9999   &  1.0000   &  0.9957   &  0.9949   &  0.9934   &  0.9997   &  0.9995   &  0.9728  &   0.9701   &  0.9958   &  0.0116 \\
      \sGV{} 80  &   0.9999   &  1.0000   &  0.9958   &  0.9948   &  0.9933   &  0.9997   &  0.9995   &  0.9710  &   0.9667   &  0.9958   &  0.0354 \\
     \sGV{} 110  &   1.0000   &  1.0000   &  0.9958   &  0.9948   &  0.9943   &  0.9996   &  0.9994   &  0.9714  &   0.9672   &  0.9945   &  0.0599 \\
     \sGV{} 140  &   1.0000   &  1.0000   &  0.9958   &  0.9946   &  0.9953   &  0.9996   &  0.9993   &  0.9753  &   0.9693   &  0.9943   &  0.0817 \\
     \sGV{} 150  &   1.0000   &  1.0000   &  0.9954   &  0.9941   &  0.9955   &  0.9995   &  0.9992   &  0.9703  &   0.9632   &  0.9910   &  0.0877 \\
     \sGV{} 160  &   1.0000   &  1.0000   &  0.9955   &  0.9940   &  0.9957   &  0.9994   &  0.9987   &  0.9635  &   0.9552   &  0.9899   &  0.0780 \\
     \sGV{} 190  &   1.0000   &  1.0000   &  0.9957   &  0.9944   &  0.9960   &  0.9994   &  0.9990   &  0.9676  &   0.9600   &  0.9900   &  0.0825 \\
      \sSA{} 50  &   1.0000   &  1.0000   &  0.9995   &  0.9999   &  1.0006   &  0.9996   &  0.9994   &  0.9830  &   0.9782   &  0.9859   &  0.0115 \\
      \sSA{} 80  &   1.0000   &  1.0000   &  0.9994   &  0.9999   &  1.0008   &  0.9993   &  0.9993   &  0.9795  &   0.9742   &  0.9922   &  0.0355 \\
     \sSA{} 110  &   1.0000   &  1.0000   &  0.9994   &  0.9999   &  1.0006   &  0.9991   &  0.9991   &  0.9748  &   0.9701   &  0.9927   &  0.0601 \\
     \sSA{} 140  &   1.0000   &  1.0000   &  0.9995   &  0.9999   &  1.0005   &  0.9990   &  0.9991   &  0.9773  &   0.9701   &  0.9915   &  0.0822 \\
     \sSA{} 150  &   1.0000   &  1.0000   &  0.9995   &  0.9998   &  1.0005   &  0.9989   &  0.9992   &  0.9756  &   0.9698   &  0.9903   &  0.0892 \\
     \sSA{} 160  &   1.0000   &  1.0000   &  0.9995   &  0.9998   &  1.0005   &  0.9989   &  0.9980   &  0.9731  &   0.9654   &  0.9902   &  0.0786 \\
     \sSA{} 190  &   1.0000   &  1.0000   &  0.9995   &  0.9998   &  1.0004   &  0.9989   &  0.9990   &  0.9759  &   0.9714   &  0.9887   &  0.0831 \\
  \lasthline
 \end{tabular}
\end{table*}
\begin{table*}
 \caption{Same as \tab{lltd} for the \sJdiv{} test cases.
 }
 \label{t:stun_div}
 \centering
 \tiny
 \strtable
 \begin{tabular}{
 @{~}l  |
 c@{\quad}  c@{\quad}  c@{\quad} c@{\quad} c@{\quad}  |
 c@{\quad}  c@{\quad}  c@{\quad} c@{\quad} c@{\quad}  |
 c@{\quad}
       }
\firsthline
 Label  & \multicolumn{5}{c}{$\vBp$ vs $\Nabla \times\vAp$}   &
          \multicolumn{5}{c}{$\vB$  vs $\Nabla \times\vA$}    & $\Hv$ \\
          \cline{2-11}
        & $C_{\rm Vec}$  &$C_{\rm CS}$  &$1-E_{\rm n}$ & $1-E_{\rm m}$& $\epsilon_{\rm E}$  
        & $C_{\rm Vec}$  &$C_{\rm CS}$  &$1-E_{\rm n}$ & $1-E_{\rm m}$& $\epsilon_{\rm E}$  \\ 
  \firsthline
  \sSY{}  divB at  0.2\%  &   1.0000   &  1.0000  &   0.9985  &   0.9981   &  1.0005   &  0.9990   &  0.9979  &   0.9697  &   0.9694  &   0.9854  &   0.0356 \\
  \sSY{}  divB at  1.1\%  &   1.0000   &  1.0000  &   0.9986  &   0.9981   &  1.0005   &  0.9996   &  0.9991  &   0.9727  &   0.9656  &   0.9970  &   0.0359 \\
  \sSY{}  divB at  2.0\%  &   1.0000   &  1.0000  &   0.9986  &   0.9982   &  1.0005   &  0.9995   &  0.9990  &   0.9455  &   0.9272  &   1.0051  &   0.0361 \\
  \sSY{}  divB at  4.0\%  &   1.0000   &  1.0000  &   0.9986  &   0.9984   &  1.0005   &  0.9976   &  0.9951  &   0.8747  &   0.8375  &   1.0137  &   0.0367 \\
  \sSY{}  divB at  8.2\%  &   1.0000   &  1.0000  &   0.9987  &   0.9986   &  1.0005   &  0.9883   &  0.9877  &   0.7532  &   0.6784  &   1.0048  &   0.0380 \\
  \sSY{}  divB at 14.4\%  &   1.0000   &  1.0000  &   0.9989  &   0.9991   &  1.0006   &  0.9529   &  0.9663  &   0.5348  &   0.2345  &   0.8928  &   0.0440 \\
  \sKM{}  divB at  0.2\%  &   0.9994   &  0.9985  &   0.9892  &   0.9879   &  0.9984   &  0.9986   &  0.9964  &   0.9708  &   0.9725  &   0.9936  &   0.0358 \\
  \sKM{}  divB at  1.1\%  &   0.9999   &  0.9997  &   0.9945  &   0.9942   &  0.9971   &  0.9996   &  0.9988  &   0.9805  &   0.9805  &   0.9873  &   0.0359 \\
  \sKM{}  divB at  2.0\%  &   0.9999   &  0.9999  &   0.9959  &   0.9960   &  0.9968   &  0.9993   &  0.9987  &   0.9606  &   0.9567  &   0.9821  &   0.0360 \\
  \sKM{}  divB at  4.0\%  &   0.9991   &  0.9983  &   0.9862  &   0.9852   &  0.9983   &  0.9958   &  0.9785  &   0.9031  &   0.8819  &   0.9737  &   0.0365 \\
  \sKM{}  divB at  8.2\%  &   0.9941   &  0.9946  &   0.9676  &   0.9662   &  1.0083   &  0.9794   &  0.9107  &   0.8044  &   0.7545  &   0.9642  &   0.0380 \\
  \sKM{}  divB at 14.4\%  &   0.9635   &  0.9865  &   0.9277  &   0.9249   &  1.0765   &  0.9196   &  0.8885  &   0.6264  &   0.4642  &   0.9778  &   0.0475 \\
  \sJT{}  divB at  0.2\%  &   0.9996   &  0.9973  &   0.9643  &   0.9381   &  1.0102   &  0.9984   &  0.9956  &   0.9524  &   0.9375  &   0.9925  &   0.0357 \\
  \sJT{}  divB at  1.1\%  &   0.9990   &  0.9927  &   0.9413  &   0.8966   &  0.9945   &  0.9989   &  0.9915  &   0.9384  &   0.8945  &   0.9975  &   0.0356 \\
  \sJT{}  divB at  2.0\%  &   0.9512   &  0.8492  &   0.5392  &   0.2158   &  1.2367   &  0.9707   &  0.8307  &   0.4435  &  -0.0145  &   1.1233  &   0.0382 \\
  \sJT{}  divB at  4.0\%  &   0.9985   &  0.9847  &   0.9554  &   0.9168   &  0.9944   &  0.9967   &  0.9617  &   0.8717  &   0.8182  &   1.0125  &   0.0367 \\
  \sJT{}  divB at  8.2\%  &   0.9866   &  0.9378  &   0.8266  &   0.7542   &  0.9362   &  0.9838   &  0.9029  &   0.7600  &   0.6830  &   0.9685  &   0.0373 \\
  \sJT{}  divB at 14.4\%  &   0.9039   &  0.8344  &   0.5738  &   0.5227   &  0.7202   &  0.9331   &  0.7435  &   0.5131  &   0.2749  &   0.7689  &   0.0395 \\
  \sGV{}  divB at  0.2\%  &   0.9976   &  0.9963  &   0.9834  &   0.9812   &  1.0007   &  0.9989   &  0.9964  &   0.9756  &   0.9752  &   1.0047  &   0.0359 \\
  \sGV{}  divB at  1.1\%  &   0.9996   &  0.9991  &   0.9937  &   0.9933   &  0.9961   &  0.9998   &  0.9988  &   0.9835  &   0.9822  &   0.9965  &   0.0360 \\
  \sGV{}  divB at  2.0\%  &   0.9999   &  0.9997  &   0.9963  &   0.9961   &  0.9952   &  0.9995   &  0.9987  &   0.9624  &   0.9574  &   0.9911  &   0.0361 \\
  \sGV{}  divB at  4.0\%  &   0.9964   &  0.9955  &   0.9797  &   0.9777   &  1.0017   &  0.9959   &  0.9780  &   0.9077  &   0.8835  &   0.9861  &   0.0366 \\
  \sGV{}  divB at  8.2\%  &   0.9760   &  0.9888  &   0.9501  &   0.9484   &  1.0438   &  0.9792   &  0.9110  &   0.8143  &   0.7566  &   0.9963  &   0.0380 \\
  \sGV{}  divB at 14.4\%  &   0.8687   &  0.9772  &   0.8864  &   0.8885   &  1.3247   &  0.9232   &  0.8897  &   0.6463  &   0.4670  &   1.0879  &   0.0475 \\
  \sSA{}  divB at  0.2\%  &   1.0000   &  1.0000  &   0.9996  &   0.9999   &  1.0004   &  0.9990   &  0.9974  &   0.9759  &   0.9780  &   0.9872  &   0.0356 \\
  \sSA{}  divB at  1.1\%  &   1.0000   &  1.0000  &   0.9997  &   0.9999   &  1.0004   &  0.9996   &  0.9992  &   0.9861  &   0.9853  &   0.9885  &   0.0358 \\
  \sSA{}  divB at  2.0\%  &   1.0000   &  1.0000  &   0.9997  &   0.9999   &  1.0004   &  0.9996   &  0.9986  &   0.9773  &   0.9676  &   0.9869  &   0.0361 \\
  \sSA{}  divB at  4.0\%  &   1.0000   &  1.0000  &   0.9996  &   0.9999   &  1.0004   &  0.9978   &  0.9842  &   0.9389  &   0.9174  &   0.9777  &   0.0366 \\
  \sSA{}  divB at  8.2\%  &   1.0000   &  1.0000  &   0.9996  &   0.9998   &  1.0004   &  0.9886   &  0.9694  &   0.8724  &   0.8678  &   0.9393  &   0.0378 \\
  \sSA{}  divB at 14.4\%  &   1.0000   &  1.0000  &   0.9995  &   0.9998   &  1.0005   &  0.9542   &  0.9256  &   0.7519  &   0.7180  &   0.8068  &   0.0429 \\
  \lasthline
 \end{tabular}
\end{table*}
\end{appendix}
\newpage
 \bibliographystyle{aps-nameyear}

\end{document}